\documentclass{article}

\usepackage[pdftex]{graphicx}
\usepackage{adjustbox}
\usepackage{algorithmic}
\usepackage{amsmath,amssymb,amsfonts}
\usepackage{array}
\usepackage{authblk}
\usepackage{bibunits}
\usepackage{booktabs}
\usepackage{caption}
\usepackage{csvsimple}
\usepackage{etoolbox}
\usepackage{hyperref,cleveref}
\usepackage{import}
\usepackage{natbib}[sort]
\usepackage{rotating}
\usepackage[moderate]{savetrees}
\usepackage{siunitx}
\usepackage{subcaption}
\usepackage{textcomp}
\usepackage{xcolor}
\usepackage{lib/alifeconf}
\usepackage{orcidlink}
\usepackage{pdflscape,longtable}
\usepackage{listings}
\usepackage{placeins}
\usepackage{pythonhighlight}

\makeatletter
\let\pragma@iinput=\@iinput
\def\@iinput#1{\xdef\@pragmafile{#1}\pragma@iinput{#1} }
\def\@pragmafile{default}
\def\pragmaonce{%
  \csname pragma@\@pragmafile\endcsname
  \global\expandafter\let \csname pragma@\@pragmafile\endcsname =  
}
\makeatother

\ifdefined\mydraft
  \mydraft
\fi

\defaultbibliography{bibl}
\defaultbibliographystyle{apalike-ejor}

\begin{document}

\title{ Ecology, Spatial Structure, and Selection Pressure Induce Strong Signatures in Phylogenetic Structure }

\author[1,2,3,*]{Matthew Andres Moreno\orcidlink{0000-0003-4726-4479}}
\author[4]{Santiago Rodriguez-Papa\orcidlink{0000-0002-6028-2105}}
\author[4,5]{Emily Dolson\orcidlink{0000-0001-8616-4898}}

\affil[1]{Department of Ecology and Evolutionary Biology, University of Michigan, Ann Arbor, United States}
\affil[2]{Center for the Study of Complex Systems, University of Michigan, Ann Arbor, United States}
\affil[3]{Michigan Institute for Data Science, University of Michigan, Ann Arbor, United States}
\affil[4]{Department of Computer Science and Engineering, Michigan State University, East Lansing, United States}
\affil[5]{Program in Ecology, Evolution, and Behavior, Michigan State University, East Lansing, United States}
\affil[*]{corresponding author: \textit{morenoma@umich.edu}}

\maketitle

\begin{bibunit}

\begin{abstract}
  Evolutionary dynamics are shaped by a variety of fundamental, generic drivers, including spatial structure, ecology, and selection pressure.
  These drivers impact the trajectory of evolution, and have been hypothesized to influence phylogenetic structure.
  For instance, they can help explain natural history, steer behavior of contemporary evolving populations, and influence efficacy of application-oriented evolutionary optimization.
  Likewise, in inquiry-oriented artificial life systems, these drivers constitute key building blocks for open-ended evolution.
  Here, we set out to assess (1) if spatial structure, ecology, and selection pressure leave detectable signatures in phylogenetic structure, (2) the extent, in particular, to which ecology can be detected and discerned in the presence of spatial structure, and (3) the extent to which these phylogenetic signatures generalize across evolutionary systems.
  To this end, we analyze phylogenies generated by manipulating spatial structure, ecology, and selection pressure within three computational models of varied scope and sophistication.
  We find that selection pressure, spatial structure, and ecology have characteristic effects on phylogenetic metrics, although these effects are complex and not always intuitive.
  Signatures have some consistency across systems when using equivalent taxonomic unit definitions (e.g., individual, genotype, species).
  Further, we find that sufficiently strong ecology can be detected in the presence of spatial structure.
  We also find that, while low-resolution phylogenetic reconstructions can bias some phylogenetic metrics, high-resolution reconstructions recapitulate them faithfully.
  Although our results suggest potential for evolutionary inference of spatial structure, ecology, and selection pressure through phylogenetic analysis, further methods development is needed to distinguish these drivers' phylometric signatures from each other and to appropriately normalize phylogenetic metrics.
   With such work, phylogenetic analysis could provide a versatile toolkit to study large-scale evolving populations.
\end{abstract}

\section{Introduction}
\label{sec:introduction}

Within evolutionary biology, the structure of ancestry relationships among organisms within an evolving system is described as their ``phylogeny.''
Phylogenies detail the sequence of historical lineage-branching events that gave rise to contemporary populations.
Most obviously, the phylogenetic record can reveal the ordering and timing of particular contingencies (e.g., trait originations, population separations) that gave rise to current populations \citep{pagel1999inferring,arbogast2002estimating}.
For instance, phylogenies reconstructed from fungal gene sequences suggest frequent switching of nutritional sources and that spore-dispersal mechanisms have multiple independent origins \citep{james2006reconstructing}.

However, phylogenetic analyses can also test more general hypotheses about underlying evolutionary dynamics.
Rather than seeking to detect specific events, researchers using phylogenies for this purpose instead measure structural patterns that may be indicative of different evolutionary regimes.
For instance, analysis of the counts of coexisting lineages over time has been used to detect density-dependent changes in speciation rate \citep{rabosky2008density}, which can be interpreted as an indication of transition away from niche expansion due to lessening availability of ecological opportunity.

Indeed, methods to infer underlying evolutionary dynamics from phylogenetic structure are increasingly harnessed for tangible, impactful applications.
In evolutionary oncology, for example, phylogenetic analysis has been used to quantify patterns of tumor evolution \citep{scottInferringTumorProliferative2020,lewinsohnStatedependentEvolutionaryModels2023}.
Likewise, pathogen phylogenies have been studied to identify dynamics of epidemiological spread of Ebola and HIV \citep{giardina2017inference,saulnier2017inferring,voznica2022deep}.
Phylogenetic information also has a long history of helping identify policy priorities for ecological conservation \citep{forestPreservingEvolutionaryPotential2007}.

In computational evolutionary systems, phylogenetic structure metrics are a powerful tool for summarizing evolutionary history \citep{dolson2020interpreting}.
These analyses can help digital systems further our knowledge of biology, but they are also directly valuable for learning about digital systems themselves.
Notably, within the realm of artificial life research, phylogeny-based metrics have been proposed to identify hallmarks of open-ended evolution \citep{dolson2019modes}.
They have also been used to understand evolutionary trajectories that lead to outcomes of interest \citep{lenskiEvolutionaryOriginComplex2003,lalejiniEvolutionaryOriginsPhenotypic2016,johnsonEndosymbiosisBustInfluence2022a}.
For evolution-inspired optimization algorithms, phylogeny-based metrics have been used to predict which runs of evolutionary computation will be successful \citep{hernandezWhatCanPhylogenetic2022a,shahbandeganUntanglingPhylogeneticDiversity2022a,ferguson2023potentiating} and even proactively guide artificial selection algorithms \citep{lalejini2024phylogeny,burke2003increased}.

Here, we assess the potential of phylogenetic analysis as a means to study fundamental, universal evolutionary dynamics --- ecology, selection pressure, and spatial structure.
Specifically, we consider the following questions,
\begin{enumerate}
  \item Do these dynamics leave detectable signatures in phylogenetic structure?
  \item If so, to what extent are these dynamics' signatures discernable from one another?
  \item To what extent do the structure of these dynamics' signatures generalize across evolutionary systems?
\end{enumerate}

We use \textit{in silico} experiments to investigate these questions, and test across two levels of taxonomic abstraction.
To quantify phylogenetic structure at organism-level abstraction --- meaning that phylogeny is represented at the granularity of individual parent-child relationships --- we use both a minimalistic, forward-time simulation and a rich artificial life system \citep{ofria2004avida}.
In contrast, at species-level abstraction (i.e. species trees), we used a population-level simulation designed to study origins of biodiversity \citep{hagen2021gen3sis}.

A major strength of simulation experiments for this purpose is that they provide exact ground-truth phylogenies due to their perfectly observable evolutionary history.
However, for \textit{in vivo} systems perfect phylogeny tracking is not typically possible. Although new techniques are making increasingly high-fidelity lineage tracking possible in some experiments \citep{nozoe2017inferring,woodworth2017building}, phylogenies are still usually estimated from naturally-occuring biosequence data.
Moreover, as digital evolution systems scale, issues of data loss and centralization overhead make perfect tracking at best inefficient and at worst untenable.
Thus, some systems will likely need to adopt a decentralized, reconstruction-based approach similar to biological data \citep{moreno2024analysis}, which can be achieved through the recently-developed ``hereditary stratigraphy'' methodology \citep{moreno2022hstrat}.

Effective use of phylogenetic metrics in scenarios calculated from estimated phylogenies (as opposed to exact phylogenies) requires consideration of potential confounding effects from inaccuracies introduced by reconstruction.
However, it is in precisely such scenarios where the new capabilities to characterize evolutionary dynamics could have the greatest impact; large-scale systems can produce an intractable quantity of data, making phylometrics valuable as summary statistics of the evolutionary process \citep{dolson2020interpreting}.
In particular, methods are needed to handle large-scale artificial life systems where complete, perfect visibility is not feasible and evolution operates according to implicit, contextually dependent fitness dynamics \citep{moreno2022exploring,kojima2023implementation}.
Thus, an additional goal of this paper is to quantify the magnitude and character of bias introduced by reconstruction error from hereditary stratigraphy.
To this end, we calculate phylometric metrics across a range of reconstruction accuracy levels, and report the level of accuracy necessary to attain metric readings statistically indistinguishable from ground truth.

This work builds on a series of recent studies developing, testing, and applying hereditary stratigraphy methodology since its introduction in \citet{moreno2022hereditary}.
We utilized the open-source \textit{hstrat} software library's hereditary stratigraphy algorithm implementations, which are publicly available via the Python Packaging Index \citep{moreno2022hstrat}.
In perfect-tracking experiments, we collected ground-truth baseline phylogenies using the Phylotrack library \citep{dolson2024phylotrack}. 
\citet{moreno2024guide} characterized trade-offs between memory use, inference precision, and inference accuracy across hereditary stratigraphy configurations, providing a foundation for best practices in applying the methodology.
The present paper focuses on establishing foundations for using herediatry stratigraphy to infer evolutionary dynamics from phylogenetic history.
Specifically, the work extends a conference paper \citet{moreno2023toward}, by adding (1) replications of experiments in full-fledged evolution simulation frameworks (i.e., Avida and Gen3sis, introduced in methods) and (2) refining the set of phylogeny metrics employed.
Findings from present work have facilitated application of hereditary stratigraphy to characterize dynamics in evolution simulations run on the 850,000-core Cerebras Wafer-Scale Engine \citep{moreno2024trackable}, which required engineering a simpler and more efficient algorithmic basis for hereditary stratigraphy \citep{moreno2024structured}.
Although present work considers non-hybridizing phylogenies (i.e., asexual ancestry trees and species trees), methods applying hereditary stratigraphy to sexual populations have been proposed in \citet{moreno2024methods}.

The utility of a model system for experimental evolution hinges on sufficient ability to observe and interpret underlying evolutionary dynamics.
Benchtop and field-based biological models continue to benefit from ongoing methodological advances that have profoundly increased visibility into genetic, phenotypic, and phylogenetic state \citep{woodworth2017building,blomberg2011measuring,schneider2019past}.
Simulation systems, in contrast, have traditionally enjoyed perfect, complete observability \citep{hindre2012new}.
However, ongoing advances in parallel and distributed computing hardware have begun to strain analytical omnipotence, as increasingly vast throughput introduces challenges centralizing, storing, and analyzing data \citep{klasky2021data}.
These trends, although unfolding opposite one another, exhibit intriguing convergence: computational and biological domains will provide data that is multimodal and high-resolution but also incomplete and imperfect.
Such convergence establishes great potential for productive interdisciplinary exchange.

\section{Methods}
\label{sec:methods}

\subsection{Model Systems}

To assess the generality of evolutionary dynamics' effects on phylogenetic structure, we replicated experiments across models differing in subject and approach.
We used three systems,
\begin{enumerate}
\item a simple agent-based model with explicitly encoded fitness values;
\item Avida, a self-replicating agent-based evolutionary platform \citep{ofria2004avida}; and
\item Gen3sis, a population-level macro-ecological/evolutionary model \citep{hagen2021gen3sis}.
\end{enumerate}

The following introduces each model and details experiment-specific configurations.

\noindent
\textbf{Simple Explicit-Fitness Model}

\noindent
Experiments testing the relationships between evolutionary dynamics, reconstruction error, and phylogenetic structure required a model system amenable to direct, interpretable tuning of ecology, spatial structure, and selection pressure.
Finally, a parsimonious and generic model system was desired so that findings would better generalize across digital evolution systems.
The core of this work relies on a simple agent-based model devised to fulfill these objectives.

Genomes in the simple explicit-fitness model comprised a single floating-point value, with higher magnitude corresponding to higher fitness.
Population size 32,768 ($2^{15}$) was used for all experiments.
Selection was performed using tournament selection with synchronous generations.
Treatments' selection pressure was controlled via tournament size.
Mutation was applied after selection, with a value drawn from a unit Gaussian distribution added to all genomes.
Evolutionary runs were ended after 262,144 ($2^{18}$) generations.
Each run required around 4 hours of compute time.

Treatments incorporating spatial structure used a simple island model.
In spatially structured treatments, individuals were evenly divided among 1,024 islands and only competed in selection tournaments against sympatric population members.
Islands were arranged in a one-dimensional closed ring and 1\% of population members migrated to a neighboring island each generation.

Treatments incorporating ecology used a simple niche model.
Population slots were split evenly between niches.
Organisms were arbitrarily assigned to a niche at simulation startup and were only allowed to occupy population slots assigned to that niche.
Therefore, individuals exclusively participated in selection tournaments with members of their own niche.
In treatments also incorporating spatial structure, an even allotment of population slots was provided for every niche on every island.
Every generation, individuals swapped niches with probability $3.0517578125 \times 10^{-8}$ (chosen so one niche swap would be expected every 1,000 generations).

For our main experiments, we defined the following ``regimes'' of evolutionary conditions:
\begin{itemize}
  \item \textit{plain}: tournament size 2 with no niching and no islands,
  \item \textit{weak selection}: tournament size 1 with no niching and no islands,
  \item \textit{strong selection}: tournament size 4 with no niching and no islands,
  \item \textit{spatial structure}: tournament size 2 with no niching and 1,024 islands,
  \item \textit{weak ecology}: tournament size 2 with 4 niches and niche swap probability increased 100$\times$,
  \item \textit{ecology}: tournament size 2 with 4 niches, and
  \item \textit{rich ecology}: tournament size 2 with 8 niches.
\end{itemize}

In follow-up experiments testing ecological dynamics with a spatial background, we defined the following additional evolutionary ``regimes:''
\begin{itemize}
  \item \textit{plain}: tournament size 2 with no niching over 1,024 islands,
  \item \textit{weak ecology}: tournament size 2 with 4 niches and niche swap probability increased 100$\times$ over 1,024 islands,
  \item \textit{ecology}: tournament size 2 with 4 niches over 1,024 islands, and
  \item \textit{rich ecology}: tournament size 2 with 8 niches over 1,024 islands.
\end{itemize}

Finally, to foster generalizability of findings, all experiments were performed with two alternate ``sensitivity'' variables: evolutionary length in generations and mutation operator.
We saved phylogenetic snapshots at 32,768 generation epochs.
This allowed us to test shorter runs of 32,768 and 98,304 generations (through epochs 0 and 2) in addition to the full-length runs (through epoch 7).
One additional mutation operator was tested to contrast the unit Gaussian distribution: the unit exponential distribution.
Under this distribution, deleterious mutations are not possible and large-effect mutations are more likely.

Across all experiments, each treatment comprised 50 independent replicates.
Testing over repeat simulations allowed us to characterize the influence of evolutionary dynamics on phylogenetic structure relative to background stochasticity, including the amount of distributional overlap in phylogeny metrics between treatments (i.e., Cliff's delta statistic, discussed below).

\noindent
\textbf{Avida Model}

\noindent
Avida is a virtual agent-based model system used for sophisticated \textit{in silico} evolution experiments \citep{ofria2004avida}.
Notably, unlike the simple model described above, fitness within Avida arises implicitly from agents' self-replication activity, in relation to other agents' self-replication and the availability of shared exogenous resources.
Within Avida, digital organisms' genomes comprise a sequence of virtual CPU instructions.
Avida conducts open-ended evaluation of each agent's genetic program, affording the opportunity of copying itself into output memory and thereby creating an offspring agent.
Replication imposes an intrinsic baseline level of copy errors (i.e., mutations), ensuring an ongoing supply of genetic variation.
Because self-replicators compete for a limited quantity of population slots, Darwinian evolution ensues.
This scheme induces a complex fitness landscape, which is thought to better reflect the character of biological evolution \citep{adami2006digital}.

Under baseline conditions, Avidians compete on the basis of self-replication efficiency.
However, Avida may be configured to award extra CPU cycles to agents that complete boolean logic tasks.
To create the potential for multiple niches, we associated each task with a depletable resource.
Avidians only recieve CPU cycles for completeing a task if the corresponding resource is available, and completing a task depletes the associated resource.
Importantly, tasks can each be associated with different resources, or they can all be associated with the same resource.
Association of each task to an independent resource creates potential for stable ecological coexistence between task-specialized clades.
We added the constraint that Avidians may harvest at most two tasks with the intention of preventing generalists from disrupting task specialization.

Our baseline ``\textit{plain}'' treatment rewarded four boolean logic tasks (echo, not, nand, and), all drawing from a single resource with inflow rate 400 units per update.
Population structure was well-mixed and, to somewhat weaken the fecundity of high-fitness Avidians, replication destroyed the parent organism with probability 0.2.

We surveyed six additional ``regimes'' of evolutionary conditions, defined accoring to pre-existing Avida configuration options.
Notably, compared to the simple model above, these manipulations are somewhat more indirect and, therefore, potentially weaker in nature.
We anticipate they might better reflect the character of variations in evolutionary drivers that might be encountered in practice.

Regimes were configured as follows,
\begin{itemize}
  \item \textit{weak selection}: the probability of parent death was increased to 0.5;
  \item \textit{strong selection}: parent death probability was set to 0.0 and offspring were protected from replacement by their own offspring;
  \item \textit{spatial structure}: population was arranged as a two-dimensional toroidal grid, with Avidians replicating exclusively between neighboring population sites;
  \item \textit{weak ecology}: tasks were assigned independent resource pools, but three resources were only supplied at an inflow rate of 33 units while the fourth was supplied at 300 units per update,
  \item \textit{ecology}: each task drew from an independent resource pool with inflow rate of 100 units, and
  \item \textit{rich ecology}: 28 tasks were rewarded, each drawing from an independent resource pool with an inflow rate of 100 units.
\end{itemize}

In follow-up experiments, we defined the following additional evolutionary ``regimes'' with two-dimensional toroidal spatial structure,
\begin{itemize}
  \item \textit{plain}: four boolean logic tasks drawing from a single resource with inflow rate 400 units per update,
  \item \textit{weak ecology}: tasks were assigned independent resource pools, but three resources were only supplied at an inflow rate of 33 units while the fourth was supplied at 300 units per update,
  \item \textit{ecology}: each task drew from an independent resource pool with inflow rate of 100 units, and
  \item \textit{rich ecology}: 28 tasks were rewarded, each drawing from an independent resource pool with an inflow rate of 100 units.
\end{itemize}

\begin{figure*}
\begin{subfigure}[b]{1\columnwidth}
\includegraphics[height=0.12\textheight,width=\textwidth]{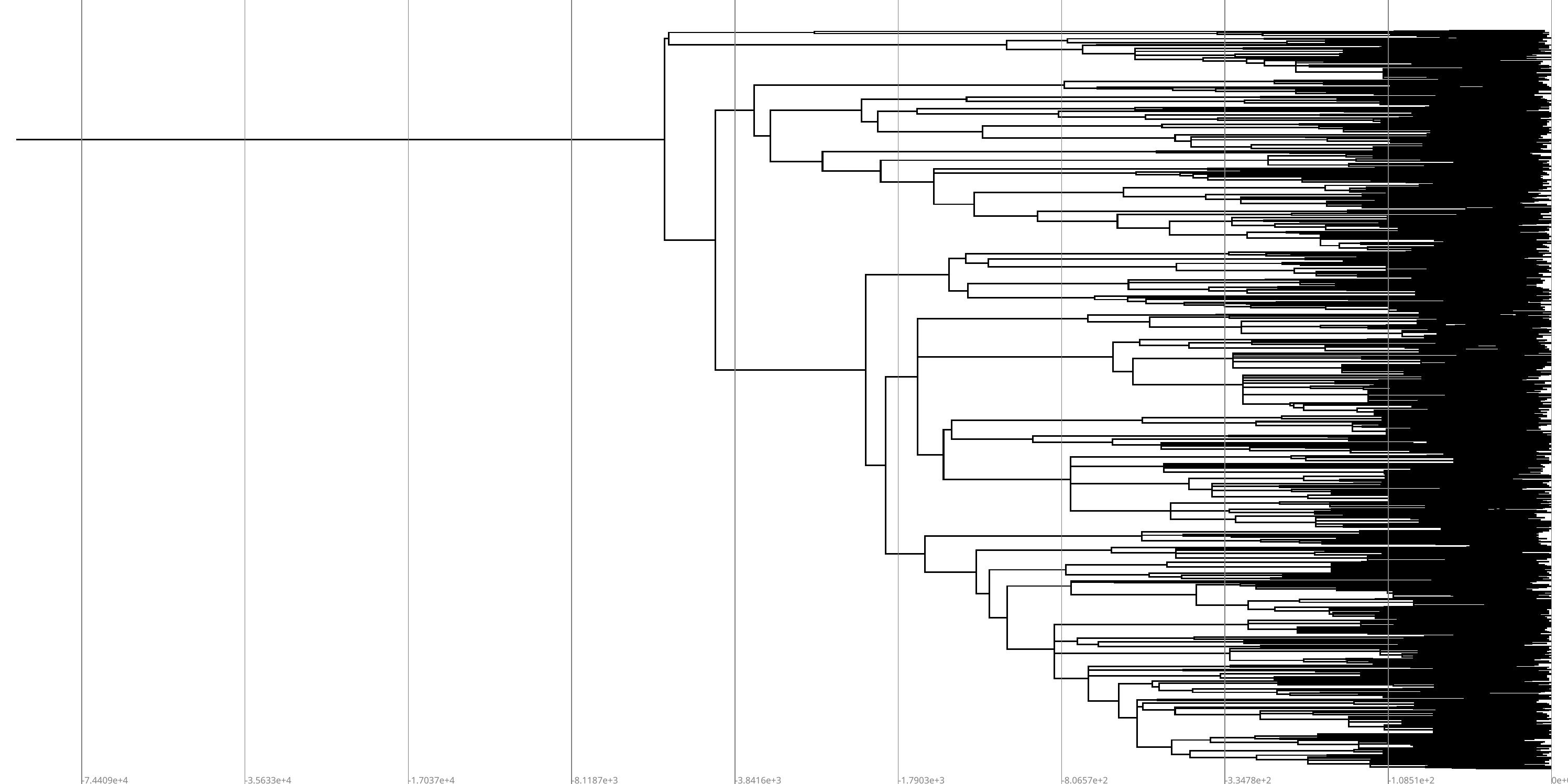}
    \caption{%
      genome-level tracking}
  \end{subfigure}
  \hfill
  \begin{subfigure}[b]{1\columnwidth}
    \includegraphics[height=0.12\textheight,width=\textwidth]{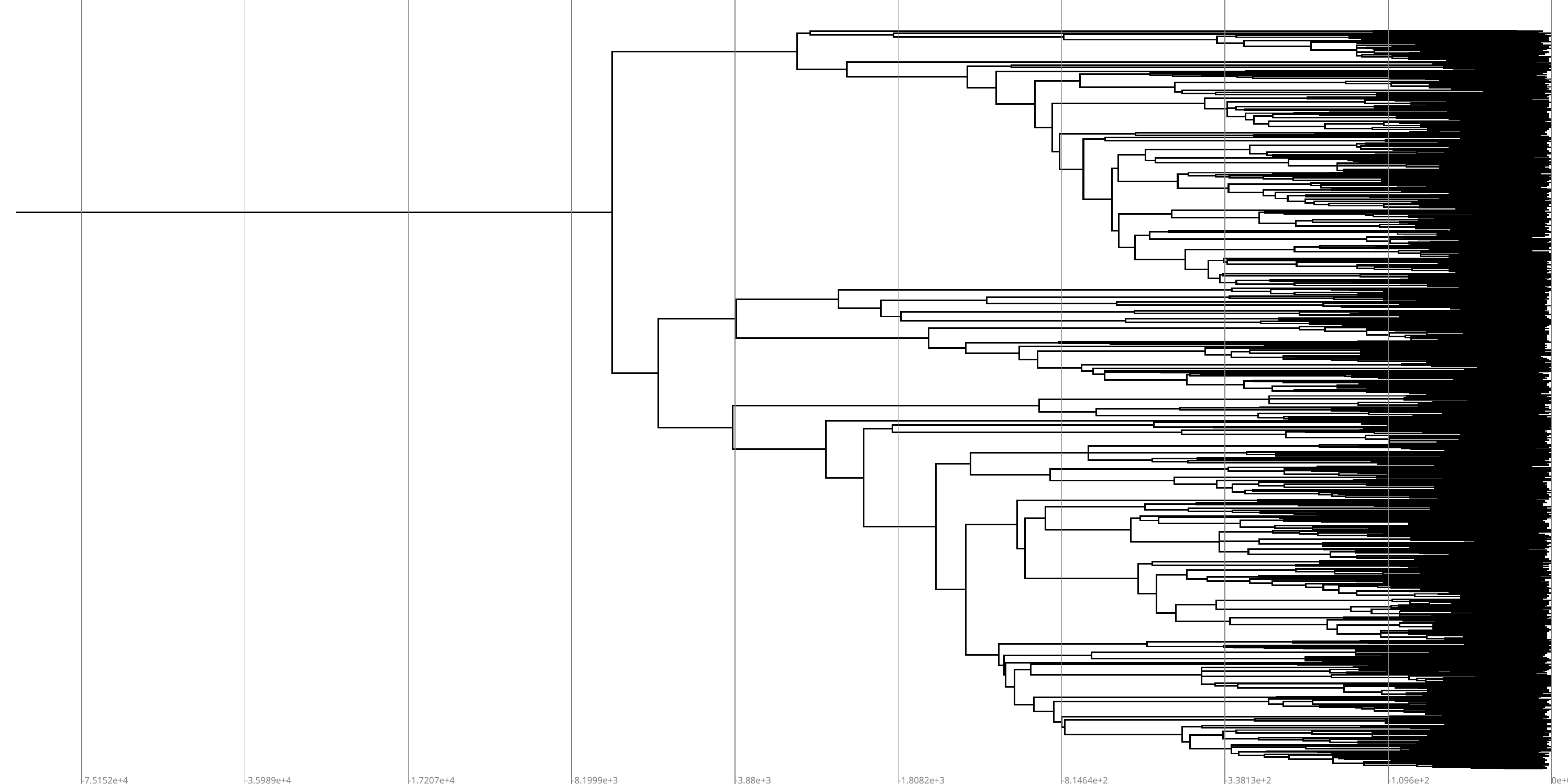}
    \caption{%
      organism-level tracking}
  \end{subfigure}
  \caption{%
    \textbf{Sample reference phylogenies from Avida under ``plain'' regime.}
    Time axis is log-scale.
  }
  \label{fig:perfect-tree-phylogenies-log-avida}
\end{figure*}

Experimental treatments using Avida comprised 30 replicates, conducted with population size 3,600 for durations of 100,000 time steps (c. 20k generations; range 9k-40k).

In order to achieve the phylogenetic tracking necessary for our experiments, we used a fork of Avida originally developed for research on MODES \citep{dolson2019modes}, linked in supplemental materials described below.
This tracking system allowed two configurations: (1) tracking on the level of individual agents, where each Avidian constituted a taxonomic unit, or (2) tracking on the level of genotypes, where clonal sets of Avidians with identical genotypes constituted a taxonomic unit.
Figure \ref{fig:perfect-tree-phylogenies-log-avida} compares example phylogenies from Avida under the ``plain'' treatment, tracked at organism and genome level.
Surprisingly, we noticed several sign-change differences between individual-level and genotype-level tracking with respect to treatments' effects on phylometrics (Figures \ref{fig:perfect-tree-phylometrics-avida-heatmap} and \ref{fig:perfect-tree-phylometrics-heatmap-avida-genome}).
Much, but not all, of the difference related to effects of spatial structure.
These differences may be in part related to occurrences of polytomies within genotype-level tracked phylogenies, as arbitrarily resolving polytomies into sets of bifurcating nodes gave somewhat more similar results to individual-level tracking.
Here, we report results from individual-level tracking, which corresponds to how phylogenies were tracked in the simple model.

\noindent
\textbf{Gen3sis Model}

\noindent
In contrast to Avida and the simple model, which are agent-based, Gen3sis is a population-level model.
Gen3sis abstracts evolution to interactions between spatially-dispersed, speciating subpopulations.
Co-located subpopulations of different species compete for shares of site-specific carrying capacity.
Interaction between subpopulation traits and environmental factors (e.g., aridity, temperature) mediates abundance determinations.
Disjoined subpopulations of the same species accumulate genetic incompatibilities absent sufficient gene flow and, past a defined incompatibility threshold, speciate.
Within this model, the taxonomic unit of phylogeny is species.

Gen3sis also stands out in its intended level of direct biological realism.
The software can be configured to reflect the particular geological and climate histories of continental biomes.
For our experiments, we used curated raster files depicting conditions over 30 million years in South America bundled with the package that had been synthesized from a number of sources \citep{straume2020global,westerhold2020astronomically,fick2017worldclim,hagen2019mountain,annan2013new,cramwinckel2018synchronous,evans2018eocene,hollis2019deepmip,hutchinson2018climate,keating2011warm,sijp2014role,zhang2019evolution}.

Among other features, Gen3sis allows differentiation of aquatic (e.g., lakes, ocean) zones from terrestrial regions with respect to suppression of inter-population migration rates.
To enhance potential for manipulable spatial structure within the model, we converted 30 randomly-selected thin linear segments spanning the map to be aquatic rather than terrestrial (i.e., ``rivers'').
Spatial structure was then induced by adjusting the water traversal cost function.

\begin{figure*}
\centering
\begin{subfigure}[b]{1\columnwidth}
\includegraphics[height=0.12\textheight,width=\textwidth]{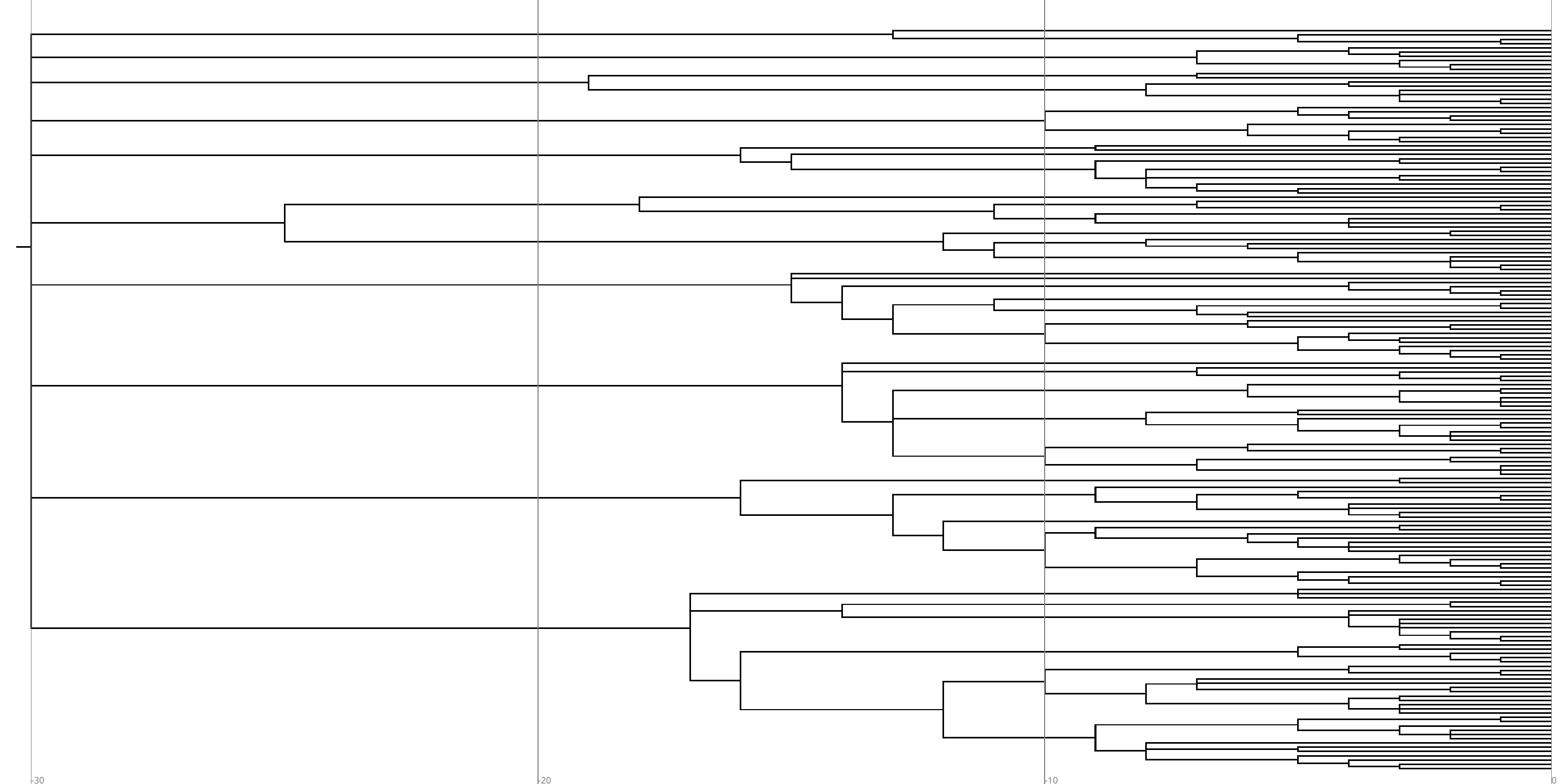}
  \end{subfigure}
  \caption{%
    \textbf{Sample reference phylogeny from Gen3sis under ``plain'' regime.}
    Time axis is linear-scale.
  }
  \label{fig:perfect-tree-phylogeny-gen3sis}
\end{figure*}

We assessed the following evolutionary regimes,
\begin{itemize}
\item \textit{plain}: aquatic terrain imposed no additional migration penalties and population traits did not influence abundances;
\item \textit{spatial structure}: aquatic terrain imposed a migration barrier $50\times$ that of terrestrial terrain;
\item \textit{ecology}: species abundances were determined according to how close populations' trait value matches to site temperature; and
\item \textit{ecology + spatial structure}: the $50\times$ aquatic dispersal penalty was applied, and site temperature was used to determine species' abundances.
\end{itemize}
Figure \ref{fig:perfect-tree-phylogeny-gen3sis} shows an example Gen3sis phylogeny from the ``plain'' regime.

Gen3sis treatments comprised 30 replicates.
Note that, unlike other surveyed models, Gen3sis does not operate with constant ``population size.''
Rather, species count increased continuously over each of the 30 simulated 1 million year time steps as a consequence of ongoing speciation.
Maximum species count per spatial site was configured as 2,500 and within the entire simulation as 25,000, although neither limit was ever reached.

Full Gen3sis configuration files are linked in supplemental materials, described below.
For more on Gen3sis itself, see \citep{hagen2021gen3sis}.

\subsection{Hereditary Stratigraphic Annotations and Tree Reconstruction}

Experiments testing the impact of phylogenetic inference error on phylometrics employ the recently-developed ``hereditary stratigraphy'' technique to facilitate phylogenetic inference \citep{moreno2022hstrat}.
This technique works by attaching heritable annotations to individual digital genomes.
Every generation, a new random ``fingerprint'' is generated and appended to the individuals' inherited annotations.
To reconstruct phylogenetic history, fingerprints from extant organisms' annotations can be compared.
Where two organisms share identical fingerprints along the record, they likely shared common ancestry.
Mismatching fingerprints indicate a split in compared organisms' ancestry.
Although extensions of hereditary stratigraphy to sexual lineages are possible \citep{moreno2024methods}, all lineages used for experiments were asexual in nature.

Hereditary stratigraphy enables a tunable trade-off between annotation size and estimation accuracy.
Fingerprints may be discarded to decrease annotation size at the cost of reduced density of reference points to test for common (or divergent) ancestry along organisms' generational histories.

We test four levels of fingerprint retention.
Each level is described as a $p\%$ ``resolution'' meaning that the generational distance between reference points any number of generations $k$ back is less than $(p / 100) \times k$.
So, a high percentage $p$ indicates coarse resolution and a low percentage $p$ indicates fine resolution.
In detail, at the conclusion of 262,144 generation evolutionary runs,
\begin{itemize}
  \item at 33\% resolution 68 fingerprints are retained per genome,
  \item at 10\% resolution 170 fingerprints are retained per genome,
  \item at 3\% resolution 435 fingerprints are retained per genome, and
  \item at 1\% resolution 1,239 fingerprints are retained per genome.
\end{itemize}

This work uses 1 byte fingerprints, which collide with probability $1/256$.
Greater space efficiency could be achieved using 1 bit fingerprints.
However, this would require careful accounting for ubiquitous generation of identical fingerprints by chance and is left to future work.

Previous work with hereditary stratigraphy used UPGMA distance-based reconstruction techniques \citep{moreno2022hereditary}.
Large-scale reconstructions required for these experiments necessitated development of a more efficient technique that did not require all pairs (i.e., $\mathcal{O}(n^2)$) distance comparison.
To accomplish this, we devised an agglomerative tree building algorithm that works by successively adding leaf organism annotations and percolating them down from the tree root along the tree path of internal nodes consistent with their fingerprint sequence, then affixing them where common ancestry ends.
This new tree-building approach reduced compute time from multiple hours to around 5 minutes in most cases.
Supplementary Listing \ref{lst:build_tree_trie} provides source code with full implementation details, see \citep{moreno2024analysis} for a more detailed discussion.

\begin{figure}
  \centering
  \begin{subfigure}[b]{\linewidth}
    \centering
    \includegraphics[width=\textwidth, height=0.13\textheight]{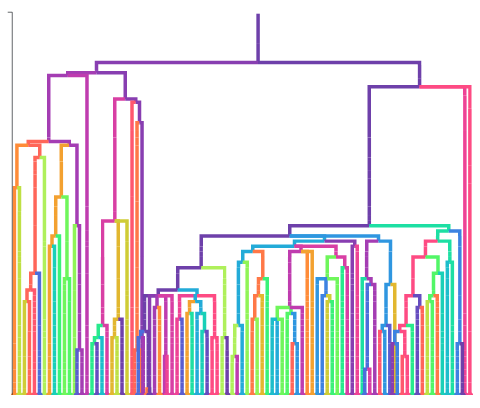}
    \caption{%
      reference tree}
    \label{fig:plain-perfect-and-reconstruction-phylogenies:reference}
  \end{subfigure}
  \begin{subfigure}[b]{\linewidth}
    \centering
    \includegraphics[width=\textwidth, height=0.13\textheight]{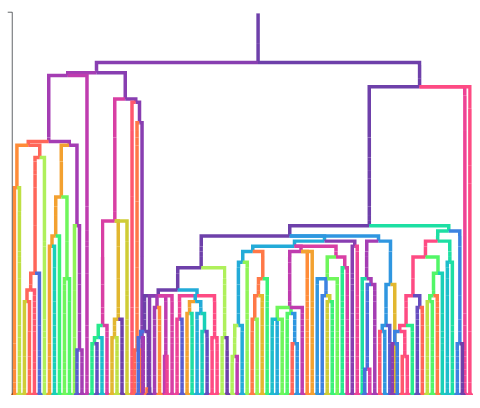}
    \caption{%
      1\% resolution}
    \label{fig:plain-perfect-and-reconstruction-phylogenies:resolution_100}
  \end{subfigure}
  \begin{subfigure}[b]{\linewidth}
    \centering
    \includegraphics[width=\textwidth, height=0.13\textheight]{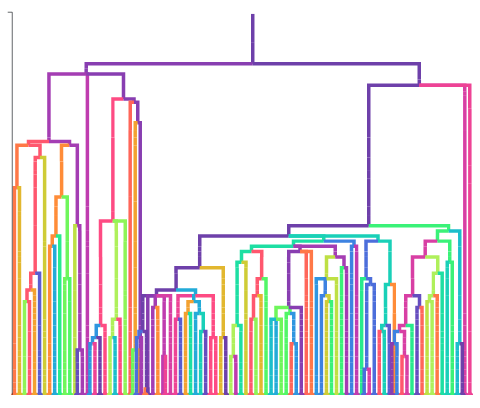}
    \caption{%
      3\% resolution}
    \label{fig:plain-perfect-and-reconstruction-phylogenies:resolution_30}
  \end{subfigure}
  \begin{subfigure}[b]{\linewidth}
    \centering
    \includegraphics[width=\textwidth, height=0.13\textheight]{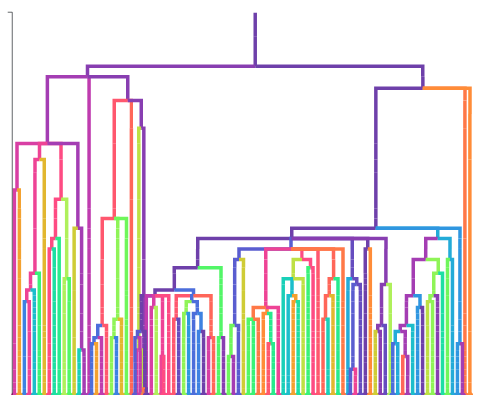}
    \caption{%
      10\% resolution}
    \label{fig:plain-perfect-and-reconstruction-phylogenies:resolution_10}
  \end{subfigure}
  \begin{subfigure}[b]{\linewidth}
    \centering
    \includegraphics[width=\textwidth, height=0.13\textheight]{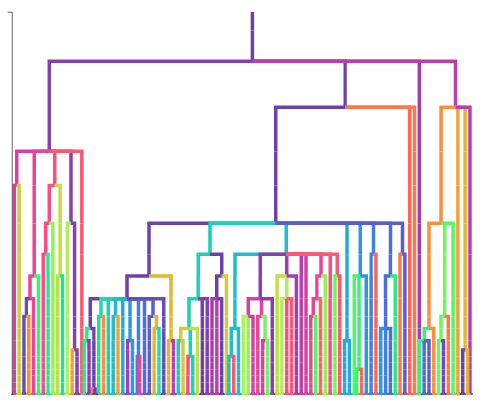} \caption{%
      33\% resolution}
    \label{fig:plain-perfect-and-reconstruction-phylogenies:resolution_3}
  \end{subfigure}
  \caption{%
  \textbf{Comparison of phylogeny reconstructions across different hereditary stratigraphy resolutions in the plain evolutionary regime.}
    To maintain visual legibility, these trees contain the same sub-sample of 100 leaf nodes out of the 32,768 in the full trees.
    Sub-figures are arranged from top to bottom in coarsening order of reconstruction resolution.
    Taxon and branch color coding is consistent across subpanels.
    Visit \url{mmore500.com/hstrat-evolutionary-inference/} for mouseover-based highlighting of corresponding clades between reconstructions and reference.
  }
  \label{fig:plain-perfect-and-reconstruction-phylogenies}
\end{figure}

To assess the efficacy of the new agglomerative tree-building approach, we calculated all reconstructed trees' quartet distance to their respective reference.
Quartet distance ranges from 0 (between identical trees) to 0.75 (between random trees), providing in this case a measure of reconstruction error.
As expected, this measure of reconstruction error varied significantly with resolution for trees across all evolutionary regimes (Kruskal-Wallis tests; all $p < 10^{-20}$; Supplementary Table \ref{tab:reconstruction-error-comparisons-between-resolutions}).
Reconstruction error also varied significantly with evolutionary regime for each reconstruction resolution level (Kruskal-Wallis tests; all $p < 10^{-8}$; Supplementary Table \ref{tab:reconstruction-error-comparisons-between-regimes}).

For 3\% and 1\% resolutions, mean reconstruction error was less than 0.01 in all cases and at 10\% resolution mean reconstruction error was less than 0.05 in all cases.
At 33\% resolution, mean reconstruction error was less than 0.12 in all cases.
The largest reconstruction errors observed at 1\%, 3\%, 10\%, and 33\% resolutions were, respectively, 0.051 (weak selection regime), 0.093 (weak 4 niche ecology regime), 0.14 (plain evolutionary regime), and 0.45 (plain evolutionary regime).
Supplementary Table \ref{tab:tree-reconstruction-quality-quartet-summary-statistics} reports mean, median, standard deviation, and maxima for reconstruction error across surveyed evolutionary conditions.

To generate reconstructed trees in experiments, we simulated the inheritance of hereditary stratigraphic annotations along a reference phylogeny to yield the set of annotations that would be attached to extant population members at the end of a run, then used our agglomerative tree building technique to infer.
Thus, each reconstruction replicate has a directly-corresponding reference tree from a perfect-tree treatment replicate.
Figure \ref{fig:plain-perfect-and-reconstruction-phylogenies} shows a reference tree and corresponding reconstructions performed using 1\%, 3\%, 10\%, and 33\% resolution hereditary stratigraph annotations.

\subsection{Phylometrics}

A wide range of metrics exists for quantifying the topology of a phylogeny.
Tucker et al. showed that these metrics can be classified into the following three dimensions: richness, divergence, and regularity \citep{tuckerGuidePhylogeneticMetrics2017}.
Richness metrics quantify the amount of phylogenetic diversity/evolutionary history represented by a phylogeny.
Divergence metrics quantify how different the units of the phylogeny are from each other.
Regularity metrics quantify the variance of other properties (i.e. how consistent they are across the phylogeny).
Here, we focus on four metrics spread across these categories:

\textbf{Sum Pairwise Distance:} This measurement sums the nodeal distance between each pair of leaf nodes.
It is a metric of phylogenetic richness \citep{tuckerGuidePhylogeneticMetrics2017}, also referred to as \textit{F}  \citep{izsakLinkEcologicalDiversity2000}.
Consequently, we would expect it to be increased by the presence of ecology or spatial structure, as both these factors increase diversity.

\textbf{Colless-like Index:}
The original Colless Index \citep{collessReviewPhylogeneticsTheory1982}, also often referred to as $I_c$ \citep{shaoTreeBalance1990}, is a measure of tree imbalance (i.e., it gets higher as the tree gets less balanced).
In the context of Tucker et al.'s framework, it is a regularity metric.
However, the traditional Colless Index only works for strictly bifurcating trees.
As our trees have multifurcations, we instead use the Colless-like Index, which is an extension of the Colless Index to multifurcating trees \citep{mirSoundCollesslikeBalance2018}.
Tree imbalance is thought to be associated with varying ecological pressures \citep{chamberlainPhylogeneticTreeShape2014, burressEcologicalOpportunityAlters} and has also been observed to increase in the presence of spatial structure \citep{scottInferringTumorProliferative2020}.

\textbf{Mean Pairwise Distance:}
This metric is calculated by computing the shortest distance between all pairs of leaf nodes and taking the mean of these values \citep{webbExploringPhylogeneticStructure2000}.
Note that these distances are measured in terms of the number of nodes in between the pair, not in terms of branch lengths.
Mean pairwise distance is a metric of evolutionary divergence \citep{tuckerGuidePhylogeneticMetrics2017}.
Mean pairwise distance should be increased by scenarios that promote the long-term maintenance of distinct phylogenetic branches, such as ecology.
Conversely, factors that act to reduce diversity should also reduce mean pairwise distance.

\textbf{Mean Evolutionary Distinctiveness:}
Evolutionary distinctiveness is a metric that can be calculated for individual taxa to quantify how evolutionarily different that taxon is from all other taxa in the phylogeny \citep{isaacMammalsEDGEConservation2007}.
To get mean evolutionary distinctiveness, we average this value across all extant taxa in the tree.
Like mean pairwise distance, mean evolutionary distinctiveness is a metric of evolutionary divergence.
However, it is known to capture substantially different information than mean pairwise distance \citep{tuckerGuidePhylogeneticMetrics2017}.
Unlike our other metrics, evolutionary distinctiveness is heavily influenced by branch length.
We generally expect mean evolutionary distinctiveness to be increased by similar factors to mean pairwise distance.

\subsection{Effect-size Analysis}
\label{sec:effect-size-analysis}

We expect statistical tests for between treatments we respect to evolutionary dynamics of interest to serve as an important use case for phylometric analyses in digital evolution.
As such, we wish to report the capability of reported phylometrics to discern between surveyed evolutionary conditions.

Cliff's delta provides useful nonparametric means for such effect size analysis.
This statistic reports the proportion of distributional non-overlap between two distributions, ranging from -1/1 if two distributions share no overlap to 0 if they overlap entirely \citep{meissel2024using,cliff1993dominance}.
When reporting effect size, we use conventional thresholds of 0.147, 0.33, and 0.474 to distinguish between negligible, small, medium, and large effect sizes \citep{hess2004robust}.

Note that the Cliff's delta statistic tops/bottoms out entirely once two distributions become completely separable.
Although this property suits most analyses performed, it is occasionally useful to distinguish the extent of divergence between phylometric distributions past the point of complete separability.
For these purposes, we perform a simple procedure to normalize phylometrics relative baseline conditions by subtracting out the baseline mean and dividing by the baseline standard deviation.

We typically pair effect-size analysis with Mann-Whitney U testing in order to assess the extent to which differences between phylometric readings under different conditions are, or are not, evidenced by available data \citep{mann1947on}.
As a final detail, note that we typically report negated Cliff's delta values where necessary to ensure positive values correspond to larger phylometric values and vice versa.

\subsection{Software and Data Availability}

Software, configuration files, and executable notebooks for this work are available at \href{https://doi.org/10.5281/zenodo.10896667}{doi.org/10.5281/zenodo.10896667}.
Data and supplemental materials are available via the Open Science Framework \url{https://osf.io/vtxwd/} \citep{foster2017open}.
Note that materials for earlier versions of this work are also contained in these repositories \citep{moreno2023toward}.

All hereditary stratigraph annotation, reference phylogeny generation, and phylogenetic reconstruction tools used in this work are published in the \texttt{hstrat} Python package \citep{moreno2022hstrat}.
This project can be visited at \url{https://github.com/mmore500/hstrat}.

This project uses data formats and tools associated with the ALife Data Standards project \citep{lalejini2019data} and benefited from many pieces of open-source scientific software \citep{ofria2020empirical,sand2014tqdist,2020SciPy-NMeth,harris2020array,reback2020pandas,mckinney-proc-scipy-2010,sukumaran2010dendropy,cock2009biopython,dolson2024phylotrackpy,torchiano2016effsize,waskom2021seaborn,hunter2007matplotlib,moreno2024apc,moreno2024qspool,moreno2023teeplot,hagen2021gen3sis,ofria2004avida,torchiano2016effsize}.

\section{Results and Discussion}
\label{sec:results}

\subsection{Phylometric Signatures of Evolutionary Dynamics}

\begin{figure*}
  \begin{minipage}{1\columnwidth}
    \centering
    Generations Ago (approx.)
  \end{minipage}
  \hfill
  \begin{minipage}{1\columnwidth}
    \centering
    Generations Ago (approx.)
  \end{minipage}
  \begin{minipage}{1\columnwidth}
    \hspace{0.02\linewidth}
    \rotatebox{30}{\makebox[0.1\linewidth][c]{200,000}}
    \hfill
    \rotatebox{30}{\makebox[0.1\linewidth][c]{50,000}}
    \hfill
    \rotatebox{30}{\makebox[0.1\linewidth][c]{10,000}}
    \hfill
    \rotatebox{30}{\makebox[0.1\linewidth][c]{2,000}}
    \hfill
    \rotatebox{30}{\makebox[0.1\linewidth][c]{30}}
    \rotatebox{90}{\makebox[0.05\linewidth][c]{0}}
  \end{minipage}
  \hfill
  \begin{minipage}{1\columnwidth}
    \hspace{0.02\linewidth}
    \rotatebox{30}{\makebox[0.1\linewidth][c]{200,000}}
    \hfill
    \rotatebox{30}{\makebox[0.1\linewidth][c]{50,000}}
    \hfill
    \rotatebox{30}{\makebox[0.1\linewidth][c]{10,000}}
    \hfill
    \rotatebox{30}{\makebox[0.1\linewidth][c]{2,000}}
    \hfill
    \rotatebox{30}{\makebox[0.1\linewidth][c]{30}}
    \rotatebox{90}{\makebox[0.05\linewidth][c]{0}}
  \end{minipage}
  \hfill
  \begin{subfigure}[b]{1\columnwidth}
    \includegraphics[height=0.12\textheight,width=\textwidth]{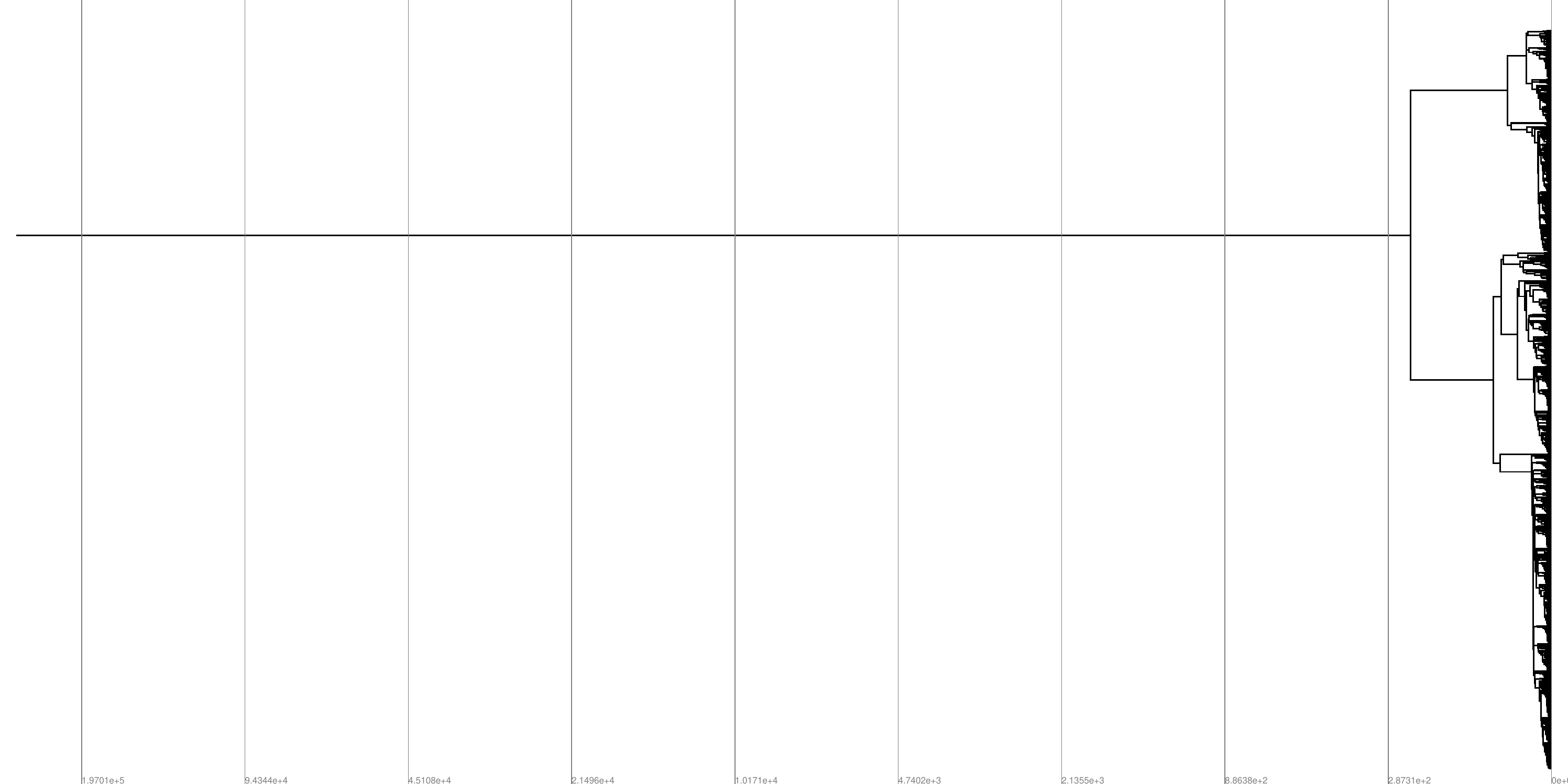}
    \caption{%
      strong selection}
  \end{subfigure}
  \hfill
  \begin{subfigure}[b]{1\columnwidth}
    \includegraphics[height=0.12\textheight,width=\textwidth]{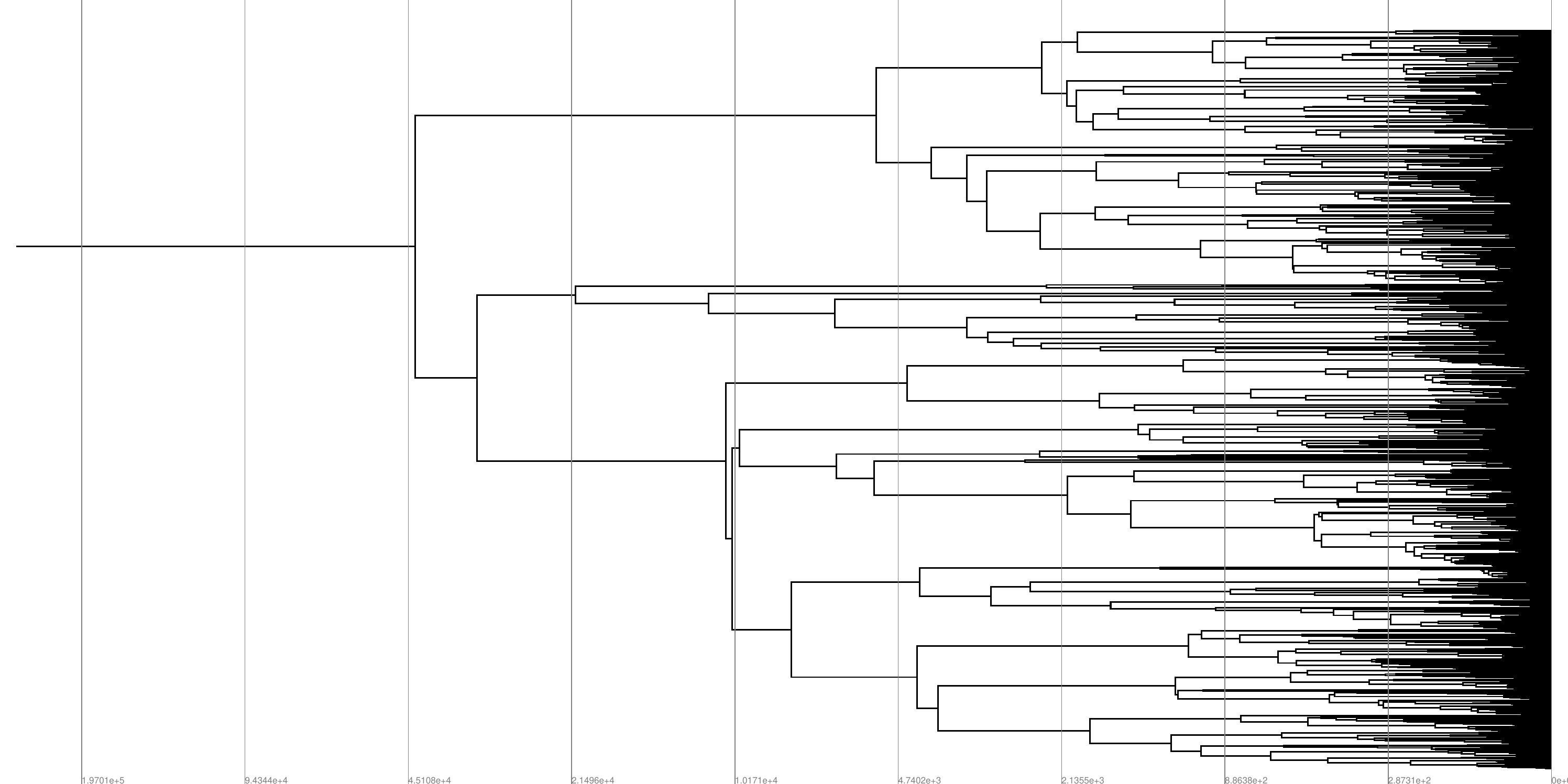}
    \caption{%
      weak selection}
  \end{subfigure}
  \hfill
  \begin{subfigure}[b]{1\columnwidth}
    \centering
    \includegraphics[height=0.12\textheight,width=\textwidth]{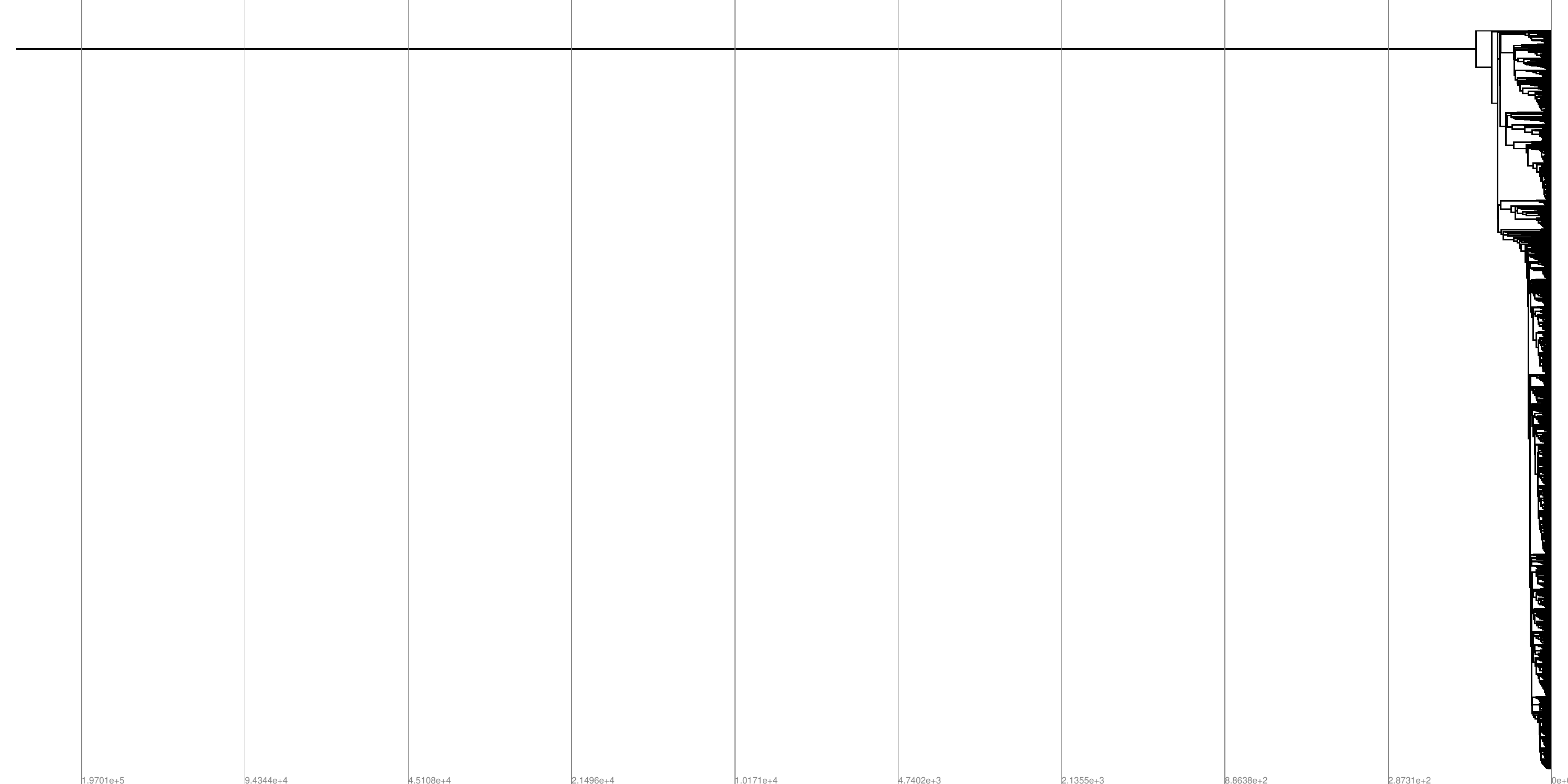}
    \caption{%
      plain}
  \end{subfigure}
  \hfill
  \begin{subfigure}[b]{1\columnwidth}
    \includegraphics[height=0.12\textheight,width=\textwidth]{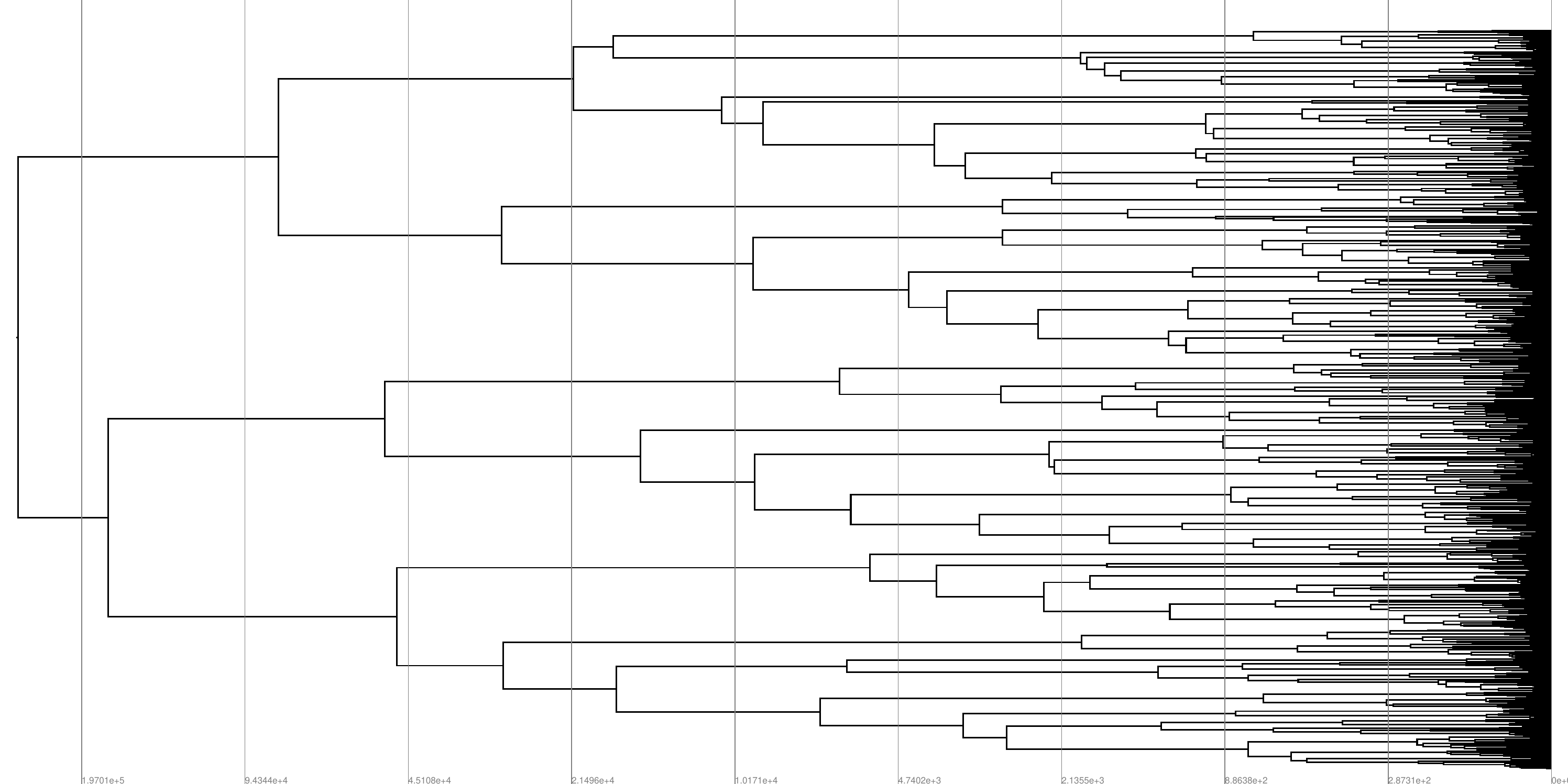}
    \caption{%
      spatial structure}
  \end{subfigure}
  \hfill
  \begin{subfigure}[b]{1\columnwidth}
    \includegraphics[height=0.12\textheight,width=\textwidth]{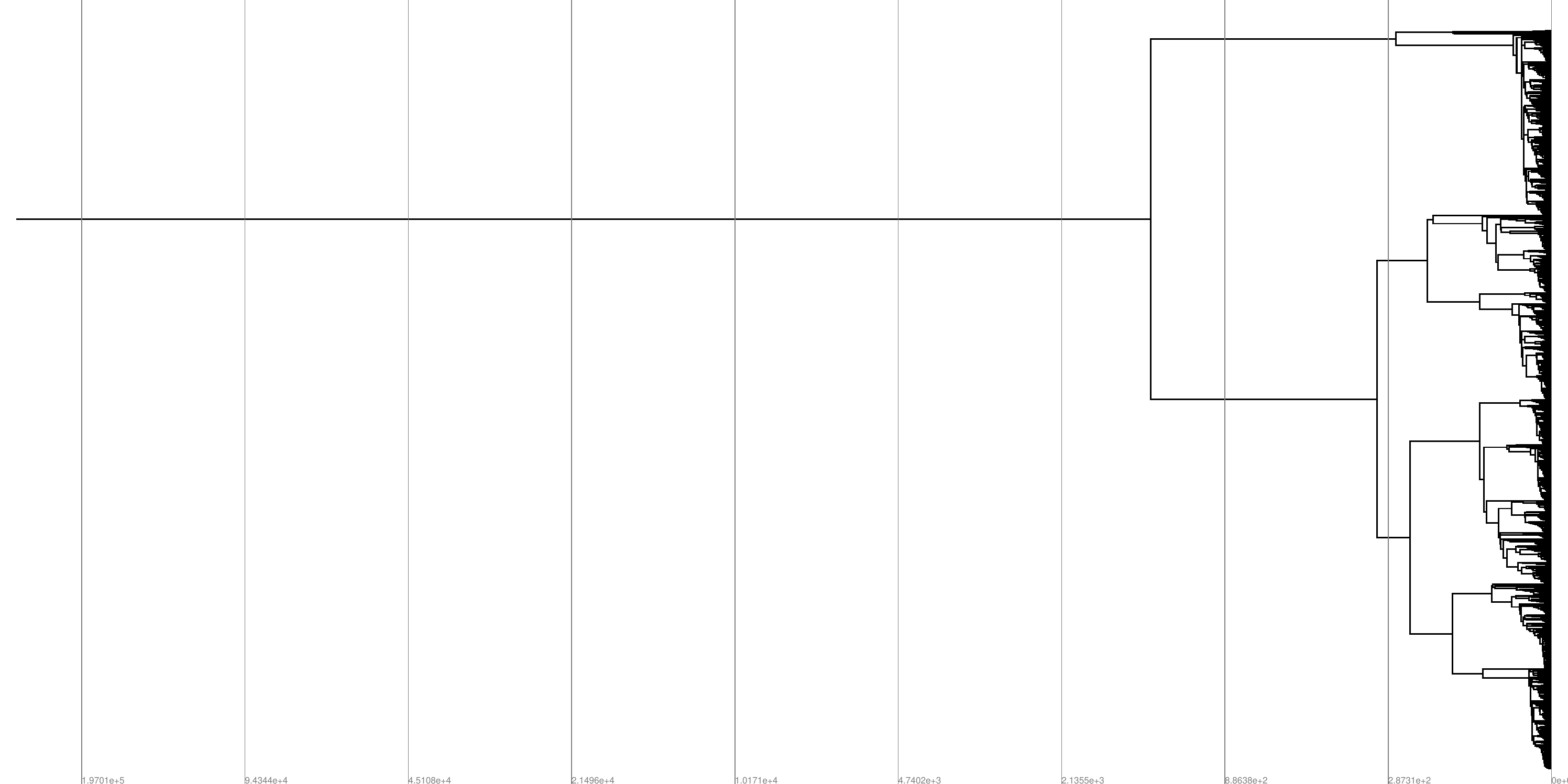}
    \caption{%
      weak 4 niche ecology}
  \end{subfigure}
  \hfill
  \begin{subfigure}[b]{1\columnwidth}
      \includegraphics[height=0.12\textheight,width=\textwidth]{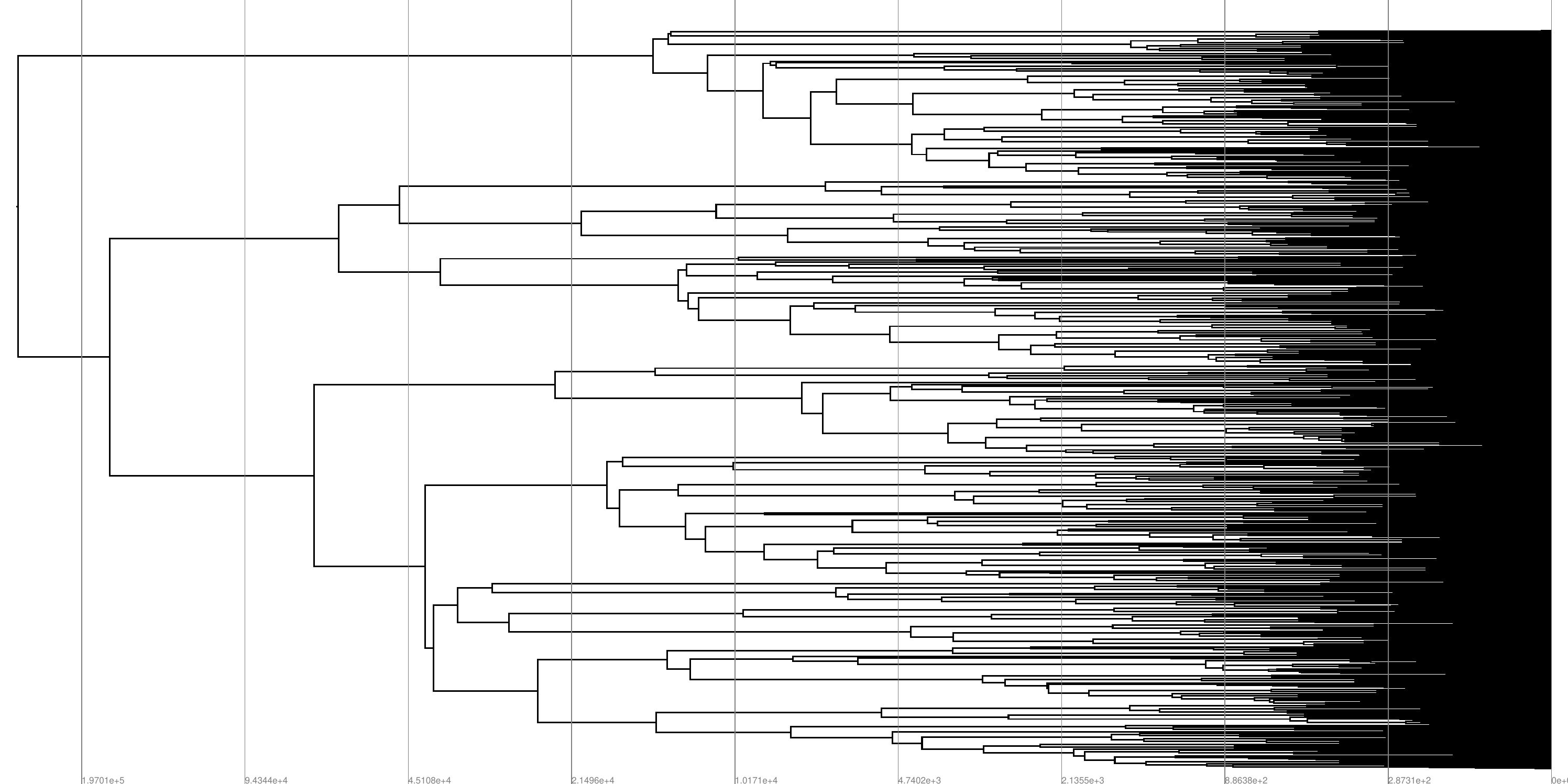}
    \caption{%
      weak 4 niche ecology with spatial structure }
  \end{subfigure}
  \hfill
  \begin{subfigure}[b]{1\columnwidth}
    \includegraphics[height=0.12\textheight,width=\textwidth]{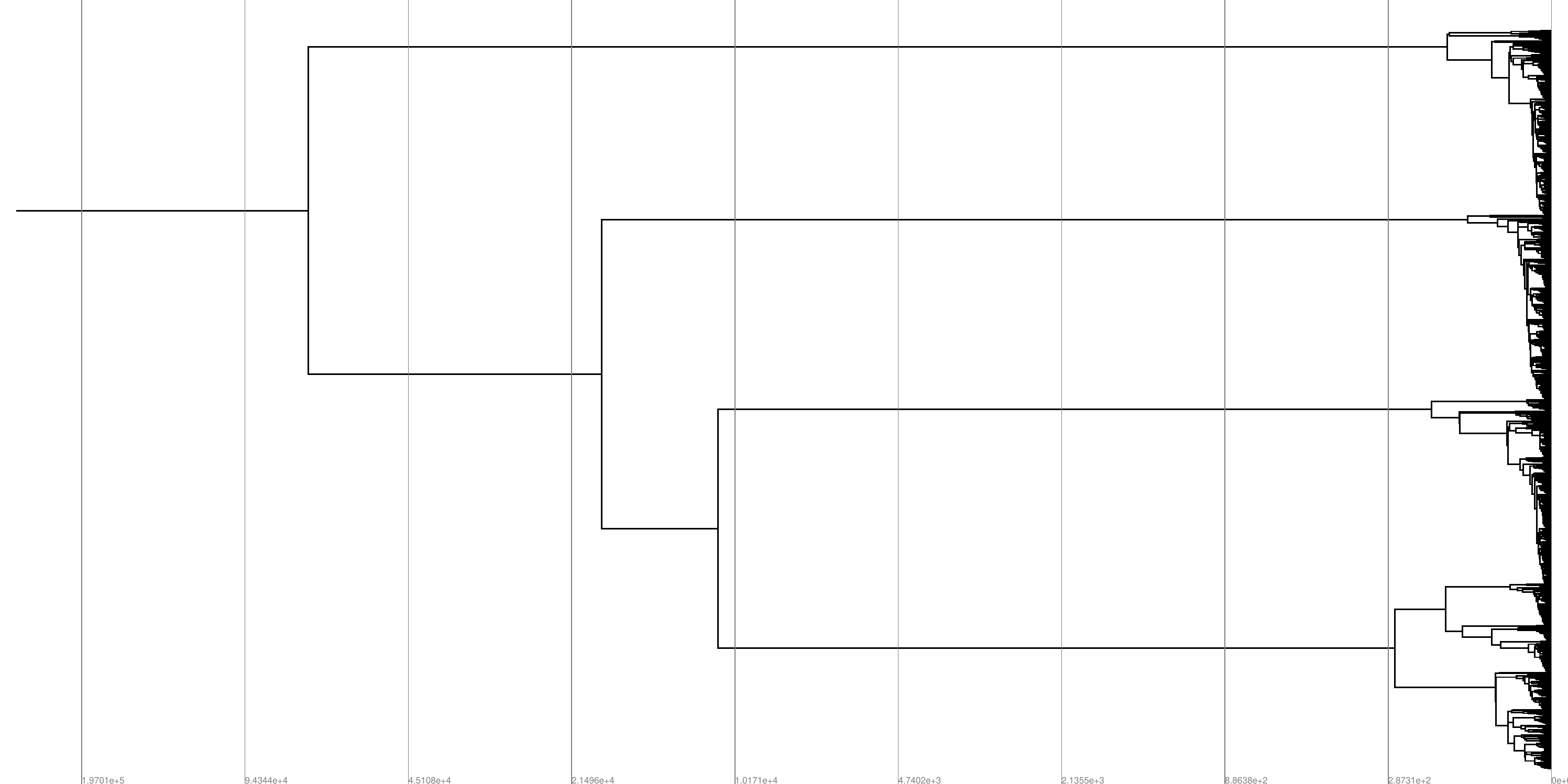}
    \caption{%
      4 niche ecology}
  \end{subfigure}
  \hfill
  \begin{subfigure}[b]{1\columnwidth}
    \includegraphics[height=0.12\textheight,width=\textwidth]{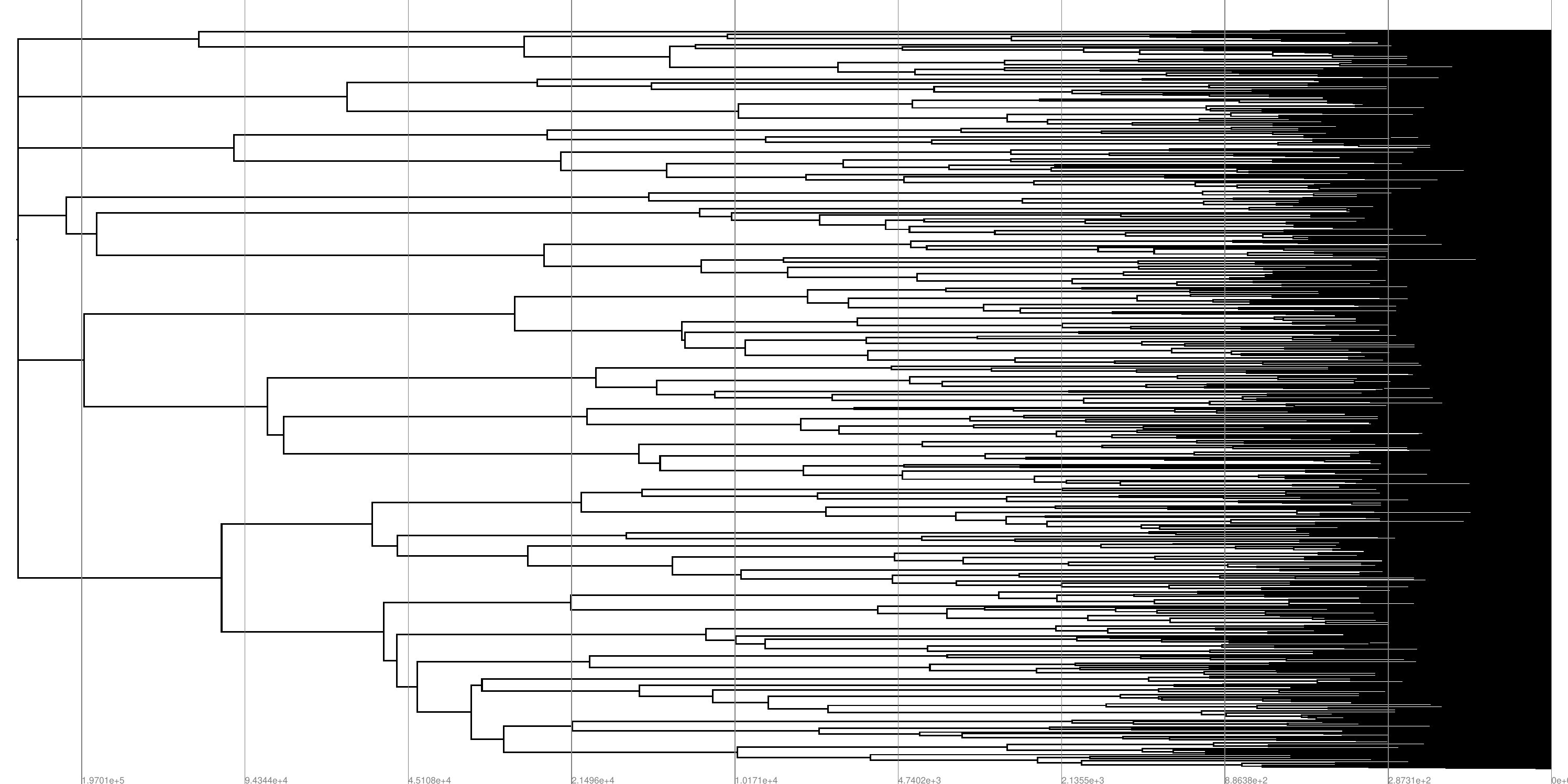}
    \caption{%
      4 niche ecology with spatial structure}
  \end{subfigure}
  \hfill
  \begin{subfigure}[b]{1\columnwidth}
    \includegraphics[height=0.12\textheight,width=\textwidth]{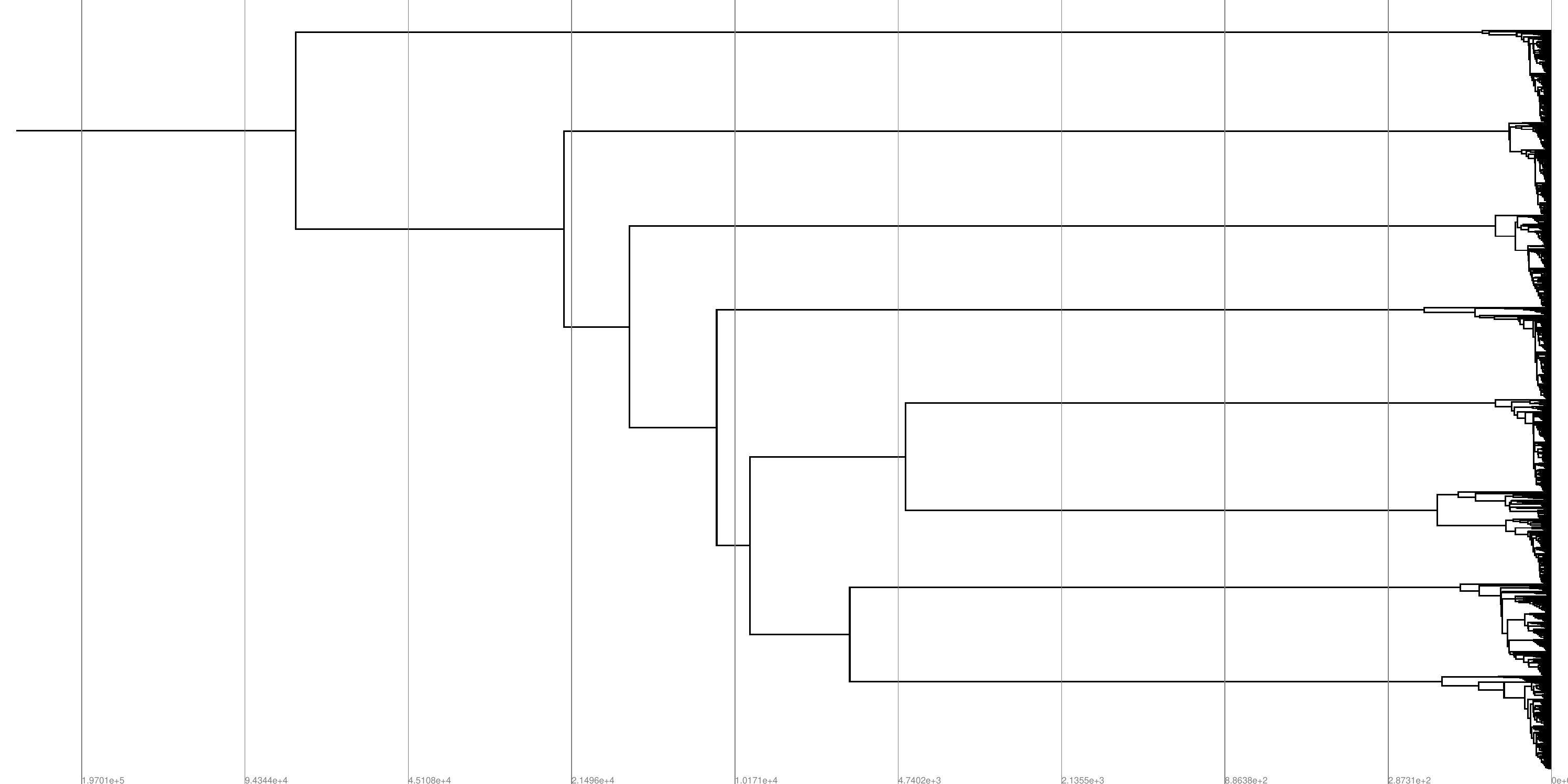}
    \caption{%
      8 niche ecology}
  \end{subfigure}
  \hfill
  \begin{subfigure}[b]{1\columnwidth}
    \includegraphics[height=0.12\textheight,width=\textwidth]{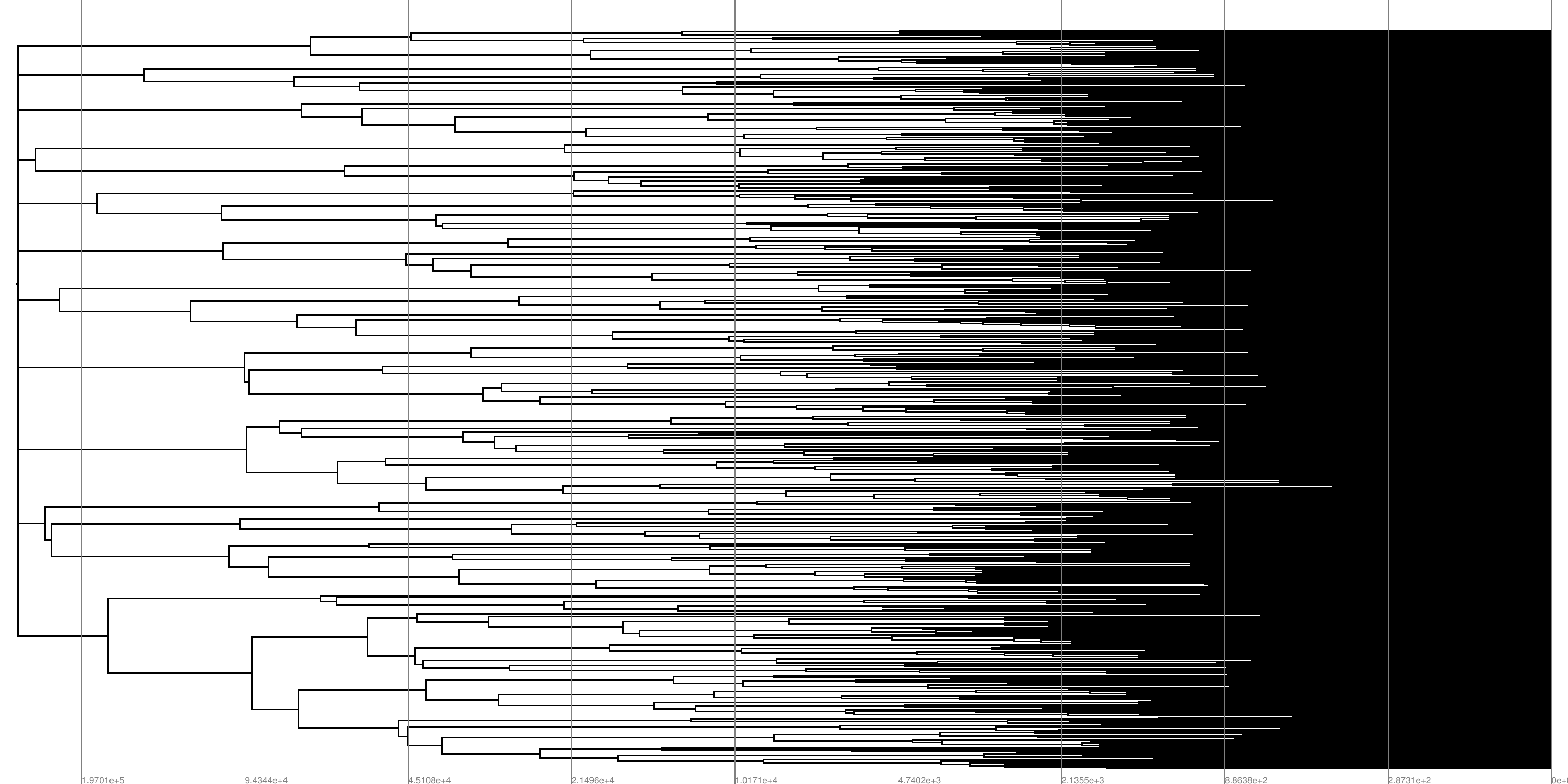}
    \caption{%
      8 niche ecology with spatial structure}
  \end{subfigure}
  \hfill
  \caption{%
    Sample reference phylogenies across surveyed evolutionary metrics.
    Each phylogeny has 32,768 leaves.
    Note log-scale $x$ axis.
  }
  \label{fig:perfect-tree-phylogenies-log}
\end{figure*}

\begin{figure*}
  \centering
  \begin{subfigure}[b]{\textwidth}
    \includegraphics[width=\textwidth]{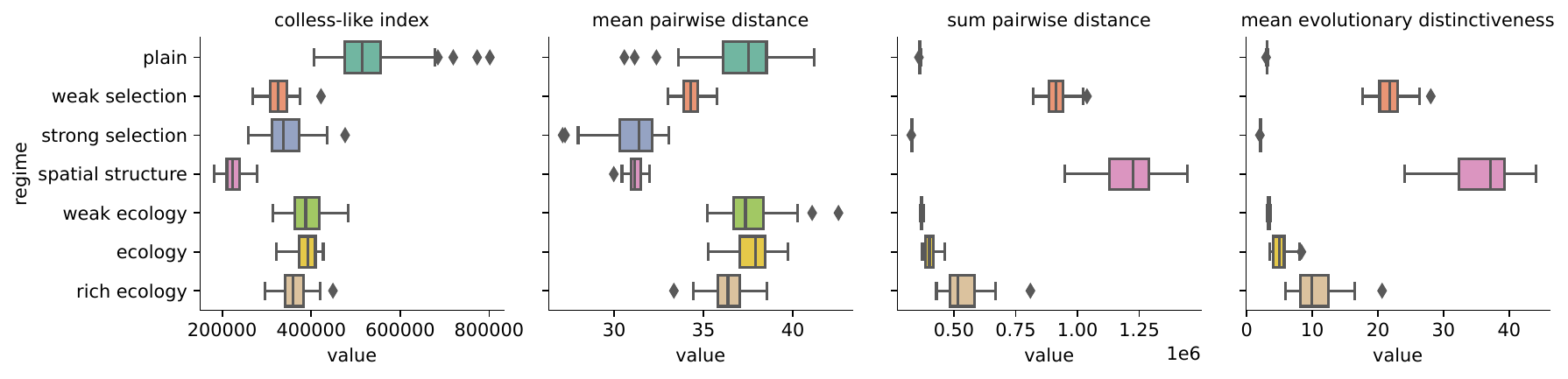}
  \caption{%
    Distribution of phylometrics under the simple model.
    Sample sizes $n=50$.
  }
  \label{fig:perfect-tree-phylometrics-simple-boxplot}
  \end{subfigure}

  \begin{subfigure}[b]{\textwidth}
    \includegraphics[width=\textwidth]{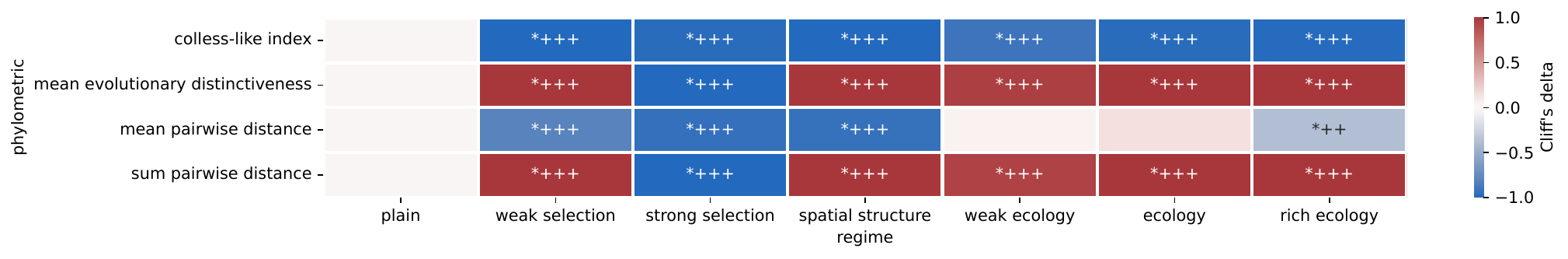}
    \caption{
 Sizes of the effect of evolutionary regimes on each phylometric relative to ``plain'' baseline in the simple model.
    Sample sizes $n=50$.
    }
\label{fig:perfect-tree-phylometrics-simple-heatmap}
  \end{subfigure}%

\begin{subfigure}[b]{\textwidth}
  \includegraphics[width=\textwidth]{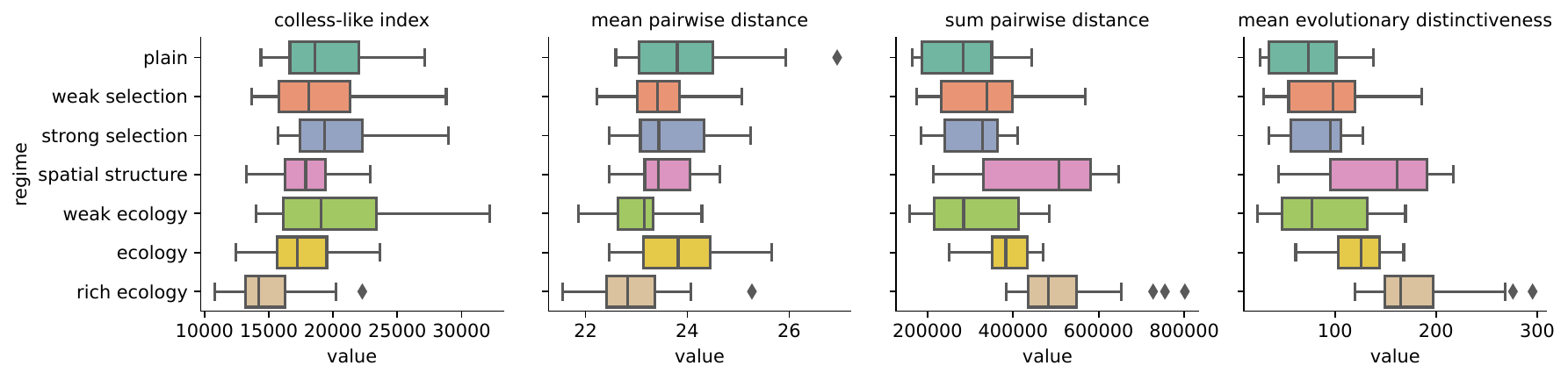}
  \caption{
  Distribution of phylometrics in Avida.
  Sample sizes $n=30$.
  }
  \label{fig:perfect-tree-phylometrics-avida-boxplot}
\end{subfigure}%

\begin{subfigure}[b]{\textwidth}
  \includegraphics[width=\textwidth]{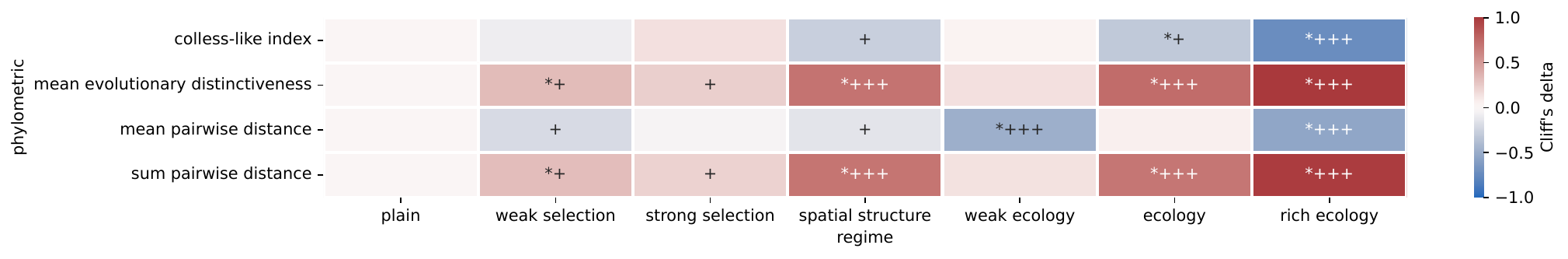}
\caption{%
   Sizes of the effect of evolutionary regimes on each phylometric relative to ``plain'' baseline in Avida. Sample sizes $n=30$.
}
\label{fig:perfect-tree-phylometrics-avida-heatmap}
\end{subfigure}

  \caption{%
  \textbf{Phylometric responses for simple model and Avida.}
  Phylometrics across surveyed evolutionary regimes, calculated on perfect-fidelity individual-level phylogenies from the simple model and Avida.
  Note that nonparametric effect size normalization caps out to 1.0/-1.0 past the point of complete disbributional nonoverlap.
  For heatmap charts, +'s indicate small, medium, and large effect sizes using the Cliff's delta statistic and *'s indicate statistical significance at $\alpha = 0.05$ via Mann-Whitney U test.
  Results from simple model are for standard experimental conditions: gaussian mutation distribution at epoch 7 (generation 262,144).
  See Figure \ref{fig:perfect-tree-phylometrics-sensitivity-analysis} for results under sensitivity analysis conditions.
  }
  \label{fig:perfect-tree-phylometrics}
\end{figure*}

The feasibility of harnessing phylogenetic analysis to identify evolutionary dynamics hinges on the premise that these dynamics induce detectable structure within the phylogenetic record.
Fortunately, as shown in Figure \ref{fig:perfect-tree-phylogenies-log}, dendrograms of phylogenetic histories from the different evolutionary conditions tested do indeed exhibit striking visual differences.

As a first step to characterizing the phylogenetic impact of spatial structure, ecology, and selection pressure, we tested whether surveyed evolutionary conditions exhibited detectable differences in a representative suite of four phylometrics: evolutionary distinctiveness, Colless-like index, mean pairwise distance, and sum pairwise distance.
Figure \ref{fig:perfect-tree-phylometrics} summarizes the distributions of each metric across surveyed conditions.
Statistical tests confirmed that each phylometric exhibited significant variation among surveyed evolutionary conditions for both the simple model and Avida (Kruskal-Wallis tests; all $p < 10^{-40}$; $n=50$ per condition simple model, $n=30$ Avida; Supplementary Table \ref{tab:phylostatistics-comparison-between-regimes-kwallis}).

To quantify the phylometric effects of surveyed evolutionary regimes, we performed nonparametric statistical comparisons against the ``plain'' baseline treatment.
We used a measure of distributional overlap --- Cliff's delta --- to assess effect sizes, binning into ``negligible'', ``small'' (+), ``medium'' (++), and ``large'' (+++) effects based on conventional thresholds \citep{hess2004robust}.
Significance at $\alpha = 0.05$ (*) was assessed through Mann-Whitney tests.
Figure \ref{fig:perfect-tree-phylometrics-simple-heatmap} shows nonparametric significance and effect size test results.

\noindent
\textbf{Summary of Phylometric Effects}

\noindent
Relative to the plain regime, all evolutionary regimes in the simple model depress the Colless-like index.
Reduction in this statistic indicates that all deviations from baseline conditions increased regularity in generated phylogenies.
This observation runs somewhat counter to prior results on similar tree balance metrics, in which the presence of spatial structure increased imbalance \citep{scottInferringTumorProliferative2020}.
One possible contributing factor is that taxa in our phylogenies were individuals, whereas Scott et al. used genotype-level abstraction (i.e., their trees were gene trees).
This possible effect of taxonomic unit is consistent with our results from the species-level phylogenies in the  Gen3sis system (Figure \ref{fig:perfect-tree-phylometrics-gen3sis}), in which ecological and spatial conditions elevated Colless imbalance.
Avida individual-level phylogenies were more consistent with the simple model than with Gen3sis; Colless index was significantly depressed under ecological regimes, and weakly but insignificantly depressed under spatial structure.
However, other modes of evolution did not meaningfully affect Colless-like index of Avida phylogenies.

Colless-like index is sensitive to changes in evolutionary conditions.
However, it appears to be the least useful metric in distinguishing different drivers of evolutionary dynamics, decreasing significantly under all non-plain evolutionary conditions.

Mean evolutionary distinctiveness was significantly higher under weak selection and with spatial structure than in the plain regime.
This metric significantly decreased under strong selection and under ecological regimes, but the numerical magnitudes of these effects were relatively smaller (Figure \ref{fig:perfect-tree-phylometrics-heatmap-parametric}).
We observed similar results in Avida, except no significant effect of strong selection and weak ecology was detected on the phylometric outcome.

Mean pairwise distance was significantly depressed under all regimes except ecology and weak ecology, although again the numerical magnitude of effects on ecological regimes were relatively smaller (Figure \ref{fig:perfect-tree-phylometrics-heatmap-parametric}).
Within Avida, weak but insignificant depressing effects were observed under weak selection and spatial structure regimes.
In contrast to results from the simple model, the strongest depressing effects were observed under the weak ecology and strong ecology regimes.

Finally, sum pairwise distance was significantly increased under all regimes compared to baseline, except for the strong selection regime where it was significantly depressed.
Effect size was again strongest under spatial structure and weak selection (Figure \ref{fig:perfect-tree-phylometrics-heatmap-parametric}).
Avida gave similar results, except that no significant effect was detected from the weak ecology and strong selection treatments, with a weak but insignificant increase effect detected under strong selection.

\noindent
\textbf{Discussion of Phylometric Effects}

\noindent
Ecological dynamics have significant influence on the surveyed phylometrics.
However, the numerical magnitudes of these effects are generally weak compared to spatial structure and selection effects (Figure \ref{fig:perfect-tree-phylometrics-heatmap-parametric}).
So, it appears careful accounting for other evolutionary dynamics (i.e., selection pressure and spatial structure) will be essential to accurate detection of ecology through phylogenetic analysis.
Mean pairwise distance may play a role in identifying ecological dynamics, as ecological dynamics --- in contrast to other factors such as spatial structure and changes in selection pressure --- have weaker effects on this phylometric.
Other phylogenetic metrics may also be better suited to detecting ecological dynamics (e.g. the ecology metric in \citep{dolson2019modes}).

Phylometric outcomes within Avida generally mirror the simple fitness model, although in many cases phylometric effects are weaker or not significant.
These less pronounced effects are not unexpected --- whereas the simple model was explicitly designed for direct manipulation of evolutionary drivers, we use more subtle configurations to impose evolutionary drivers on the Avida model (particularly, with respect to selection pressure).
Notably, the strong selection treatment resulted in no significant effects on phylometric values.
Weak selection significantly increased mean evolutionary distinctiveness and sum pairwise distance, as with the simple model, but the effect size was small.
However, spatial structure's effect size on these metrics was large and agreed with the simple model.
The ecology and rich ecology treatments agreed in sign with the simple model and generally had large effect size.
However, unlike the simple model, Avida's sole phylometric outcome under weak ecology was a strong, significant decrease in mean pairwise distance.
In contrast, the simple model exhibited no effect on this metric under the weak ecology treatment.

\begin{figure*}
  \centering
  \begin{subfigure}[b]{\textwidth}
    \centering
    \includegraphics[width=\textwidth]{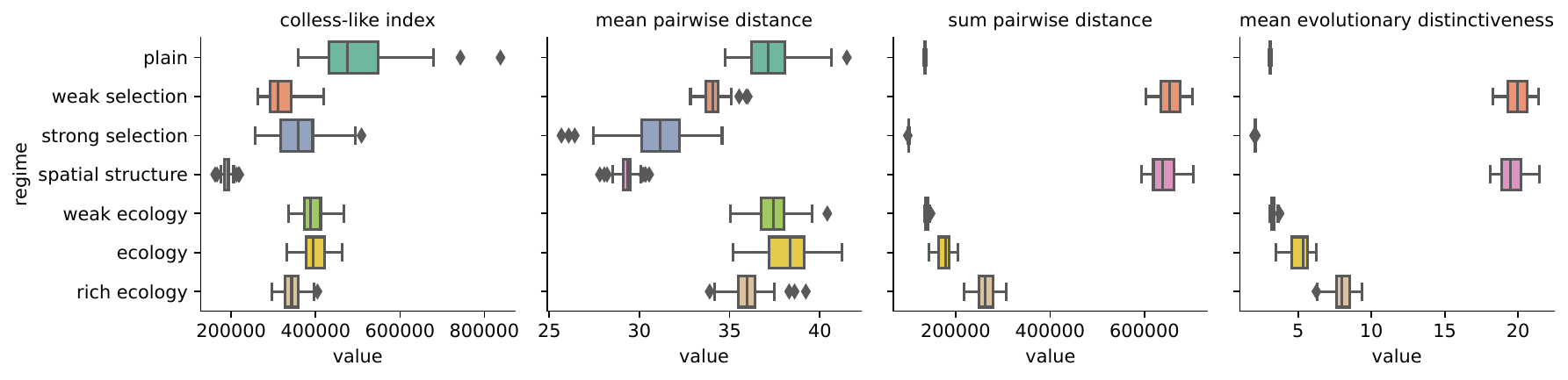}
    \caption{%
      gaussian mutation distribution at epoch 0 (generation 32,768)}
    \label{fig:perfect-tree-phylometrics-sensitivity-analysis:epoch0}
  \end{subfigure}
  \begin{subfigure}[b]{\textwidth}
    \centering
    \includegraphics[width=\textwidth]{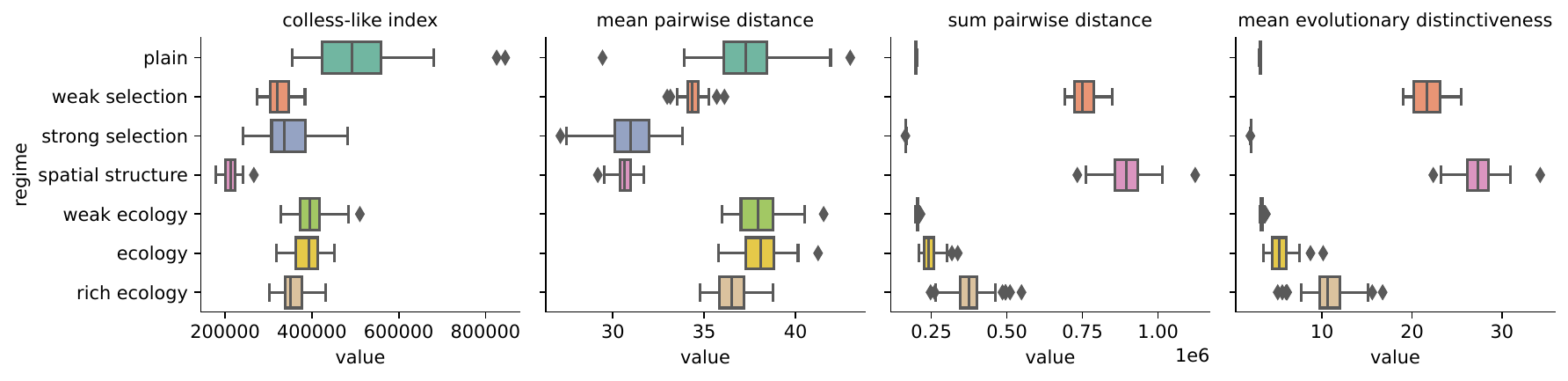}
    \caption{%
      gaussian mutation distribution at epoch 2 (generation 98,304)}
    \label{fig:perfect-tree-phylometrics-sensitivity-analysis:epoch2}
  \end{subfigure}
  \begin{subfigure}[b]{\textwidth}
    \centering
    \includegraphics[width=\textwidth]{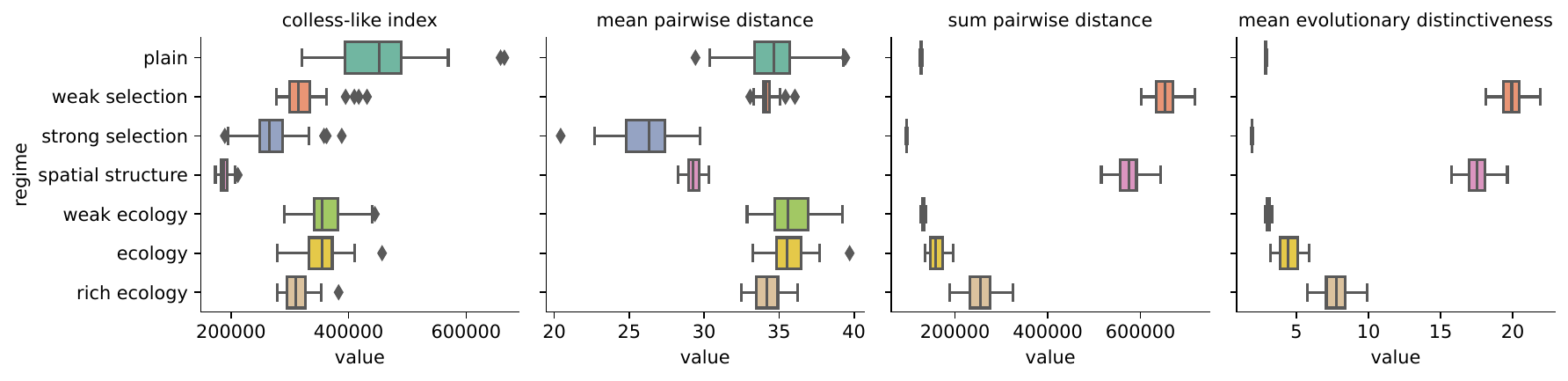}
    \caption{%
      exponential mutation distribution at epoch 7 (generation 262,144)}
    \label{fig:perfect-tree-phylometrics-sensitivity-analysis:exponential}
  \end{subfigure}
  \caption{
    Distribution of tree phylometrics across surveyed evolutionary regimes for sensitivity analysis conditions.
    Phylometrics were calculated on perfect-fidelity simulation phylogenetic records.
    Sample sizes of $n=50$ replicates define each depicted distribution.
  }
  \label{fig:perfect-tree-phylometrics-sensitivity-analysis}
\end{figure*}

\begin{figure*}
  \centering
  \begin{subfigure}[b]{\textwidth}
    \centering
    \includegraphics[width=\textwidth]{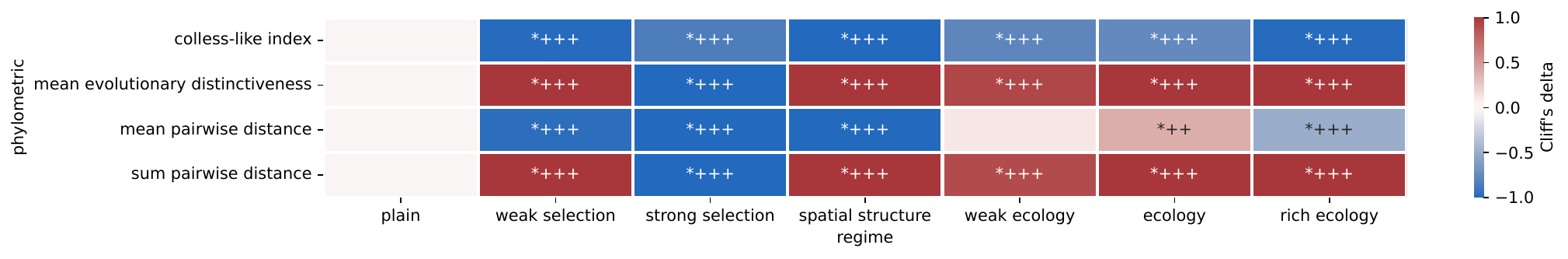}
    \caption{%
      gaussian mutation distribution at epoch 0 (generation 32,768)}
    \label{fig:perfect-tree-phylometrics-heatmap-sensitivity-analysis:epoch0}
  \end{subfigure}
  \begin{subfigure}[b]{\textwidth}
    \centering
    \includegraphics[width=\textwidth]{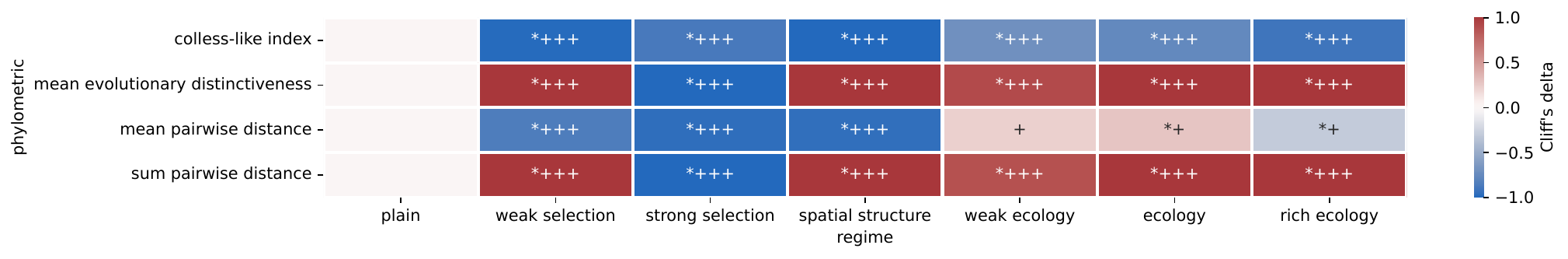}
    \caption{%
      gaussian mutation distribution at epoch 2 (generation 98,304)}
    \label{fig:perfect-tree-phylometrics-heatmap-sensitivity-analysis:epoch2}
  \end{subfigure}
  \begin{subfigure}[b]{\textwidth}
    \centering
    \includegraphics[width=\textwidth]{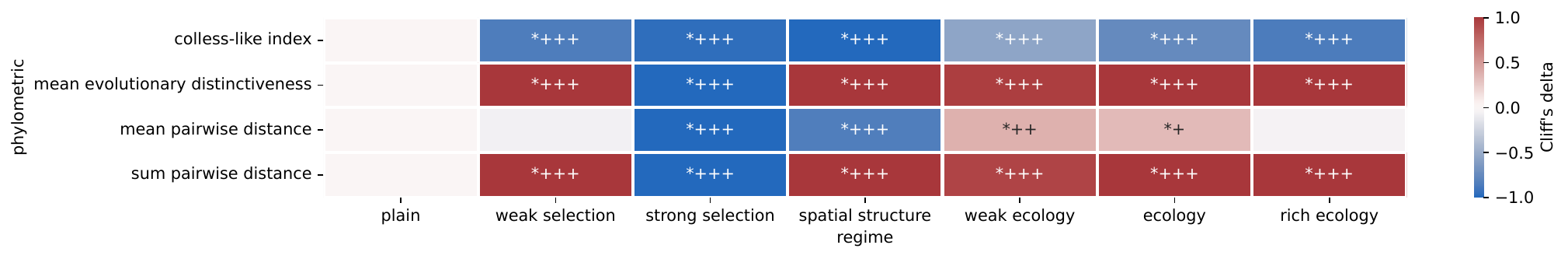}
    \caption{%
      exponential mutation distribution at epoch 7 (generation 262,144)}
    \label{fig:perfect-tree-phylometrics-heatmap-sensitivity-analysis:exponential}
  \end{subfigure}
  \caption{%
    Sensitivity analysis results for normalized tree phylometrics across surveyed evolutionary regimes, calculated on perfect-fidelity simulation phylogenetic records.
    Normalized tree phylometrics are depicted as a heatmap for each sensitivity analysis condition.
  }
  \label{fig:perfect-tree-phylometrics-heatmap-sensitivity-analysis}
\end{figure*}

We additionally performed a sensitivity analysis for results from the simple model over an alternate exponential mutation operator and earlier phylogeny sampling time points.
The effects of evolutionary conditions on phylometrics were generally consistent with Figure \ref{fig:perfect-tree-phylometrics-simple-heatmap} across surveyed conditions (Supplementary Figures \labelcref{fig:perfect-tree-phylometrics-sensitivity-analysis,fig:perfect-tree-phylometrics-heatmap-sensitivity-analysis}).

\subsection{Phylometric Signatures of Ecological Dynamics in Spatially Structured Populations}

\begin{figure*}
  \centering
\includegraphics[width=\textwidth]{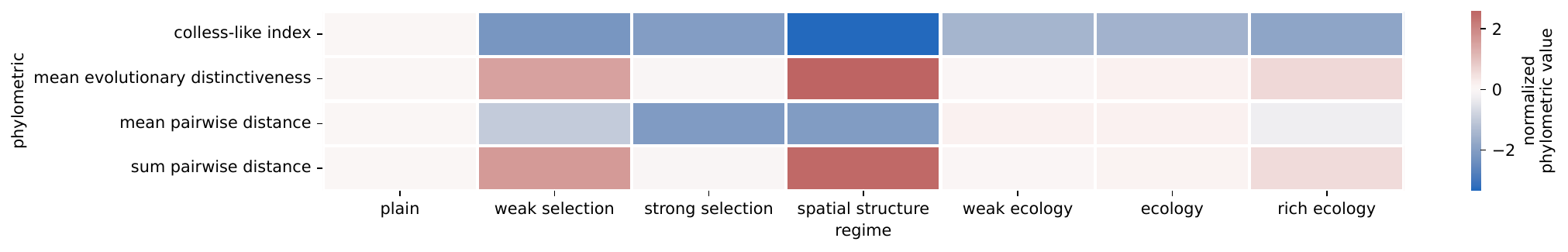}
\caption{%
  \textbf{Normalized phylometric responses.}
  Heatmap of normalized tree phylometrics across surveyed evolutionary regimes, calculated on perfect-fidelity phylogenies from the simple model.
  Note that normalization shows magnitude of phylometric effect beyond the point of distributional nonoverlap, unlike nonparametric normalization which tops out with complete distributional nonoverlap.
}
  \label{fig:perfect-tree-phylometrics-heatmap-parametric}
\end{figure*}

The effects of spatial structure are of particular interest, as evolution within very large populations typically entail elements of spatial structure owing to dispersal across geographic terrain.
Even within \textit{in silico} contexts, populations too large for a single processor will almost inevitably integrate spatial structure that reflects practical limitations of distributed computing hardware \citep{ackley2014indefinitely,moreno2021conduit}.
Therefore, understanding the background effects of spatial structure on the phylogenetic signatures of other evolutionary dynamics will be essential to applications of phylogenetic inference in such applications.
For this analysis, we chose to focus on ecological dynamics due to interest in how their relatively weak phylometric signatures would respond to the relatively strong influence of spatial structure (Figure \ref{fig:perfect-tree-phylometrics-heatmap-parametric}).

\noindent
\textbf{Summary of Phylometric Effects with Background Spatial Structure}

\begin{figure*}
  \centering
\begin{subfigure}[b]{\textwidth}
  \includegraphics[width=\textwidth]{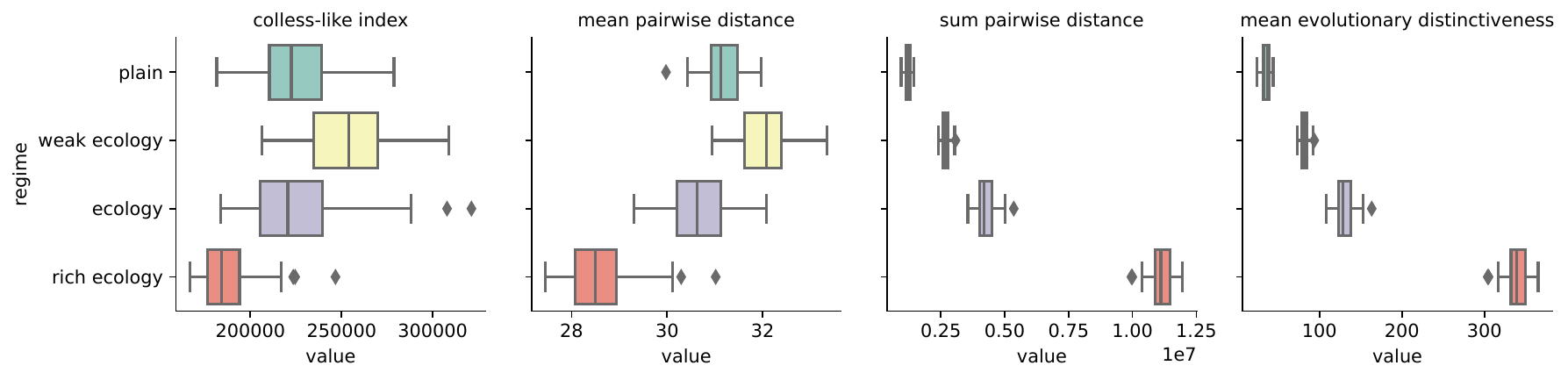}
  \caption{ Distribution of phylometrics under the simple model.
    Sample sizes $n=50$.}
\end{subfigure}

\begin{subfigure}[b]{\textwidth}
  \includegraphics[width=\textwidth]{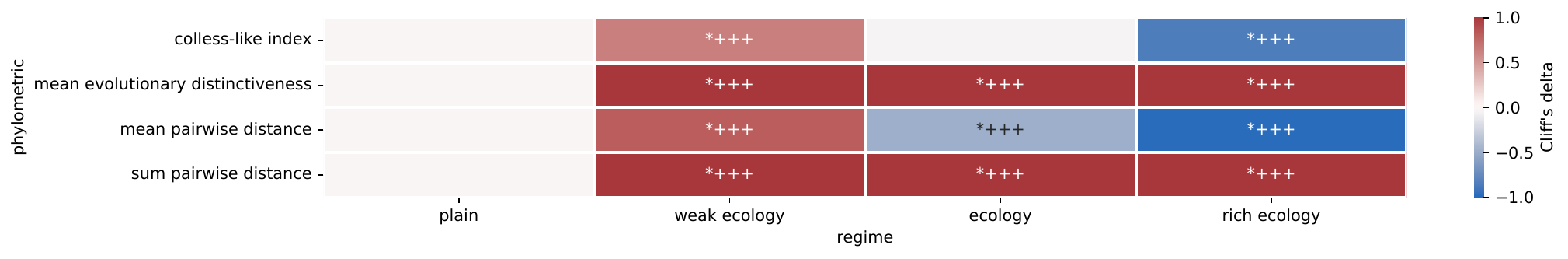}
  \caption{Sizes of the effect of evolutionary regimes on each phylometric relative to ``plain'' baseline in the simple model.
    Sample sizes $n=50$.}
\end{subfigure}

\begin{subfigure}[b]{\textwidth}
  \includegraphics[width=\textwidth]{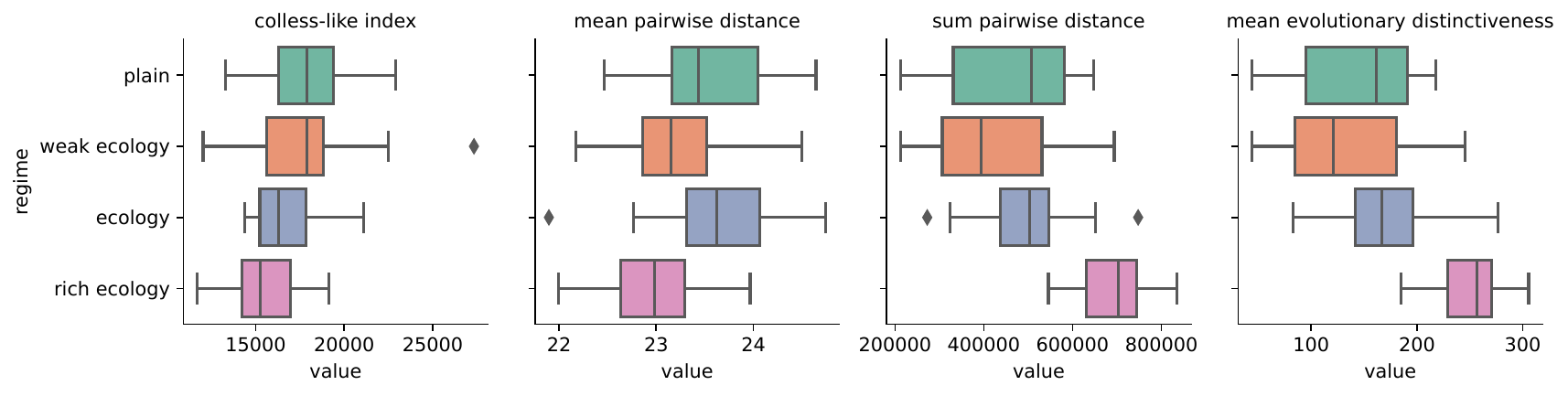}
  \caption{ Distribution of phylometrics in Avida.
  Sample sizes $n=30$.}
\end{subfigure}%

\begin{subfigure}[b]{\textwidth}
  \includegraphics[width=\textwidth]{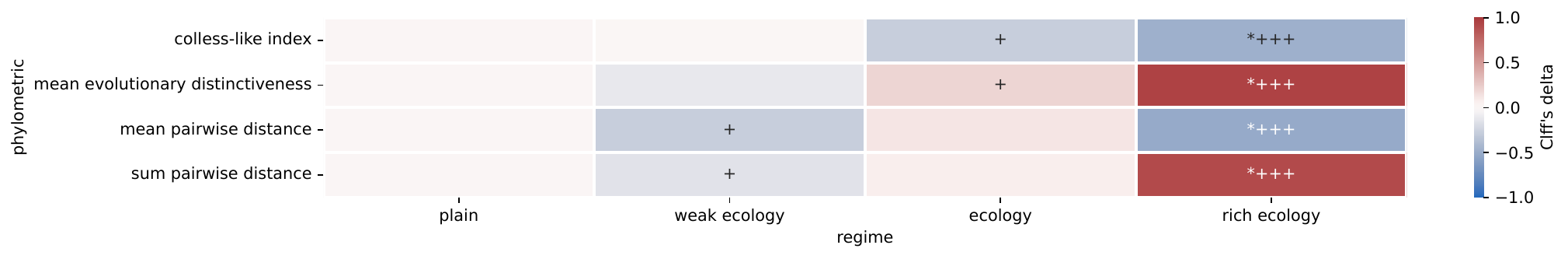}
\caption{%
Sizes of the effect of evolutionary regimes on each phylometric relative to ``plain'' baseline in Avida.
    Sample sizes $n=30$.
}
\end{subfigure}
\vspace{1cm}

  \caption{%
    \textbf{Phylometric responses under spatial structure.}
    Distribution of phylometrics across the three surveyed ecological regimes and the control non-ecological regime---all with spatial population structure (i.e., island count 1,024 for simple model, toroidal population grid for Avida).
    Phylometrics were calculated on perfect-fidelity phylogenies.
    Note that nonparametric effect size normalization caps out to 1.0/-1.0 past the point of complete disbributional nonoverlap.
    For heatmap charts, +'s indicate small, medium, and large effect sizes using the Cliff's delta statistic and *'s indicate statistical significance at $\alpha = 0.05$ via Mann-Whitney U test.
    Results from simple model are for standard experimental conditions: gaussian mutation distribution at epoch 7 (generation 262,144).
    See Figure \ref{fig:perfect-tree-phylometrics-with-spatial-nuisance-sensitivity-analysis} for results under sensitivity analysis conditions.
  }
  \label{fig:perfect-tree-phylometrics-with-spatial-nuisance}
\end{figure*}

\noindent
Figure \ref{fig:perfect-tree-phylometrics-with-spatial-nuisance} summarizes the distribution of surveyed phylometrics under the three surveyed ecological regimes and the control non-ecological regime, all with spatial population structure.
Statistical tests confirmed that each phylometric exhibited significant variation among these evolutionary regimes, indicating the presence of detectable structural signatures in phylogenetic structure (Kruskal-Wallis tests; all $p < 1\times10^{-8}$; $n=50$ per condition for simple model, $n=30$ Avida; Supplementary Tables \labelcref{tab:phylostatistics-comparison-between-regimes-spatial-nuisance-kwallis,tab:phylostatistics-comparison-between-regimes-spatial-nuisance-kwallis-avida}).

As in prior experiments, all ecology treatments drove significant increases in mean evolutionary distinctiveness and sum pairwise distance.
Also consistent with spatially unstructured results, rich ecology drove significant, large-effect depression of both Colless-like index and mean pairwise distance and ecology.
However, in the presence of spatial structure, the ecology treatment depressed only mean pairwise distance.
Without spatial structure, the ecology treatment depressed Colless-like index instead.
Results under weak ecology differed notably from non-spatial baseline, with all phylometrics seeing significant, large-effect increases.
Spatial-background rich ecology results from Avida agreed with the simple model.
However, significant phylometric effects were not detected from the ecology and weak ecology treatments under spatial structure conditions.

\begin{figure*}
  \centering
  \begin{subfigure}[b]{\textwidth}
    \centering
    \includegraphics[width=\textwidth]{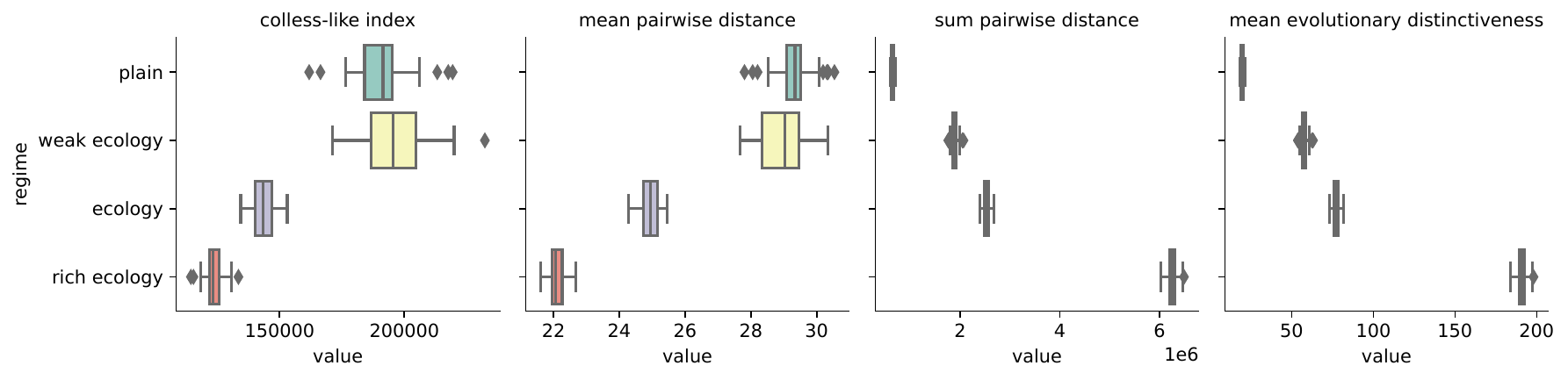}
    \caption{%
      gaussian mutation distribution at epoch 0 (generation 32,768)}
    \label{fig:perfect-tree-phylometrics-with-spatial-nuisance-sensitivity-analysis:epoch0}
  \end{subfigure}
  \begin{subfigure}[b]{\textwidth}
    \centering
    \includegraphics[width=\textwidth]{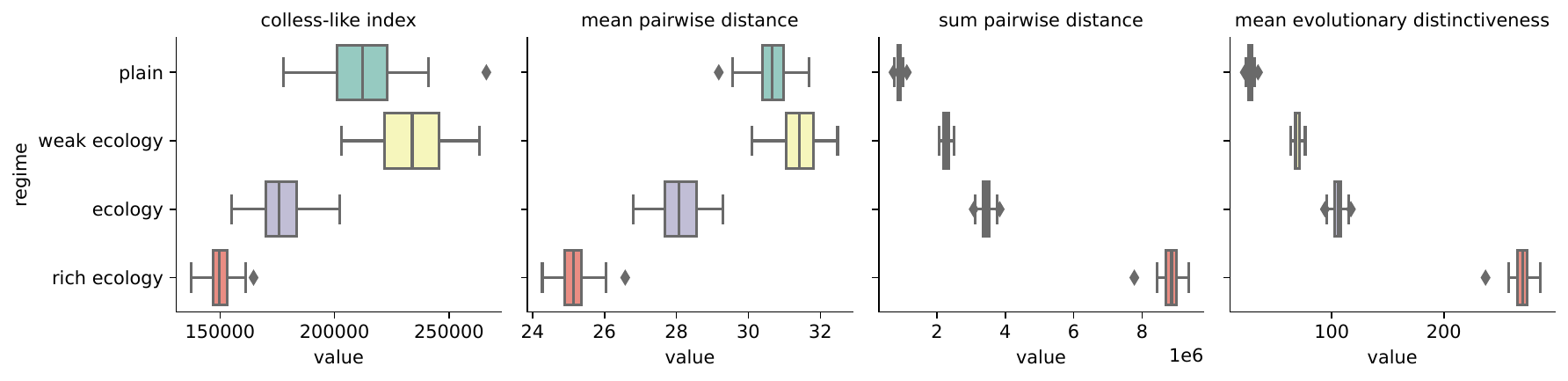}
    \caption{%
      gaussian mutation distribution at epoch 2 (generation 98,304)}
    \label{fig:perfect-tree-phylometrics-with-spatial-nuisance-sensitivity-analysis:epoch2}
  \end{subfigure}
  \begin{subfigure}[b]{\textwidth}
    \centering
    \includegraphics[width=\textwidth]{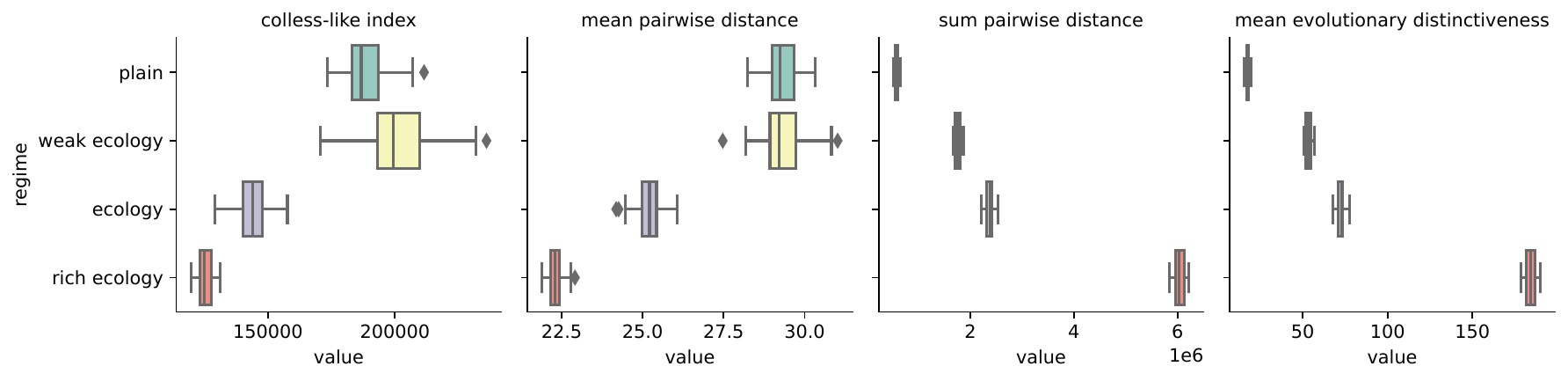}
    \caption{%
      exponential mutation distribution at epoch 7 (generation 262,144)}
    \label{fig:perfect-tree-phylometrics-with-spatial-nuisance-sensitivity-analysis:exponential}
  \end{subfigure}
  \caption{%
    Distribution of phylometrics across surveyed evolutionary regimes with spatial structure (i.e., island count 1,024) for sensitivity analysis conditions.
    Phylometrics were calculated on perfect-fidelity simulation phylogenetic records.
    Sample sizes of $n=50$ replicates define each depicted distribution.
  }
  \label{fig:perfect-tree-phylometrics-with-spatial-nuisance-sensitivity-analysis}
\end{figure*}

For these experiments, we again performed a sensitivity analysis over an alternate exponential mutation operator and earlier phylogeny sampling time points.
We found the effects of evolutionary conditions on phylometrics to be generally consistent across surveyed conditions (Supplementary Figure \ref{fig:perfect-tree-phylometrics-with-spatial-nuisance-sensitivity-analysis} and Supplementary Tables \labelcref{tab:phylostatistics-comparison-between-regimes-spatial-nuisance-kwallis,tab:phylostatistics-comparison-between-resolutions-allpairs-wilcox-spatial-nuisance}).

\subsection{Species-level Phylogenies from Gen3sis Model}

\begin{figure*}
  \begin{subfigure}[b]{0.5\textwidth}
    \includegraphics[width=\textwidth]{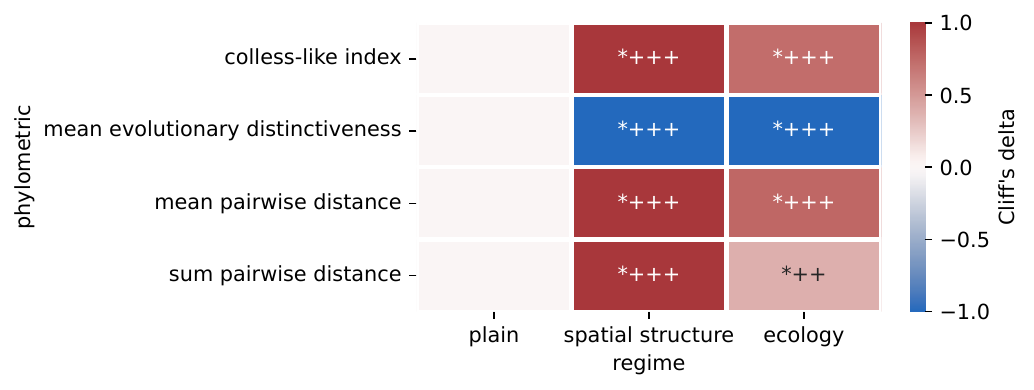}
    \caption{non-spatial baseline}
    \label{fig:perfect-tree-phylometrics-heatmap-gen3sis}
  \end{subfigure}%
  \begin{subfigure}[b]{0.5\textwidth}
    \includegraphics[width=\textwidth]{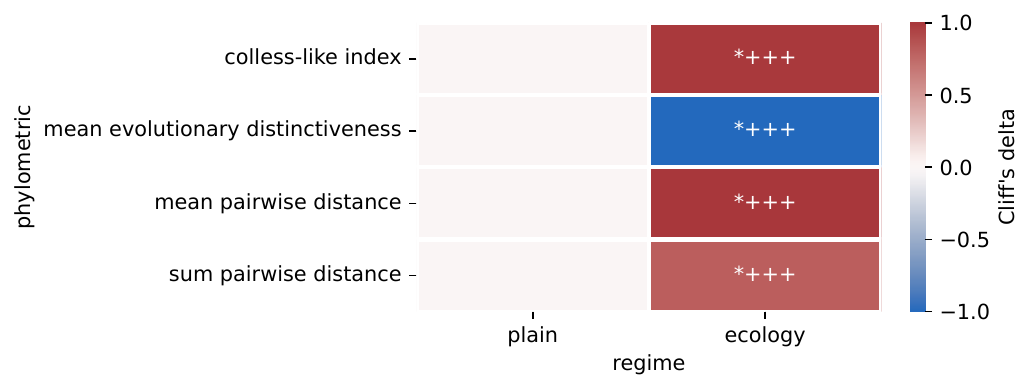}
    \caption{spatial baseline}
    \label{fig:perfect-tree-phylometrics-spatial-heatmap-gen3sis}
  \end{subfigure}
  \caption{%
    Evolutionary regimes' effect sizes relative to ``plain'' baseline under the Gen3sis model with perfect phylogenetic tracking, normalized via Cliff's delta.
    Sample sizes $n=30$.
    Annotated +'s indicate small, medium, and large effect sizes using the Cliff's delta statistic and *'s indicate statistical significance at $\alpha = 0.05$ via Mann-Whitney U test.
  }
  \label{fig:perfect-tree-phylometrics-gen3sis}
\end{figure*}

\begin{figure*}
  \centering
\includegraphics[width=\textwidth]{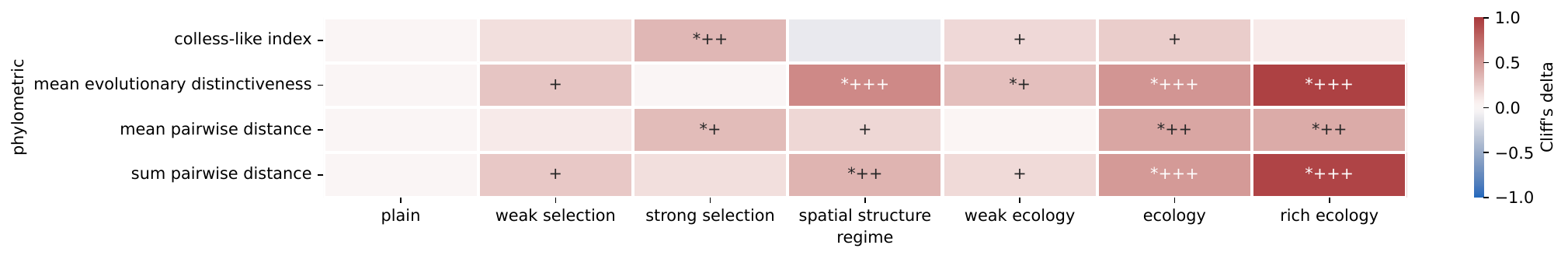}
\caption{%
  \textbf{Phylometrics for genotype-level phylogenies in Avida.}
  Tree phylometrics across surveyed evolutionary regimes, calculated on perfect-fidelity simulation phylogenies in which the taxonomic unit is genotype rather than individual.
  Note that nonparametric effect size normalization caps out to 1.0/-1.0 past the point of complete disbributional nonoverlap.
  For heatmap charts, +'s indicate small, medium, and large effect sizes using the Cliff's delta statistic and *'s indicate statistical significance at $\alpha = 0.05$ via Mann-Whitney U test.
}
  \label{fig:perfect-tree-phylometrics-heatmap-avida-genome}
\end{figure*}

Significant, large-effect changes were detected across all four phylometrics in each treatment in Gen3sis (see Figure \ref{fig:perfect-tree-phylometrics-gen3sis}).
However, with the exception of sum pariwise distance, effect signs of treatments were opposite to those for individual-level phylogenies from Avida and the simple model across all phylometrics.
Tip count effects seem likely to play a role in this discrepancy.
Unlike Avida and the simple model, which held population size constant, species richness grew freely under the Gen3sis model.
It is also possible that phylometric outcomes may be sensitive to granularity level of the taxonomic unit of the phylogeny.
As noted in our methods and shown in Figure \ref{fig:perfect-tree-phylometrics-heatmap-avida-genome}, in Avida experiments, phylogenies using genome-level tracking (as opposed to individual-level tracking) were notably different than those using individual-level tracking.
These differences also included changes in the sign of some treatment effects, lending credence to the idea that differences between Gen3sis and the individual-level phylogenies could be partially the result of Gen3sis having a more abstract taxonomic unit.

\subsection{Phylometric Bias of Reconstruction Error}
\label{sec:phylometric-bias-reconstruction-error}

Shifting from perfect phylogenetic tracking to approximate phylogenetic reconstruction will facilitate efficiency and robust digital evolution simulations at scale, but introduces a complicating factor into phylogenetic analyses: tree reconstruction error.
A clear understanding of the impact of these errors on the computed phylometrics will be necessary for informative future phylogenetic analyses.

To explore this question, we compared phylometrics computed on reconstructed trees to corresponding true reference trees under the simple model (Wilcoxon tests; $n=50$ per condition; Supplementary Table \ref{tab:phylostatistics-comparison-between-resolutions-allpairs-wilcox}).
To err towards conservatism in detecting phylometric biases, we did not correct for multiple comparisons.
Reconstructions were performed across a range of precisions, ranging from 1\% relative resolution for MRCA estimates (most precise) to 33\% relative resolution for MRCA estimates (least precise).
Precision was manipulated by adjusting the information content of underlying hereditary stratigraphic genome annotations used to perform phylogenetic reconstruction \citep{moreno2022hereditary}.
Note that important differences exist the between nature of reconstruction error under hereditary stratigraphy versus traditional biosequence-based methods, discussed further below.

\begin{figure*}
  \centering
  \includegraphics[width=\textwidth]{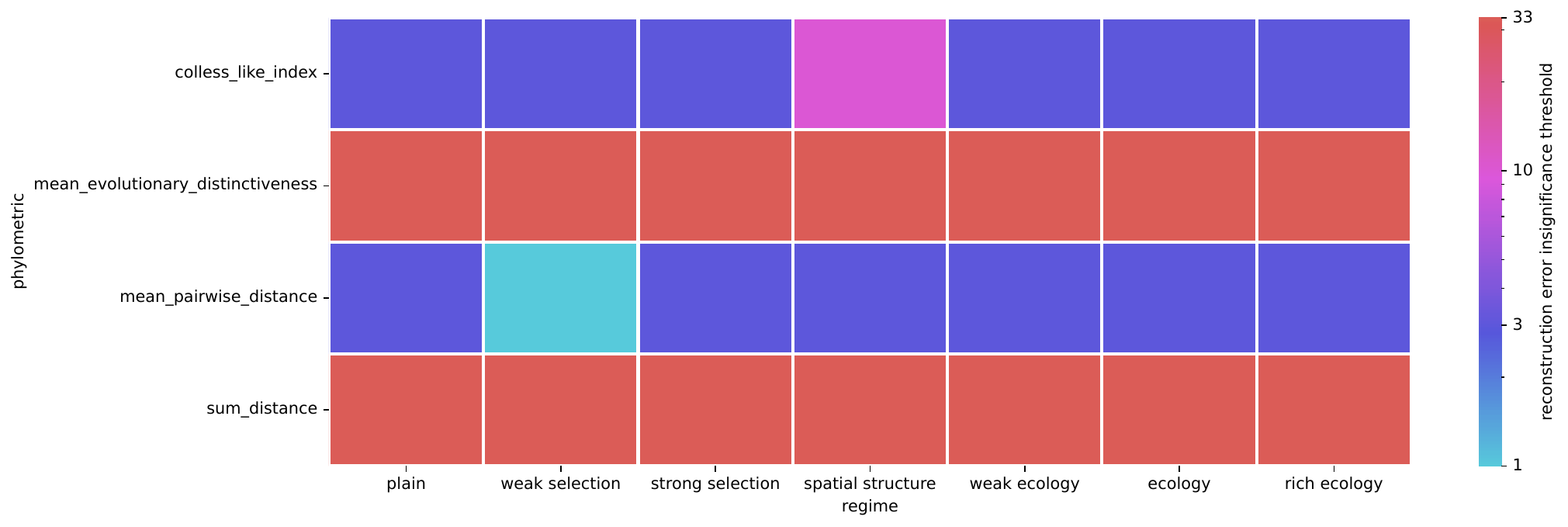}
  \caption{%
    \textbf{Reconstruction resolutions required to achieve statistical indistinguishability between reconstructions and ground-truth reference trees.}
    Significance level $p<0.05$ under the Wilcoxon signed-rank test between samples of 50 replicates each is used as the threshold for statistical distinguishability.
    Phylometrics with looser reconstruction resolution thresholds (i.e., higher resolution percentages) are less sensitive to reconstruction error.
    White heat map tiles indicate that no surveyed reconstruction resolution threshold was sufficient to achieve indistinguishability from the reference tree with respect a particular phylometric.
    See Supplementary Figure \ref{fig:reconstructed-tree-phylometrics-error-sensitivity-analysis} for sensitivity analysis results.
  }
  \label{fig:reconstructed-tree-phylometrics-error}
\end{figure*}

\noindent
\textbf{Phylometric Sensitivites to Reconstruction Error}

\noindent
For each phylometric, we sought to determine the minimum resolution required to achieve statistical non-detection (i.e., $p > 0.05$) of bias between reconstructions and their corresponding references.
For nearly all cases, 3\% reconstruction resolution was sufficient to achieve statistical indistinguishability between reference and reconstruction.
Mean evolutionary distinctiveness and sum pairwise distance were particularly robust to reconstruction error, showing no detectable bias even at only 33\% reconstruction resolution.

Phylometric sensitivity to reconstruction error was broadly consistent across evolutionary regimes.
Figure \ref{fig:reconstructed-tree-phylometrics-error} summarizes these results.

Where detectable, estimation uncertainty bias decreased all surveyed phylometrics' numerical value.
So, when testing for expected increases in phylometric values, the potential for systematic false positives due to reconstruction error can be discounted.
Supplementary Figure \ref{fig:reconstructed-tree-phylometrics} provides a full comparison of the distribution of phylometric estimates on reference trees with the distributions of phylometric estimates for reconstructed trees across reconstruction resolutions.

\begin{figure*}
  \centering
  \begin{subfigure}[b]{\textwidth}
    \centering
    \includegraphics[width=\textwidth]{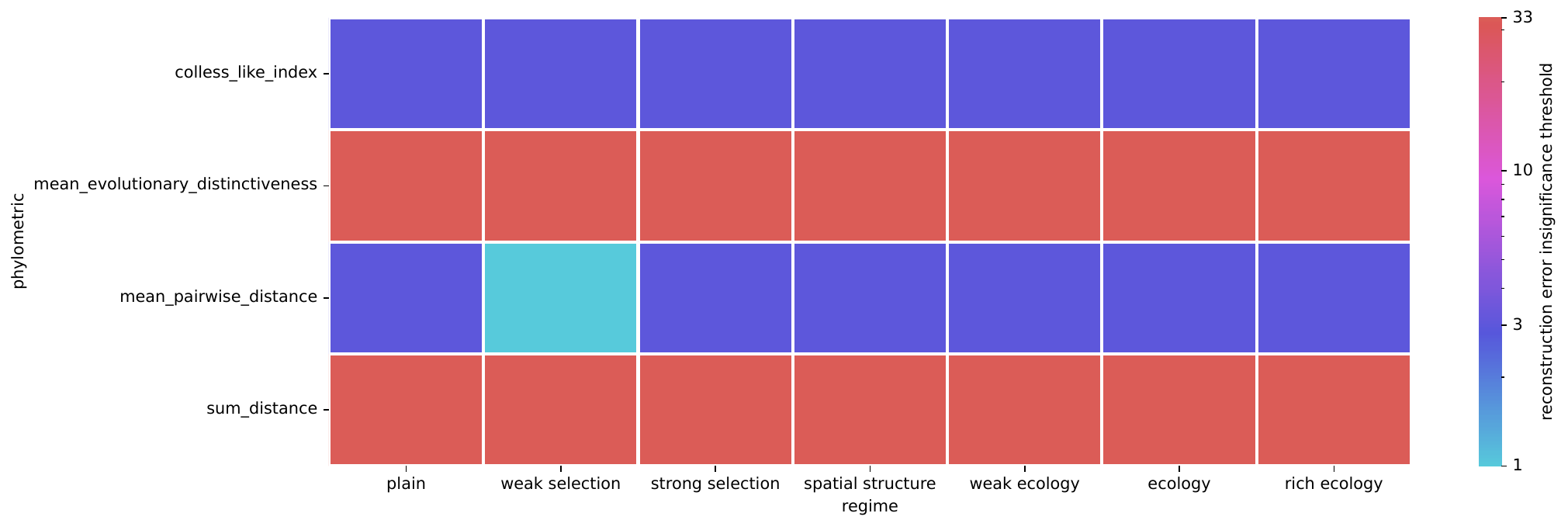}
    \caption{%
      gaussian mutation distribution at epoch 0 (generation 32,768)}
    \label{fig:reconstructed-tree-phylometrics-error-sensitivity-analysis:epoch0}
  \end{subfigure}
  \begin{subfigure}[b]{\textwidth}
    \centering
    \includegraphics[width=\textwidth]{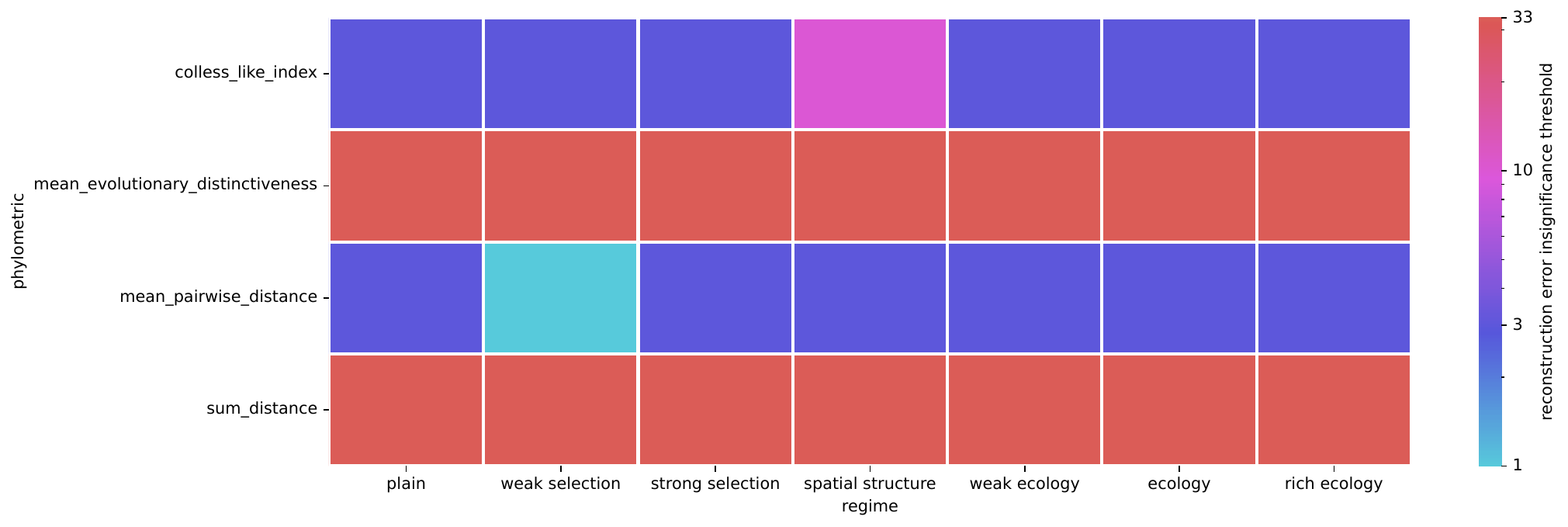}
    \caption{%
      gaussian mutation distribution at epoch 2 (generation 98,304)}
    \label{fig:reconstructed-tree-phylometrics-error-sensitivity-analysis:epoch2}
  \end{subfigure}
  \begin{subfigure}[b]{\textwidth}
    \centering
    \includegraphics[width=\textwidth]{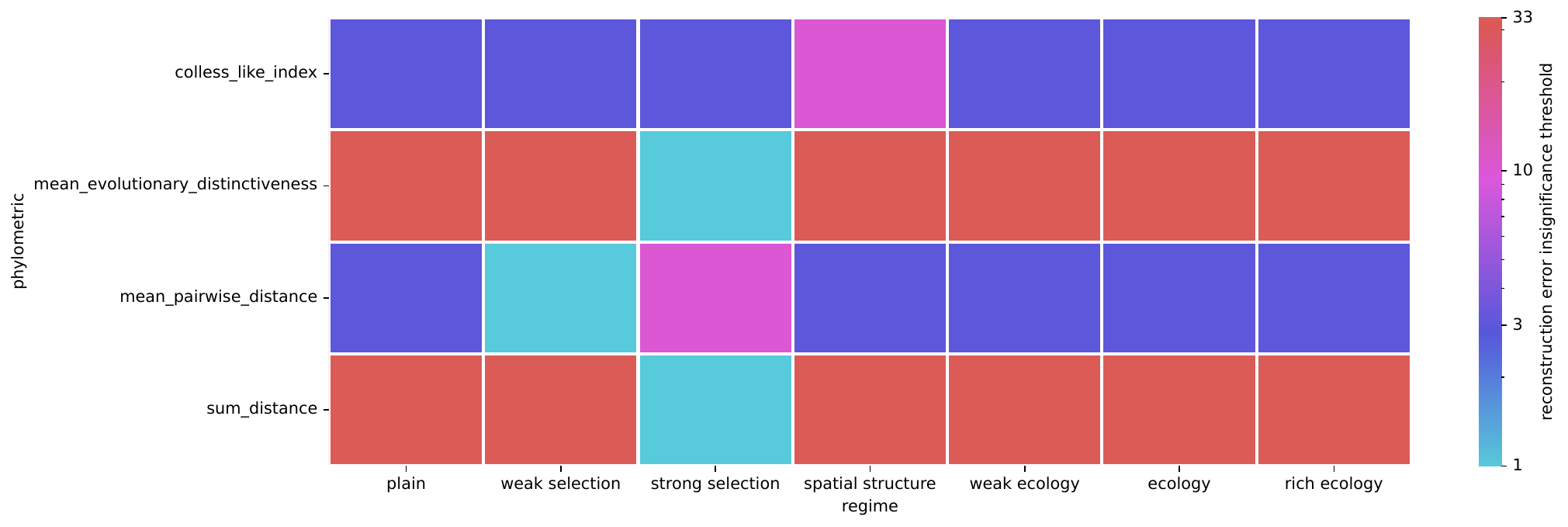}
    \caption{%
      exponential mutation distribution}
    \label{fig:reconstructed-tree-phylometrics-error-sensitivity-analysis:exponential}
  \end{subfigure}
  \caption{%
    Sensitivity analysis results for reconstruction resolutions required to achieve statistical indistinguishability between reconstructions corresponding reference trees for each phylometric across surveyed evolutionary conditions.
    Significance level $p<0.05$ under the Wilcoxon signed-rank test between samples of 50 replicates each is used as the threshold for statistical distinguishability.
    Phylometrics with looser reconstruction resolution thresholds (i.e., higher resolution percentages) are less sensitive to reconstruction error.
    White heat map tiles indicate that no surveyed reconstruction resolution threshold was sufficient to achieve indistinguishability from the reference tree with respect a particular phylometric.
  }
  \label{fig:reconstructed-tree-phylometrics-error-sensitivity-analysis}
\end{figure*}

\begin{sidewaysfigure*}
  \centering
  \includegraphics[height=\textwidth,angle=-90]{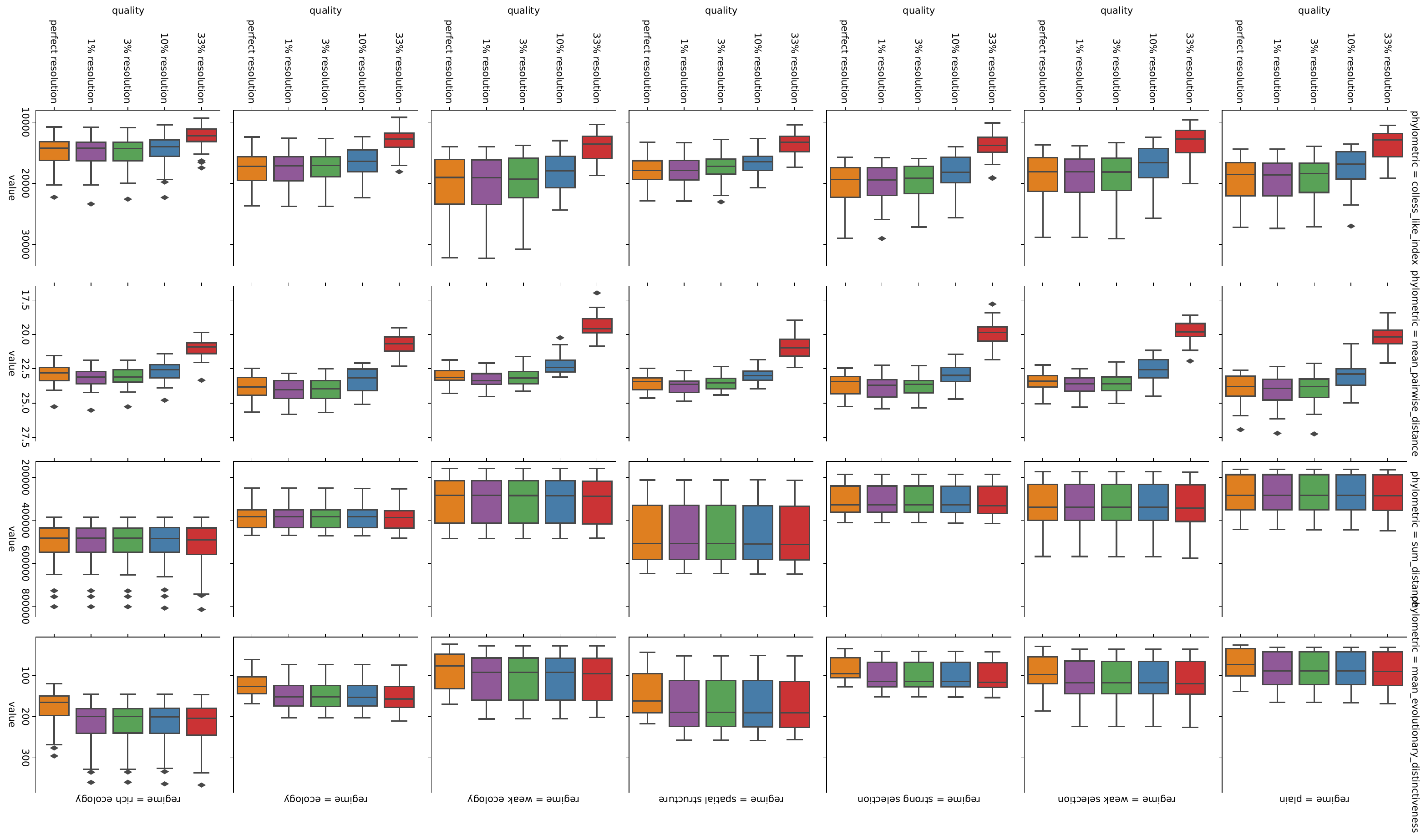}
  \caption{%
    Distributions of phylometrics across surveyed reconstruction fidelities and evolutionary regimes under Avida model.
    Sample sizes $n=30$.
  }
  \label{fig:reconstructed-tree-phylometrics-avida}
\end{sidewaysfigure*}

\begin{figure*}
  \centering
  \includegraphics[width=\textwidth]{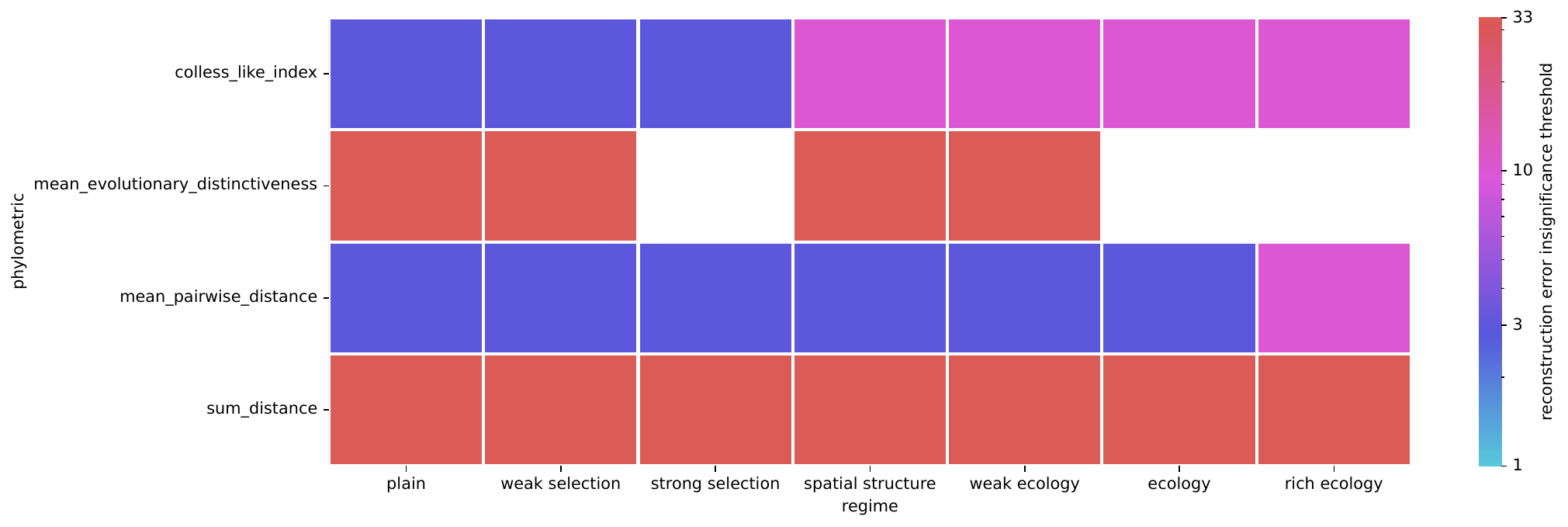}
  \caption{%
    Reconstruction resolutions required to achieve statistical indistinguishability between reconstructions corresponding reference trees for each phylometric across surveyed evolutionary conditions.
    Significance level $p<0.05$ under the Wilcoxon signed-rank test between samples of 30 replicates each is used as the threshold for statistical distinguishability.
    Phylometrics with looser reconstruction resolution thresholds (i.e., higher resolution percentages) are less sensitive to reconstruction error.
    White heat map tiles indicate that no surveyed reconstruction resolution threshold was sufficient to achieve indistinguishability from the reference tree with respect a particular phylometric.
  }
  \label{fig:reconstructed-tree-phylometrics-error-avida}
\end{figure*}

The relationship between reconstruction error and phylometric bias was similar under spatially structured regimes, the alternate exponential mutation operator, at earlier phylogeny sampling time points, and in the Avida model system (Supplementary Figures \labelcref{fig:reconstructed-tree-phylometrics-error-sensitivity-analysis,fig:reconstructed-tree-phylometrics-error-spatial-nuisance,fig:reconstructed-tree-phylometrics-with-spatial-nuisance,fig:reconstructed-tree-phylometrics-avida,fig:reconstructed-tree-phylometrics-error-avida}; Supplementary Table \labelcref{tab:phylostatistics-comparison-between-resolutions-allpairs-wilcox,tab:phylostatistics-comparison-between-resolutions-allpairs-wilcox-spatial-nuisance}).
Notably, however, reconstruction bias persisted at even 1\% relative resolution in some conditions of the sensitivity analysis and Avida experiments.
Forthcoming work has found that the byte-differentia hereditary stratigraphy configuration used for this experiment tends to lump closely contemporaneous lineage splitting events into a polytomy rather than a double-branching event --- i.e., indicating uncertainty rather than introducing error.
In contrast, reconstructions on single-bit differentia contain erroneously- sequenced branching rather than artifactual polytomies.
So, future work should explore whether working with single bit differentia could lessen phylometric bias.

\noindent
\textbf{Detection of Evolutionary Drivers' Signatures in Reconstructed Phylogenies}

\begin{figure*}
  \centering
  \begin{subfigure}[b]{\textwidth}
    \includegraphics[width=\textwidth]{binder/binder/teeplots/epoch=7+mut_distn=np.random.standard_normal+viz=heatmap+x=regime+y=phylometric+ext=.pdf}
    \caption{Reference phylogeny.}
  \end{subfigure}

\begin{subfigure}[b]{\textwidth}
  \includegraphics[width=\textwidth]{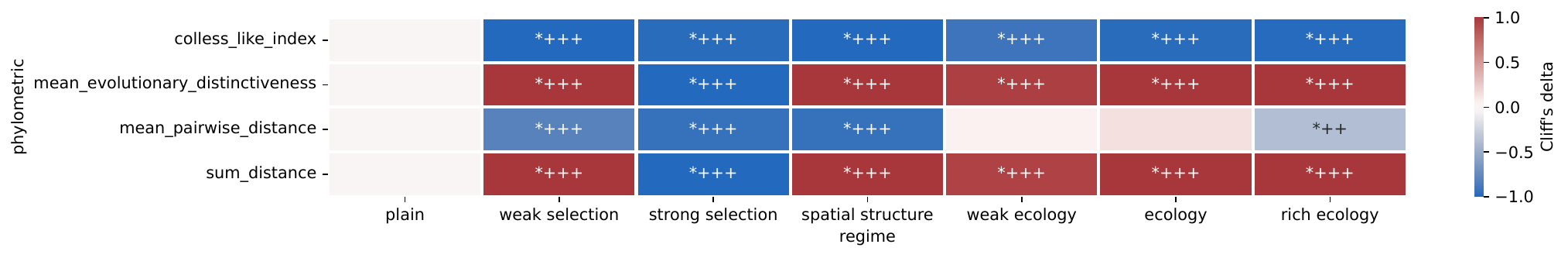}
  \caption{1\% resolution reconstruction.}
\end{subfigure}

\begin{subfigure}[b]{\textwidth}
  \includegraphics[width=\textwidth]{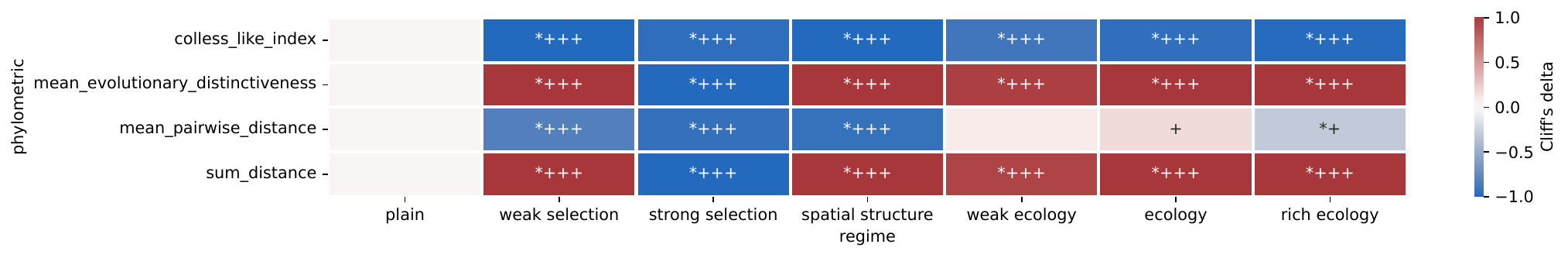}
  \caption{3\% resolution reconstruction.}
\end{subfigure}

\begin{subfigure}[b]{\textwidth}
  \includegraphics[width=\textwidth]{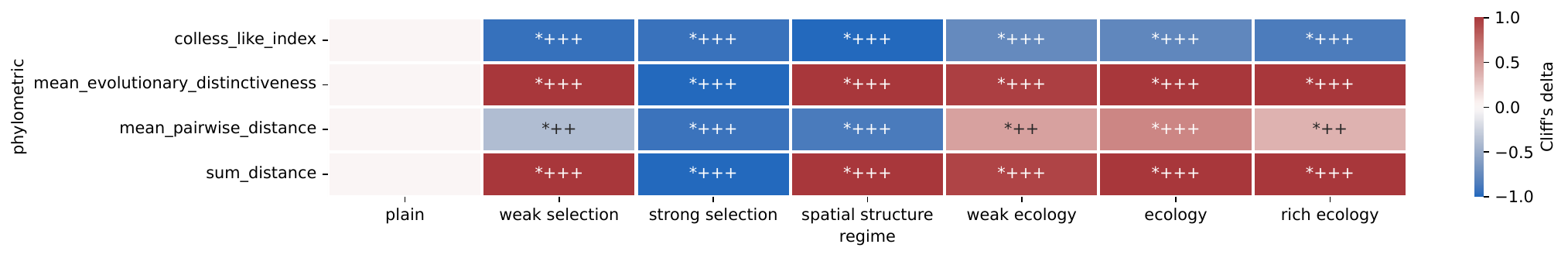}
  \caption{10\% resolution reconstruction.}
\end{subfigure}

\begin{subfigure}[b]{\textwidth}
  \includegraphics[width=\textwidth]{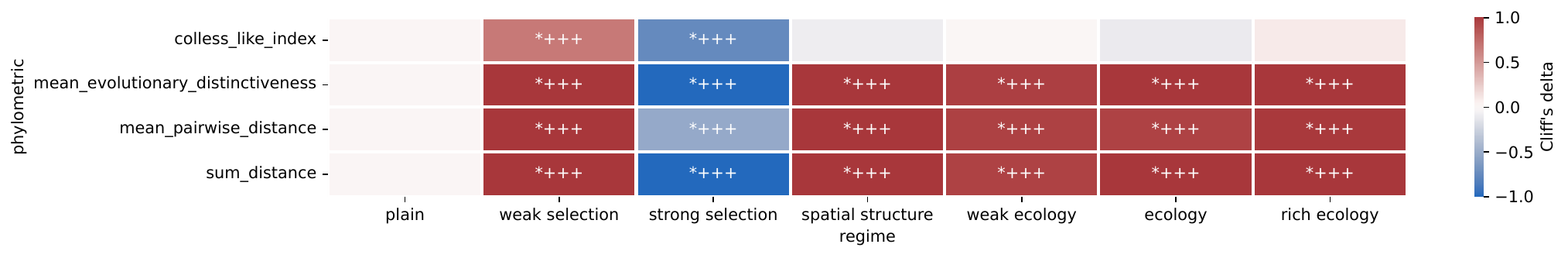}
  \caption{30\% resolution reconstruction.}
\end{subfigure}

  \caption{
Tree phylometrics across surveyed evolutionary regimes, calculated on reconstructed and perfect-fidelity simulation phylogenetic records from simple model.
Note that nonparametric effect size normalization caps out to 1.0/-1.0 past the point of complete disbributional nonoverlap.
For heatmap charts, +'s indicate small, medium, and large effect sizes using the Cliff's delta statistic and *'s indicate statistical significance at $\alpha = 0.05$ via Mann-Whitney U test.
Results from simple model are for standard experimental conditions: gaussian mutation distribution at epoch 7 (generation 262,144).
  }
  \label{fig:reconstructed-tree-phylometrics-progressive-heatmap}
\end{figure*}

\begin{figure*}
  \centering
  \begin{subfigure}[b]{\textwidth}
    \includegraphics[width=\textwidth]{binder/binder/avida-individual/teeplots/epoch=0+mut_distn=default+viz=heatmap+x=regime+y=phylometric+ext=.pdf}
    \caption{Reference phylogeny.}
  \end{subfigure}

\begin{subfigure}[b]{\textwidth}
  \includegraphics[width=\textwidth]{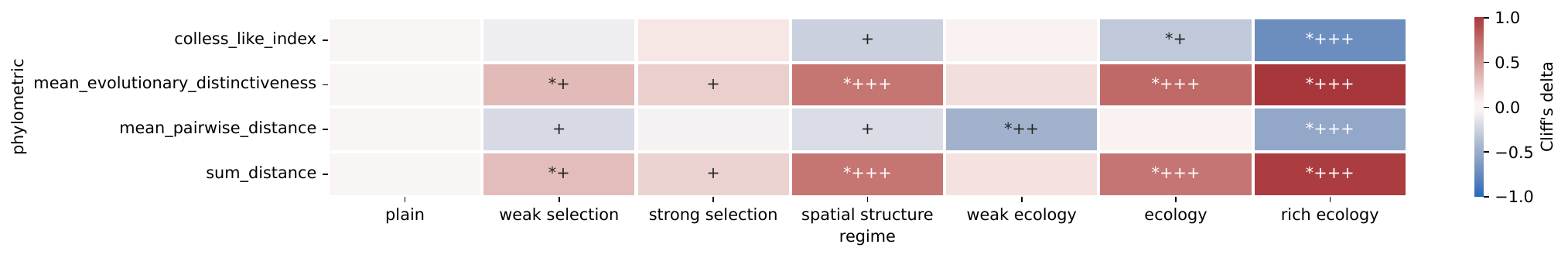}
  \caption{1\% resolution reconstruction.}
\end{subfigure}

\begin{subfigure}[b]{\textwidth}
  \includegraphics[width=\textwidth]{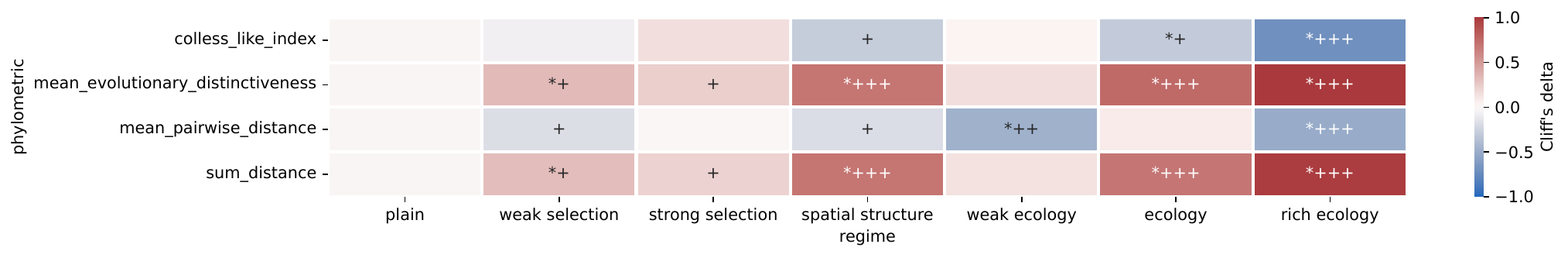}
  \caption{3\% resolution reconstruction.}
\end{subfigure}

\begin{subfigure}[b]{\textwidth}
  \includegraphics[width=\textwidth]{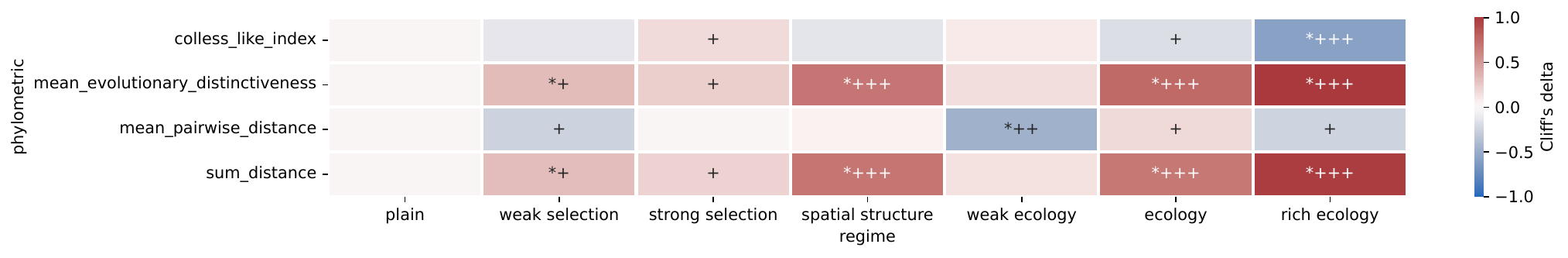}
  \caption{10\% resolution reconstruction.}
\end{subfigure}

\begin{subfigure}[b]{\textwidth}
  \includegraphics[width=\textwidth]{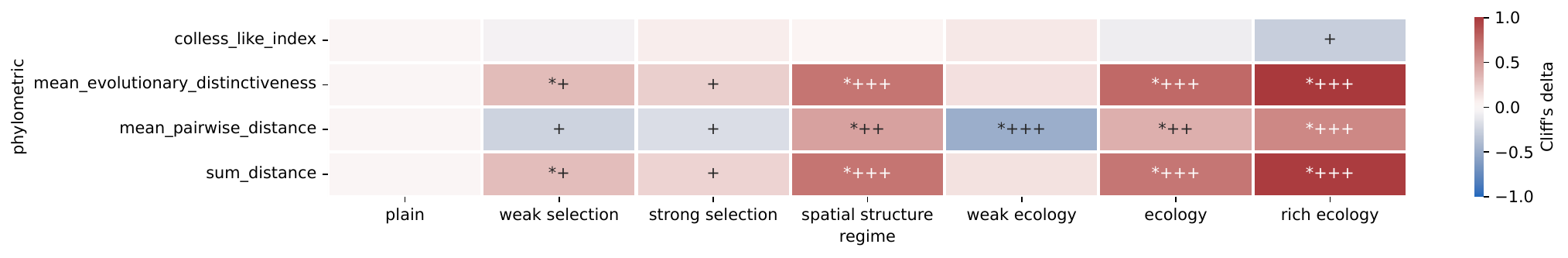}
  \caption{30\% resolution reconstruction.}
\end{subfigure}

  \caption{
Tree phylometrics across surveyed evolutionary regimes, calculated on reconstructed and perfect-fidelity simulation phylogenetic records from Avida model.
Note that nonparametric effect size normalization caps out to 1.0/-1.0 past the point of complete disbributional nonoverlap.
For heatmap charts, +'s indicate small, medium, and large effect sizes using the Cliff's delta statistic and *'s indicate statistical significance at $\alpha = 0.05$ via Mann-Whitney U test.
  }
  \label{fig:reconstructed-tree-phylometrics-progressive-heatmap-avida}
\end{figure*}

\noindent
Our last objective was to assess how reconstruction error might affect detection of phylometric signatures induced by treatment conditions.
That is, we sought to perform a sort of ``integration test'' for detection of treatment conditions when working with imperfect reconstructions rather than perfect phylogenies.
To this end, we compared the phylometric outcomes of strong/weak selection, spatial structure, ecology, and weak/rich ecology relative to plain conditions for phylogenies reconstructed at 1\%, 3\%, 10\%, and 33\% resolution levels.
Supplemental Figures \labelcref{fig:reconstructed-tree-phylometrics-progressive-heatmap,fig:reconstructed-tree-phylometrics-progressive-heatmap-avida} show heatmaps with sign, effect size, and significance of phylometric effects across gradations of reconstruction precision for the simple model and Avida, respectively.
In most cases, 3\% resolution sufficed to fully recover phylometric effects of treatments observed with perfectly-tracked phylogenies.

\section{Conclusion}
\label{sec:conclusion}

Because phylogenies are an abstraction that generalizes across all evolutionary processes, phylogenetic analysis can be applied across a breadth of biological and artificial life systems.
Consequently, a broad set of cross-disciplinary use cases exist for inferring the processes that shaped a phylogeny by quantifying its topology.
Indeed, extraction of information about evolutionary dynamics from phylogenetic history has been a longstanding and productive theme in evolutionary biology \citep{pagel1997inferring}.
This work seeks to contribute in that vein, by establishing foundations necessary to assess three fundamental evolutionary drivers --- spatial structure, ecology, and selection pressure --- in complex, distributed evolving populations.

First, we investigated the strength and character of structural signatures in phylogenetic histories left by selection pressure, ecology, and spatial population structure.
These drivers induced readily detectable effects across our suite of four surveyed phylometrics.
These effects were generally consistent across surveyed individual-level phylogenies, but differed notably under genotype-level tracking and under species-level simulation.
Although the directions of phylometric effects were generally consistent across treatments expected to increase phylogenetic richness (i.e., weak selection, ecologies, and spatial structure), effect significance/size for particular phylometrics differed among some treatments.
Compared to results from the simple model, which was amenable to strong treatment manipulations, the more sophisticated Avida model system expressed fewer phylometric effects between treatments.

Compared to spatial structure and selection pressure, drivers designed to induce ecological coexistence exerted relatively muted, sparse effects on phylogenetic metrics.
However, follow-up experiments with both individual-level model systems revealed that phylometric signatures can remain detectable under background conditions of spatial population structure.
Interactions, though, are sometimes surprising.
A notable example is the reduced mean pairwise distance observed in Avida under conditions combining spatial structure and ecology, but not in the baseline ecology treatment.
These results highlight the complexity of how interacting evolutionary dynamics impact phylogenetic structure.
Even a single dynamic in isolation can influence phylometrics in opposite directions.
For instance, we found that changes to selection pressure in either direction significantly reduced trees' Colless-like index under the simple model.

Comparison of phylometrics from perfect-fidelity trees against corresponding reconstructions revealed that phylometric statistics differed in sensitivity to reconstruction error introduced by hereditary stratigraphy-based tracking.
Colless-like index and mean pairwise distance were more sensitive, and in some conditions, reconstruction persisted at even 1\% reconstruction resolution.
On the other hand, mean evolutionary distinctiveness and sum pairwise distance were particularly robust to reconstruction error, with no bias detected even at 33\% resolution reconstruction.
Results suggest 3\% reconstruction resolution as a reasonable ballpark parameterization for applications of hereditary stratigraphy with downstream phylogenetic analysis.
Where it did occur, the sign effect of phylometric bias was consistent, which might simplify considerations to account for it in experiments.

Our findings identify key next steps necessary for development of rigorous phylogenetic assays to test for key evolutionary drivers.
Because treatments did not induce widespread opposite-sign effects, our results do not indicate a straightforward, qualitative approach to disambiguate drivers' signatures.
Future work should survey comprehensive panels of phylometrics, which might provide a stronger foothold to this end \citep{tuckerGuidePhylogeneticMetrics2017}.
Alternately, more sophisticated quantitative approaches leveraging differential effect size ratios may be fruitful.
Possibilities include application of machine learning techniques over broad phylometric panels \citep{voznica2022deep} or even inference directly over phylogenetic structure itself via graph neural networks and related methods \citep{lajaaiti2023comparison}.
The capability of artificial life approaches to generate high-quality data sets with high replication counts will be a significant asset to such work.
More information could be harnessed by incorporating trait data \citep{nozoe2017inferring} or by using time-sampled phylogenies \citep{volz2013viral}, which would include information about ``ghost'' lineages that have extincted at earlier time points.

Additionally, robust normalization methods will be crucial to enable meaningful comparisons between phylogenies differing in (1) population size, (2) population subsampling level, (3) depth of evolutionary history, and (4) demographic characteristics of organism life histories.
Despite some substantial work on this front \citep{shao1990tree,mir2018sound}, significant challenges remain.
It is worth noting that most analysis reported here normalizes to phylometric values sampled from ``plain'' evolutionary regimes, thereby implicitly depending on the availability of such data.
However, stripped-down baseline condition experiments may not always be possible, especially in biological model systems.
Extensive, likely sophisticated, considerations would be necessary to establish \textit{a priori} phylometric value predictions that could be compared against instead of testing against empirical values.
Finally, as evidenced in our experiments, reconstruction error can systematically bias phylometric values.
In scenarios where reconstruction error is unavoidable, additional normalizations may be necessary.
However, it is promising that some phylometrics appear thoroughly robust to reconstruction error and, in other cases, sufficient reconstruction accuracy might be achieved to obviate error bias.

The overriding objective of work presented here is to equip biologists and computational researchers with capability to investigate fundamental evolutionary drivers within large-scale evolving populations.
To this end, methodology must advance hand-in-hand with the software infrastructure required to put it into practice.
Artificial life experiments, specifically, tend to require particular instrumentation to collect phylogenetic histories, and we have been active in making general-use plug-and-play solutions available freely through the Python Packaging Index \citep{moreno2022hstrat,dolson2024phylotrackpy}.
As each field develops, artificial life will benefit greatly from domain-specific software and methods developed within traditional evolutionary bioinformatics and, we hope, vice versa.

  \section*{Acknowledgment}

This research was supported in part by NSF grants DEB-1655715 and DBI-0939454 as well as by Michigan State University through the computational resources provided by the Institute for Cyber-Enabled Research.
This material is based upon work supported by the National Science Foundation Graduate Research Fellowship under Grant No. DGE-1424871.
Any opinions, findings, and conclusions or recommendations expressed in this material are those of the author(s) and do not necessarily reflect the views of the National Science Foundation.
This material is based upon work supported by the Eric and Wendy Schmidt AI in Science Postdoctoral Fellowship, a Schmidt Futures program.

  \putbib

\end{bibunit}

\clearpage
\newpage

\setcounter{secnumdepth}{2}
\makeatletter
\renewcommand \thesection{S\@arabic\c@section}
\makeatother

\begin{bibunit}

  \onecolumn

\section{Supplemental Material}

\providecommand{\thecaption}{
  Kruskal-Wallis one-way analysis of variance tests for nonequivalence of reconstruction error distributions among surveyed evolutionary regimes.
  Comparisons were performed independently within each sensitivity analysis condition and reconstruction fidelities.
  Error reported as both quartet and triplet distance between reconstructed tree and corresponding reference tree.
  Sample size $n=50$ for each population, with $N=7$ populations (i.e., evolutionary regimes) compared within each sensitivity analysis condition.
}

\csvstyle{myTableStyle}{
  longtable=c c c c c c c c,
  before reading=\scriptsize,
  table head=\toprule $n$ & $N$ & Tree Distance Metric & Statistic & $p$ & Reconstruction Fidelity & Epoch & Mutation Distribution\\ \midrule\endhead\bottomrule\\[-1em]
  \caption{%
    \thecaption
  }\endfoot
  \label{tab:reconstruction-error-comparisons-between-regimes},
  head to column names,
  respect all,
}

\begin{filecontents*}{binder/binder/outdata/a=reconstruction-error-comparisons-between-regimes+ext=.csv}
n,N,phylometric,statistic,p,quality,epoch,mut_distn
50,7,quartet_distance,65.93556630036619,2.778141406820147e-12,1% resolution,0,np.random.exponential
50,7,quartet_distance,117.93292503052521,4.426885939536874e-23,10% resolution,0,np.random.exponential
50,7,quartet_distance,52.91400439560425,1.2210174512398157e-09,3% resolution,0,np.random.exponential
50,7,quartet_distance,115.72124835164846,1.2888289018380268e-22,33% resolution,0,np.random.exponential
50,7,quartet_distance,38.789497435897374,7.870770259134266e-07,1% resolution,0,np.random.standard_normal
50,7,quartet_distance,88.34049719169707,6.698730980178156e-17,10% resolution,0,np.random.standard_normal
50,7,quartet_distance,38.149798290598255,1.0500974929537594e-06,3% resolution,0,np.random.standard_normal
50,7,quartet_distance,120.2027643467643,1.4773716218562115e-23,33% resolution,0,np.random.standard_normal
50,7,quartet_distance,107.74271062271055,6.050249062560348e-21,1% resolution,2,np.random.exponential
50,7,quartet_distance,147.54797557997563,2.552189708112316e-29,10% resolution,2,np.random.exponential
50,7,quartet_distance,70.23442832722844,3.660319257589615e-13,3% resolution,2,np.random.exponential
50,7,quartet_distance,194.43400048840022,2.9012262667888492e-39,33% resolution,2,np.random.exponential
50,7,quartet_distance,96.98896019536028,1.065063929831718e-18,1% resolution,2,np.random.standard_normal
50,7,quartet_distance,102.78477753357765,6.580118842694302e-20,10% resolution,2,np.random.standard_normal
50,7,quartet_distance,50.74469352869369,3.332892857192771e-09,3% resolution,2,np.random.standard_normal
50,7,quartet_distance,161.91563565323577,2.3263865526015013e-32,33% resolution,2,np.random.standard_normal
50,7,quartet_distance,63.672302808302675,8.050303019123108e-12,1% resolution,7,np.random.exponential
50,7,quartet_distance,95.97084444444431,1.735723660297524e-18,10% resolution,7,np.random.exponential
50,7,quartet_distance,38.62639902319893,8.471531632118808e-07,3% resolution,7,np.random.exponential
50,7,quartet_distance,171.43860512820493,2.227610248382613e-34,33% resolution,7,np.random.exponential
50,7,quartet_distance,83.14988229548226,7.975356755011296e-16,1% resolution,7,np.random.standard_normal
50,7,quartet_distance,139.95126935286953,1.0262254227010028e-27,10% resolution,7,np.random.standard_normal
50,7,quartet_distance,45.43054847374833,3.8430812080812616e-08,3% resolution,7,np.random.standard_normal
50,7,quartet_distance,178.3534456654454,7.590207763912434e-36,33% resolution,7,np.random.standard_normal
50,7,triplet_distance,65.05051037851058,4.212710048929408e-12,1% resolution,0,np.random.exponential
50,7,triplet_distance,125.22499438339423,1.2998707659411192e-24,10% resolution,0,np.random.exponential
50,7,triplet_distance,60.08189499389505,4.3318429983592146e-11,3% resolution,0,np.random.exponential
50,7,triplet_distance,77.06523272283266,1.4409743904455206e-14,33% resolution,0,np.random.exponential
50,7,triplet_distance,38.7933186813184,7.857214967772178e-07,1% resolution,0,np.random.standard_normal
50,7,triplet_distance,93.79757167277148,4.919538564490169e-18,10% resolution,0,np.random.standard_normal
50,7,triplet_distance,39.04988913308921,6.998226471410047e-07,3% resolution,0,np.random.standard_normal
50,7,triplet_distance,57.02958730158707,1.8018981162602906e-10,33% resolution,0,np.random.standard_normal
50,7,triplet_distance,107.76035164835162,5.9990450071640534e-21,1% resolution,2,np.random.exponential
50,7,triplet_distance,122.68244395604393,4.4512105719062216e-24,10% resolution,2,np.random.exponential
50,7,triplet_distance,63.88389743589755,7.28880836629865e-12,3% resolution,2,np.random.exponential
50,7,triplet_distance,122.46991550671555,4.933403805938994e-24,33% resolution,2,np.random.exponential
50,7,triplet_distance,88.77935824175825,5.431281121975844e-17,1% resolution,2,np.random.standard_normal
50,7,triplet_distance,112.39306862026842,6.426838894270088e-22,10% resolution,2,np.random.standard_normal
50,7,triplet_distance,47.078368742368866,1.8049824615952433e-08,3% resolution,2,np.random.standard_normal
50,7,triplet_distance,101.31699340659338,1.3326333085157124e-19,33% resolution,2,np.random.standard_normal
50,7,triplet_distance,63.5612717948718,8.4811338878685e-12,1% resolution,7,np.random.exponential
50,7,triplet_distance,101.18588717948705,1.4193108390120854e-19,10% resolution,7,np.random.exponential
50,7,triplet_distance,39.763793894993796,5.068816766548348e-07,3% resolution,7,np.random.exponential
50,7,triplet_distance,134.86405079365068,1.2140176167431311e-26,33% resolution,7,np.random.exponential
50,7,triplet_distance,83.32480390720389,7.337496155382785e-16,1% resolution,7,np.random.standard_normal
50,7,triplet_distance,136.2934153846154,6.065473786179155e-27,10% resolution,7,np.random.standard_normal
50,7,triplet_distance,44.40904517704507,6.13224450123164e-08,3% resolution,7,np.random.standard_normal
50,7,triplet_distance,114.5295277167279,2.2916006964936045e-22,33% resolution,7,np.random.standard_normal
\end{filecontents*}
\csvreader[myTableStyle]{binder/binder/outdata/a=reconstruction-error-comparisons-between-regimes+ext=.csv}{}{\csvlinetotablerow}

\providecommand{\thecaption}{
  Kruskal-Wallis one-way analysis of variance tests for nonequivalence of reconstruction error distributions among reconstruction fidelities.
  Comparisons were performed independently within each sensitivity analysis condition and evolutionary regime.
  Error reported as both quartet and triplet distance between reconstructed tree and corresponding reference tree.
  Sample size $n=50$ for each population, with $N=4$ populations (i.e., reconstruction fidelities) compared within each sensitivity analysis condition.
}

\csvstyle{myTableStyle}{
  longtable=c c c c c c c c,
  before reading=\scriptsize,
  table head=\toprule $n$ & $N$ & Tree Distance Metric & Statistic & $p$ & Evolutionary Regime & Epoch & Mutation Distribution\\ \midrule\endhead\bottomrule\\[-1em]
  \caption{%
    \thecaption
  }\endfoot\bottomrule\\[-1em]
  \caption{%
    \thecaption
  }
  \label{tab:reconstruction-error-comparisons-between-resolutions}\endlastfoot,
  head to column names,
  respect all,
}

\begin{filecontents*}{binder/binder/outdata/a=reconstruction-error-comparisons-between-resolutions+ext=.csv}
n,N,phylometric,statistic,p,regime,epoch,mut_distn
50,4,quartet_distance,157.59420895522396,6.0574876696256864e-34,ecology,0,np.random.exponential
50,4,quartet_distance,148.94575522388072,4.448169282884224e-32,plain,0,np.random.exponential
50,4,quartet_distance,159.37024477611953,2.5063089697324245e-34,rich ecology,0,np.random.exponential
50,4,quartet_distance,119.44328358208952,1.0170278372302383e-25,spatial structure,0,np.random.exponential
50,4,quartet_distance,128.51634626865666,1.129249600155447e-27,strong selection,0,np.random.exponential
50,4,quartet_distance,141.11611940298508,2.171690322070411e-30,weak ecology,0,np.random.exponential
50,4,quartet_distance,131.51036417910439,2.5560703732339783e-28,weak selection,0,np.random.exponential
50,4,quartet_distance,164.675976119403,1.794464925237934e-35,ecology,0,np.random.standard_normal
50,4,quartet_distance,158.3551880597015,4.150314891195873e-34,plain,0,np.random.standard_normal
50,4,quartet_distance,127.67363582089547,1.7154395620953673e-27,rich ecology,0,np.random.standard_normal
50,4,quartet_distance,103.03172537313424,3.463586894009112e-22,spatial structure,0,np.random.standard_normal
50,4,quartet_distance,147.31100895522388,1.001837319410491e-31,strong selection,0,np.random.standard_normal
50,4,quartet_distance,166.15134328358215,8.619396637671411e-36,weak ecology,0,np.random.standard_normal
50,4,quartet_distance,118.02004776119406,2.0597967850456918e-25,weak selection,0,np.random.standard_normal
50,4,quartet_distance,157.12259104477607,7.6570304442438404e-34,ecology,2,np.random.exponential
50,4,quartet_distance,159.19157014925372,2.739004378156511e-34,plain,2,np.random.exponential
50,4,quartet_distance,142.60543283582092,1.0366826614830815e-30,rich ecology,2,np.random.exponential
50,4,quartet_distance,158.5428417910448,3.78082471573984e-34,spatial structure,2,np.random.exponential
50,4,quartet_distance,128.93210746268664,9.187540229907655e-28,strong selection,2,np.random.exponential
50,4,quartet_distance,131.25214328358197,2.905523060052889e-28,weak ecology,2,np.random.exponential
50,4,quartet_distance,119.54937313432833,9.6490692410515e-26,weak selection,2,np.random.exponential
50,4,quartet_distance,167.32167164179089,4.81780811986523e-36,ecology,2,np.random.standard_normal
50,4,quartet_distance,159.96269850746273,1.8671566175398913e-34,plain,2,np.random.standard_normal
50,4,quartet_distance,151.14236417910445,1.493933342937063e-32,rich ecology,2,np.random.standard_normal
50,4,quartet_distance,149.13343283582094,4.052261612555958e-32,spatial structure,2,np.random.standard_normal
50,4,quartet_distance,151.29153432835813,1.387237319182693e-32,strong selection,2,np.random.standard_normal
50,4,quartet_distance,138.01727761194024,1.0114604713018828e-29,weak ecology,2,np.random.standard_normal
50,4,quartet_distance,130.6005850746269,4.014614923822426e-28,weak selection,2,np.random.standard_normal
50,4,quartet_distance,164.35755223880597,2.1021541955249874e-35,ecology,7,np.random.exponential
50,4,quartet_distance,155.21479402985062,1.9756672960998725e-33,plain,7,np.random.exponential
50,4,quartet_distance,141.44481194029856,1.844681586782929e-30,rich ecology,7,np.random.exponential
50,4,quartet_distance,121.9610149253732,2.917902761677283e-26,spatial structure,7,np.random.exponential
50,4,quartet_distance,129.15149850746275,8.239914105393665e-28,strong selection,7,np.random.exponential
50,4,quartet_distance,136.83397014925367,1.8199427924754968e-29,weak ecology,7,np.random.exponential
50,4,quartet_distance,139.50888358208954,4.8233864509901185e-30,weak selection,7,np.random.exponential
50,4,quartet_distance,152.8280716417911,6.466389658952402e-33,ecology,7,np.random.standard_normal
50,4,quartet_distance,165.87327761194035,9.896887847019728e-36,plain,7,np.random.standard_normal
50,4,quartet_distance,144.04705671641796,5.067053462084274e-31,rich ecology,7,np.random.standard_normal
50,4,quartet_distance,147.2463641791046,1.0345240633683704e-31,spatial structure,7,np.random.standard_normal
50,4,quartet_distance,147.61000597014925,8.635853611613597e-32,strong selection,7,np.random.standard_normal
50,4,quartet_distance,154.6612656716419,2.601027928883277e-33,weak ecology,7,np.random.standard_normal
50,4,quartet_distance,116.14027462686568,5.230985835224136e-25,weak selection,7,np.random.standard_normal
50,4,triplet_distance,158.15284776119404,4.589279309153939e-34,ecology,0,np.random.exponential
50,4,triplet_distance,141.99424477611933,1.404247992252263e-30,plain,0,np.random.exponential
50,4,triplet_distance,159.00796417910442,3.000645940090509e-34,rich ecology,0,np.random.exponential
50,4,triplet_distance,117.663223880597,2.458448194891575e-25,spatial structure,0,np.random.exponential
50,4,triplet_distance,121.52734328358213,3.618097844103964e-26,strong selection,0,np.random.exponential
50,4,triplet_distance,137.18217313432842,1.5310536507092998e-29,weak ecology,0,np.random.exponential
50,4,triplet_distance,116.75844776119402,3.8501919298874024e-25,weak selection,0,np.random.exponential
50,4,triplet_distance,157.14539701492538,7.570755884510063e-34,ecology,0,np.random.standard_normal
50,4,triplet_distance,151.68709253731333,1.1397669204142987e-32,plain,0,np.random.standard_normal
50,4,triplet_distance,118.7833432835821,1.4107617315585413e-25,rich ecology,0,np.random.standard_normal
50,4,triplet_distance,101.24856119402978,8.375608253540304e-22,spatial structure,0,np.random.standard_normal
50,4,triplet_distance,145.4597014925373,2.5124303554920072e-31,strong selection,0,np.random.standard_normal
50,4,triplet_distance,164.7576119402985,1.7231152189043794e-35,weak ecology,0,np.random.standard_normal
50,4,triplet_distance,122.51954626865677,2.2118903890553837e-26,weak selection,0,np.random.standard_normal
50,4,triplet_distance,157.0349014925373,7.998014921199757e-34,ecology,2,np.random.exponential
50,4,triplet_distance,149.04416716417916,4.2359708683736253e-32,plain,2,np.random.exponential
50,4,triplet_distance,137.0171104477613,1.661787976127462e-29,rich ecology,2,np.random.exponential
50,4,triplet_distance,153.74838208955225,4.093605816724842e-33,spatial structure,2,np.random.exponential
50,4,triplet_distance,132.52994626865666,1.5410812732030535e-28,strong selection,2,np.random.exponential
50,4,triplet_distance,126.1195820895523,3.708647071792477e-27,weak ecology,2,np.random.exponential
50,4,triplet_distance,115.13484179104489,8.61101425310164e-25,weak selection,2,np.random.exponential
50,4,triplet_distance,166.78955223880598,6.276452107300717e-36,ecology,2,np.random.standard_normal
50,4,triplet_distance,146.79188059701482,1.2964868937117187e-31,plain,2,np.random.standard_normal
50,4,triplet_distance,151.9403223880597,1.0050447817317768e-32,rich ecology,2,np.random.standard_normal
50,4,triplet_distance,140.2913552238806,3.270679757340413e-30,spatial structure,2,np.random.standard_normal
50,4,triplet_distance,145.33868656716413,2.6680489601506224e-31,strong selection,2,np.random.standard_normal
50,4,triplet_distance,130.37295522388058,4.494692231234135e-28,weak ecology,2,np.random.standard_normal
50,4,triplet_distance,128.90074029850746,9.331651541390125e-28,weak selection,2,np.random.standard_normal
50,4,triplet_distance,160.9352835820896,1.15156046876798e-34,ecology,7,np.random.exponential
50,4,triplet_distance,143.78054925373135,5.784025470695082e-31,plain,7,np.random.exponential
50,4,triplet_distance,142.34337910447766,1.1807451541876186e-30,rich ecology,7,np.random.exponential
50,4,triplet_distance,122.8982208955224,1.833136881871484e-26,spatial structure,7,np.random.exponential
50,4,triplet_distance,128.11715820895517,1.3765984800207162e-27,strong selection,7,np.random.exponential
50,4,triplet_distance,121.30964776119401,4.030595523779953e-26,weak ecology,7,np.random.exponential
50,4,triplet_distance,139.27674029850755,5.4125856059843884e-30,weak selection,7,np.random.exponential
50,4,triplet_distance,140.299952238806,3.256749278036472e-30,ecology,7,np.random.standard_normal
50,4,triplet_distance,160.11278805970142,1.7329672944501506e-34,plain,7,np.random.standard_normal
50,4,triplet_distance,144.6158447761194,3.82022423147428e-31,rich ecology,7,np.random.standard_normal
50,4,triplet_distance,144.14121791044784,4.83558044383688e-31,spatial structure,7,np.random.standard_normal
50,4,triplet_distance,141.42565970149258,1.8623069805183808e-30,strong selection,7,np.random.standard_normal
50,4,triplet_distance,148.28660298507464,6.171120066413331e-32,weak ecology,7,np.random.standard_normal
50,4,triplet_distance,117.50394029850747,2.660478631136631e-25,weak selection,7,np.random.standard_normal
\end{filecontents*}
\csvreader[myTableStyle]{binder/binder/outdata/a=reconstruction-error-comparisons-between-resolutions+ext=.csv}{}{\csvlinetotablerow}

\begin{figure*}
  \begin{minipage}{0.5\columnwidth}
    \centering
    Generations Ago (approx.)
  \end{minipage}
  \hfill
  \begin{minipage}{0.5\columnwidth}
    \centering
    Generations Ago (approx.)
  \end{minipage}
  \begin{minipage}{0.5\columnwidth}
    \hspace{0.02\linewidth}
    \rotatebox{30}{\makebox[0.1\linewidth][c]{200,000}}
    \hfill
    \rotatebox{30}{\makebox[0.1\linewidth][c]{150,000}}
    \hfill
    \rotatebox{30}{\makebox[0.1\linewidth][c]{100,000}}
    \hfill
    \rotatebox{30}{\makebox[0.1\linewidth][c]{50,000}}
    \hfill
    \rotatebox{90}{\makebox[0.05\linewidth][c]{0}}
  \end{minipage}
  \hfill
  \begin{minipage}{0.5\columnwidth}
    \hspace{0.02\linewidth}
    \rotatebox{30}{\makebox[0.1\linewidth][c]{200,000}}
    \hfill
    \rotatebox{30}{\makebox[0.1\linewidth][c]{150,000}}
    \hfill
    \rotatebox{30}{\makebox[0.1\linewidth][c]{100,000}}
    \hfill
    \rotatebox{30}{\makebox[0.1\linewidth][c]{50,000}}
    \hfill
    \rotatebox{90}{\makebox[0.05\linewidth][c]{0}}
  \end{minipage}
  \hfill
  \begin{subfigure}[b]{0.5\columnwidth}
    \includegraphics[height=0.12\textheight,width=\textwidth]{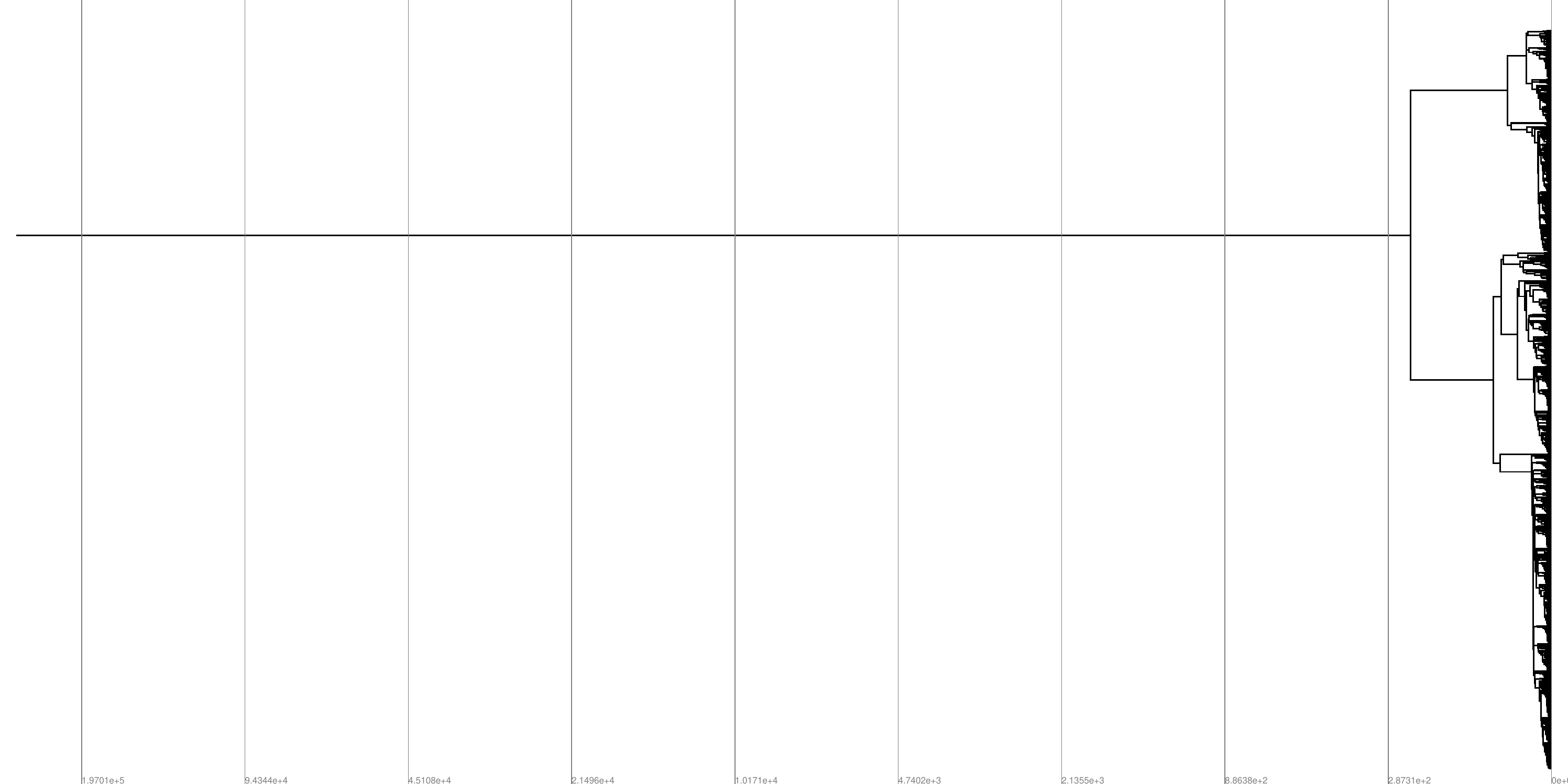}
    \caption{%
      strong selection}
  \end{subfigure}
  \hfill
  \begin{subfigure}[b]{0.5\columnwidth}
    \includegraphics[height=0.12\textheight,width=\textwidth]{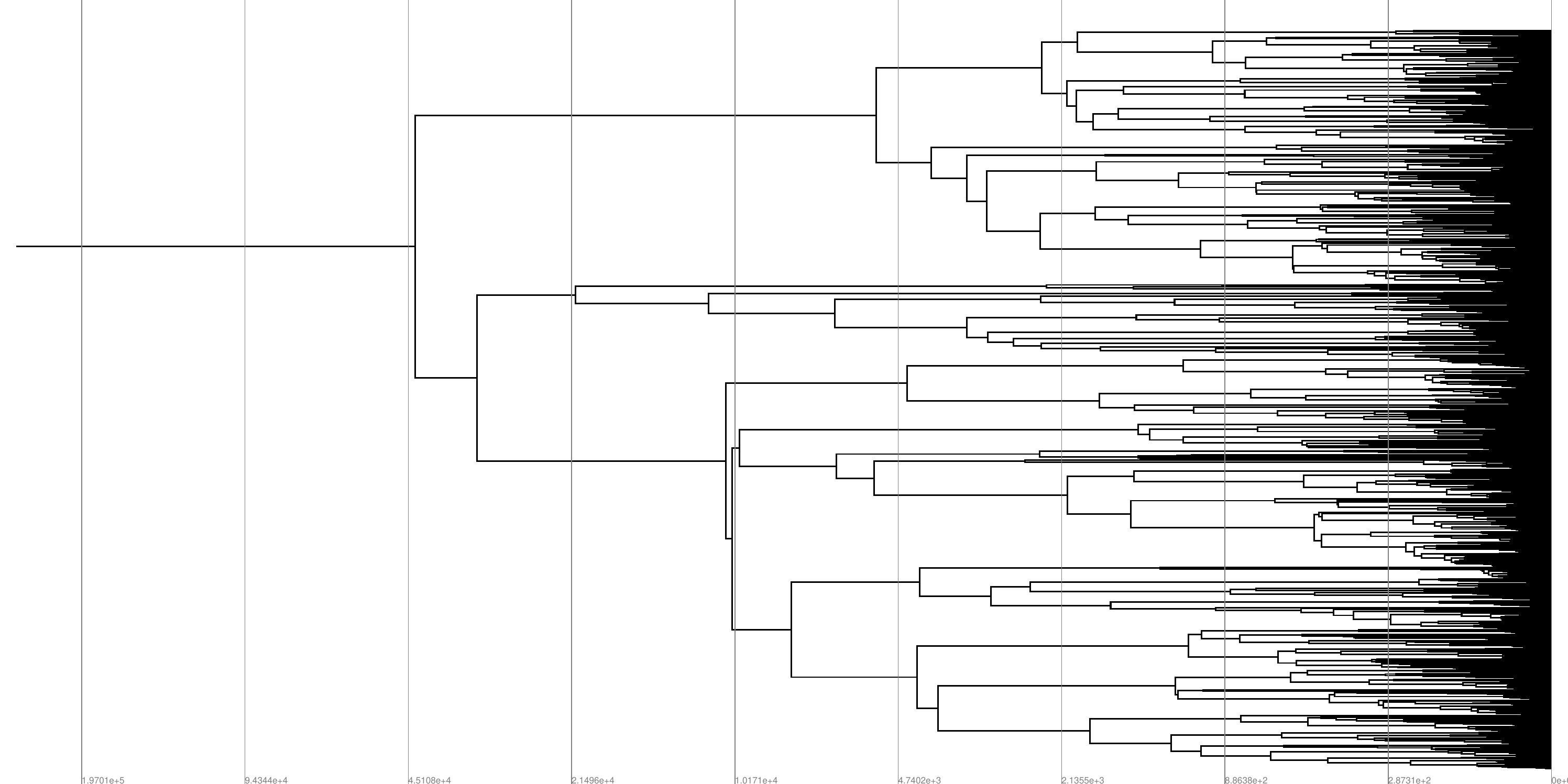}
    \caption{%
      weak selection}
  \end{subfigure}
  \hfill
  \begin{subfigure}[b]{0.5\columnwidth}
    \centering
    \includegraphics[height=0.12\textheight,width=\textwidth]{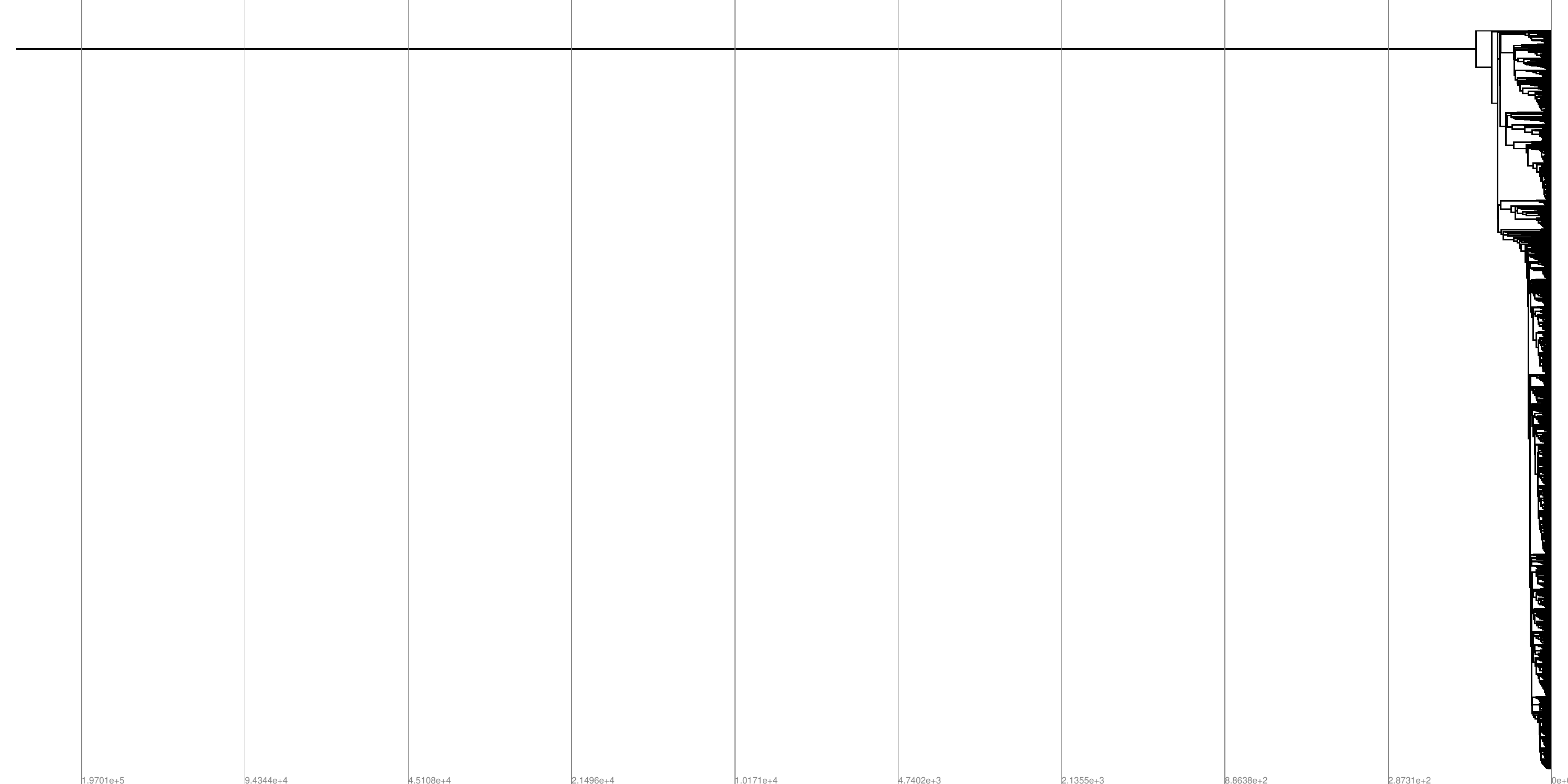}
    \caption{%
      plain}
  \end{subfigure}
  \hfill
  \begin{subfigure}[b]{0.5\columnwidth}
    \includegraphics[height=0.12\textheight,width=\textwidth]{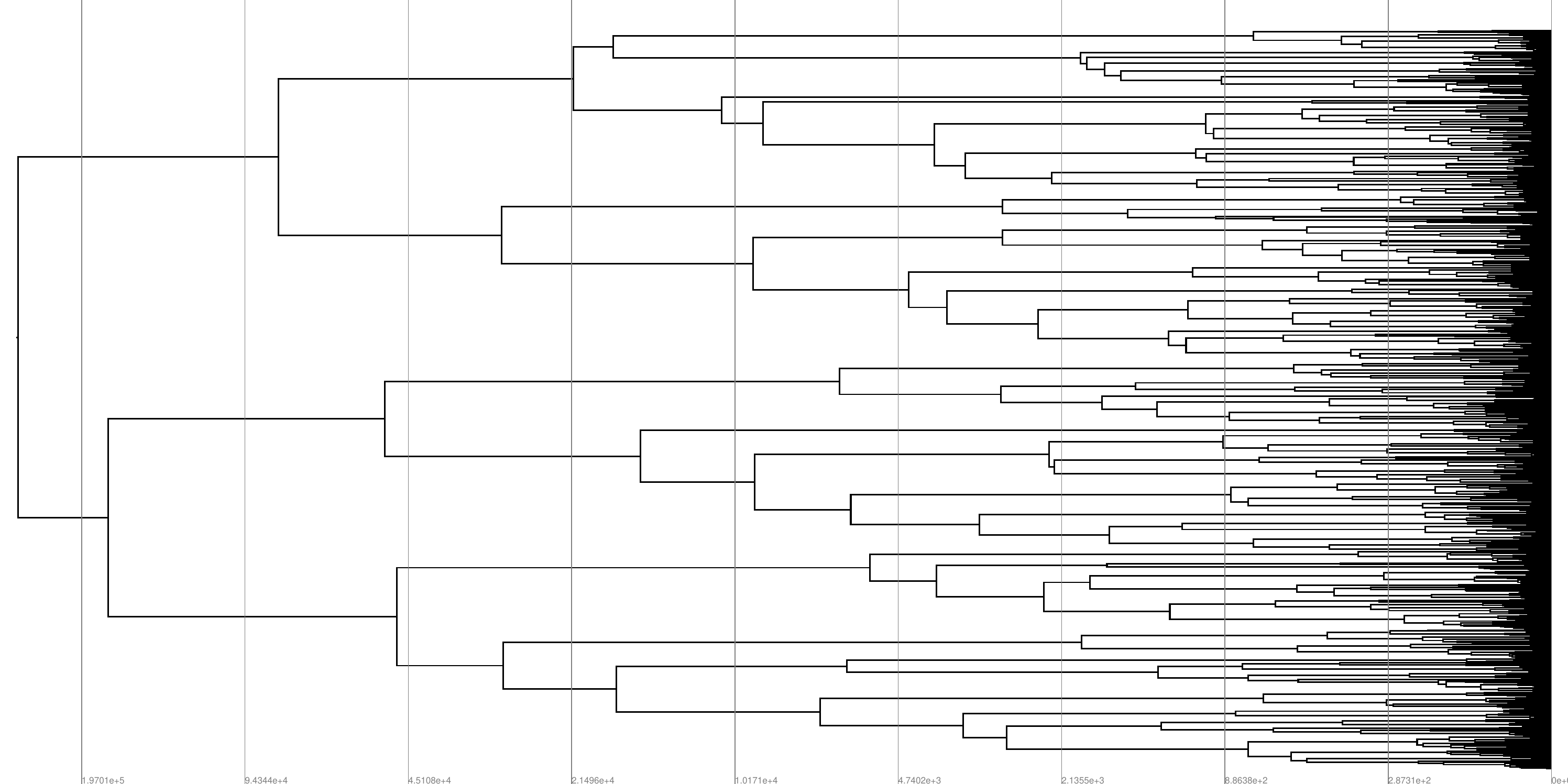}
    \caption{%
      spatial structure}
  \end{subfigure}
  \hfill
  \begin{subfigure}[b]{0.5\columnwidth}
    \includegraphics[height=0.12\textheight,width=\textwidth]{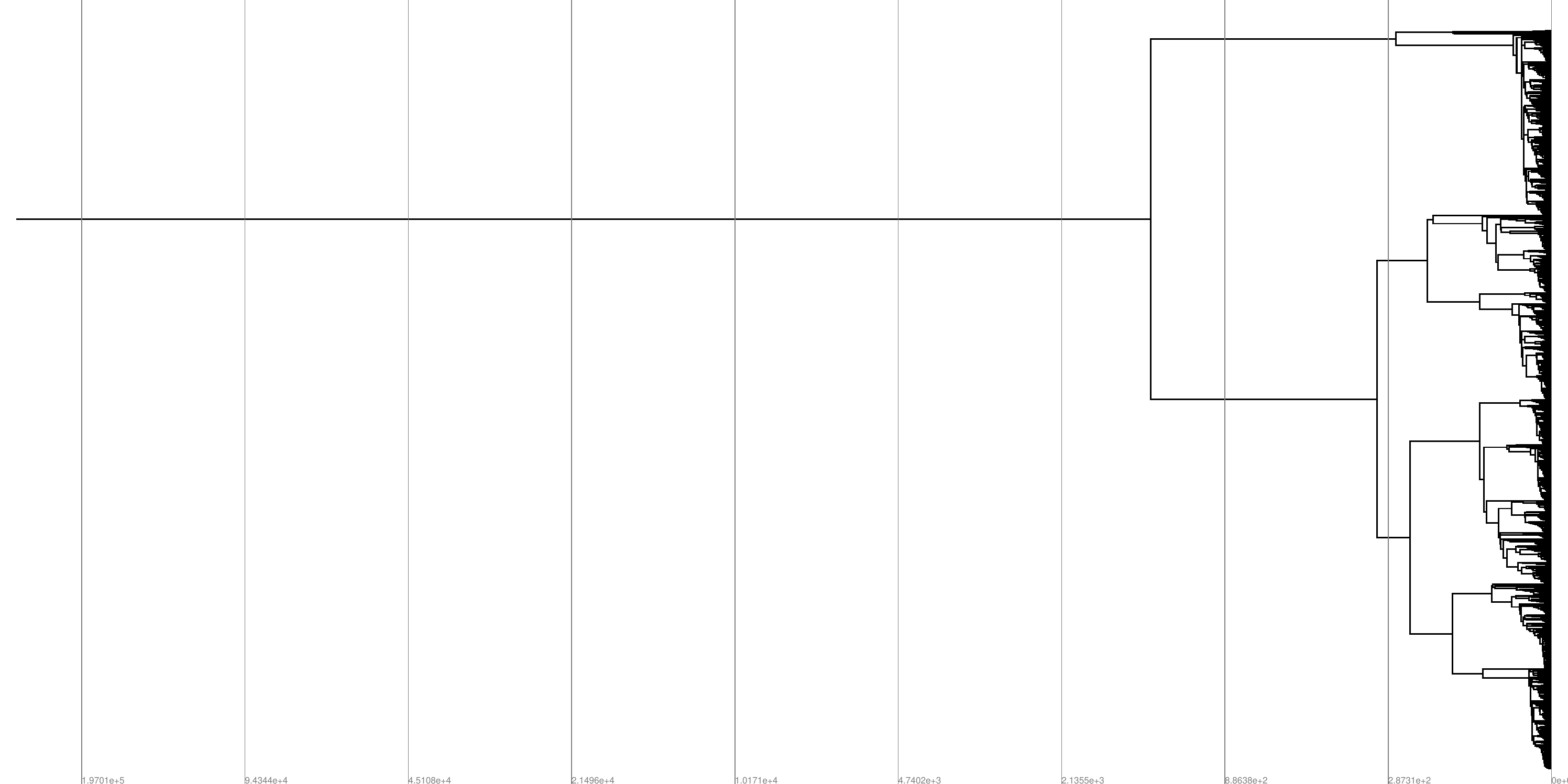}
    \caption{%
      weak 4 niche ecology}
  \end{subfigure}
  \hfill
  \begin{subfigure}[b]{0.5\columnwidth}
      \includegraphics[height=0.12\textheight,width=\textwidth]{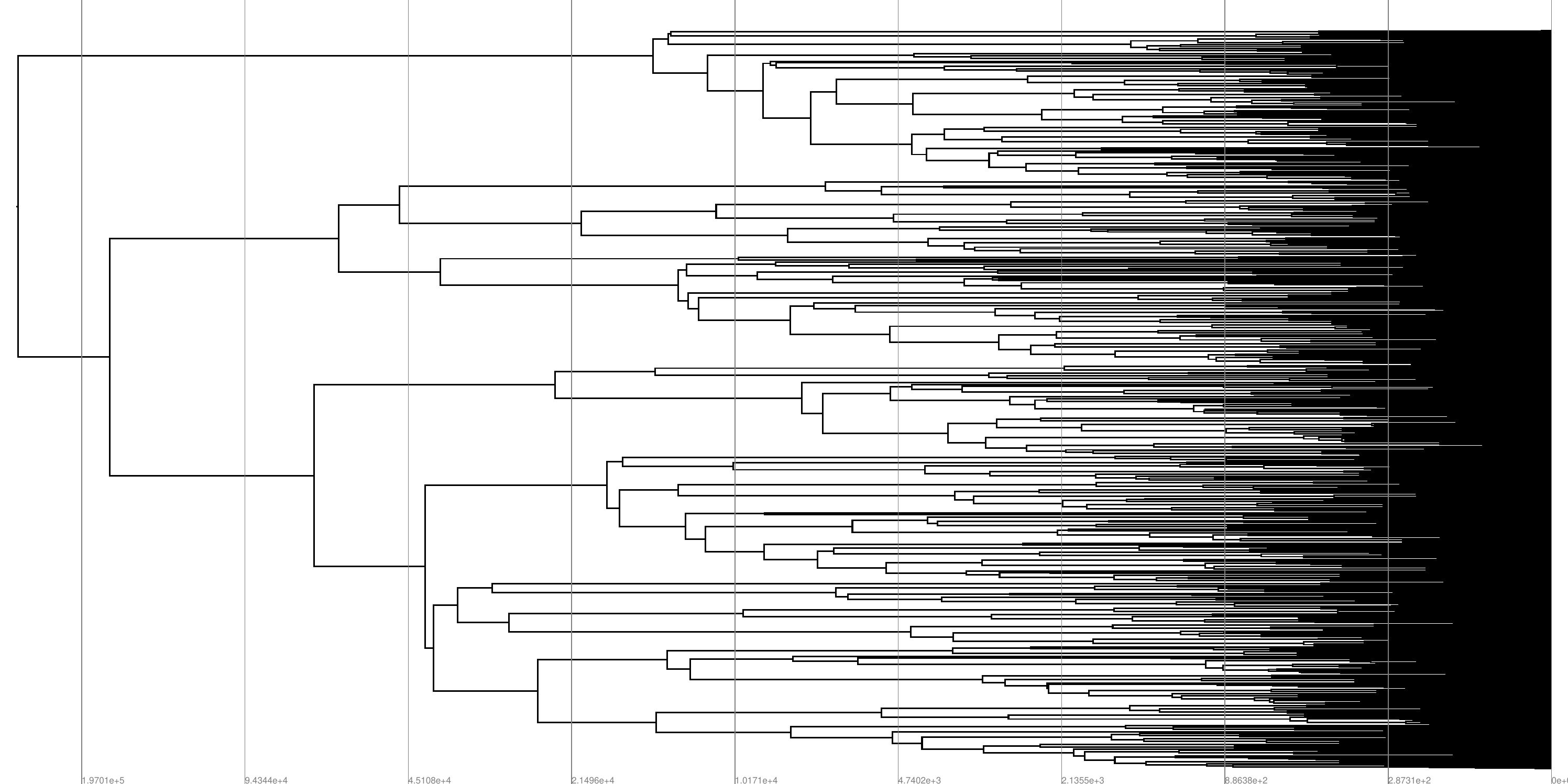}    
    \caption{%
      weak 4 niche ecology with spatial structure }
  \end{subfigure}
  \hfill
  \begin{subfigure}[b]{0.5\columnwidth}
    \includegraphics[height=0.12\textheight,width=\textwidth]{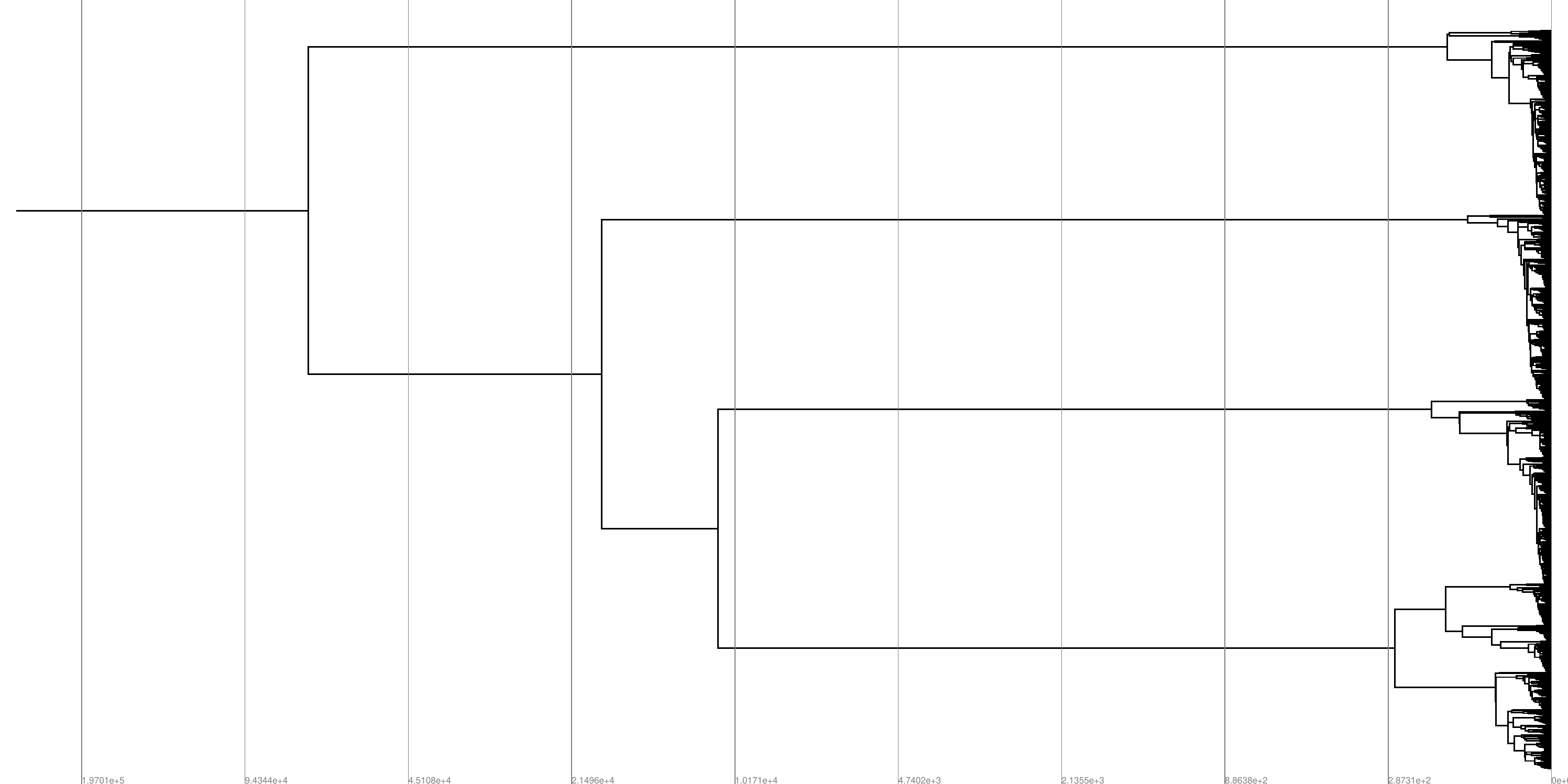}
    \caption{%
      4 niche ecology}
  \end{subfigure}
  \hfill
  \begin{subfigure}[b]{0.5\columnwidth}
    \includegraphics[height=0.12\textheight,width=\textwidth]{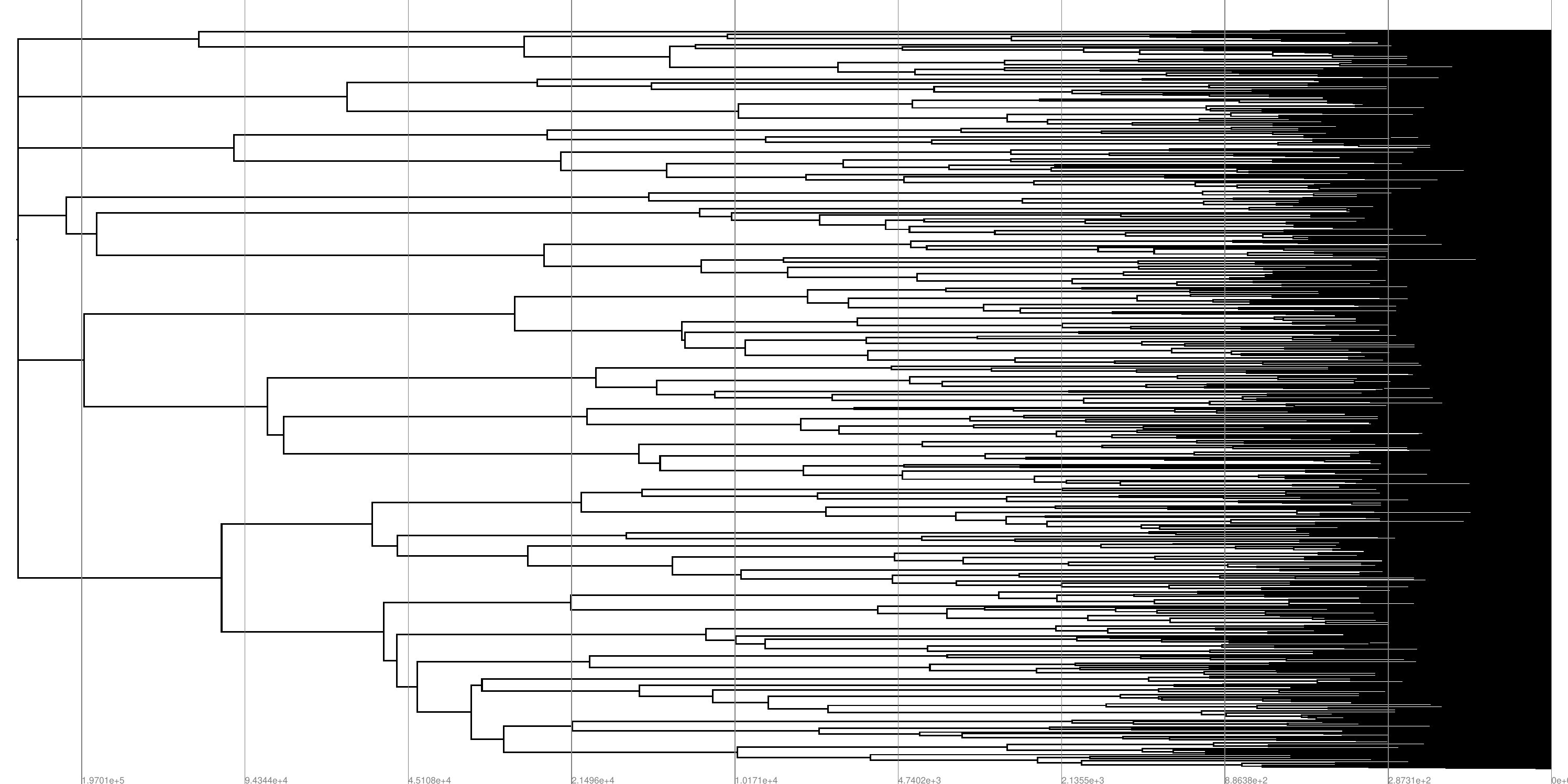}
    \caption{%
      4 niche ecology with spatial structure}
  \end{subfigure}
  \hfill
  \begin{subfigure}[b]{0.5\columnwidth}
    \includegraphics[height=0.12\textheight,width=\textwidth]{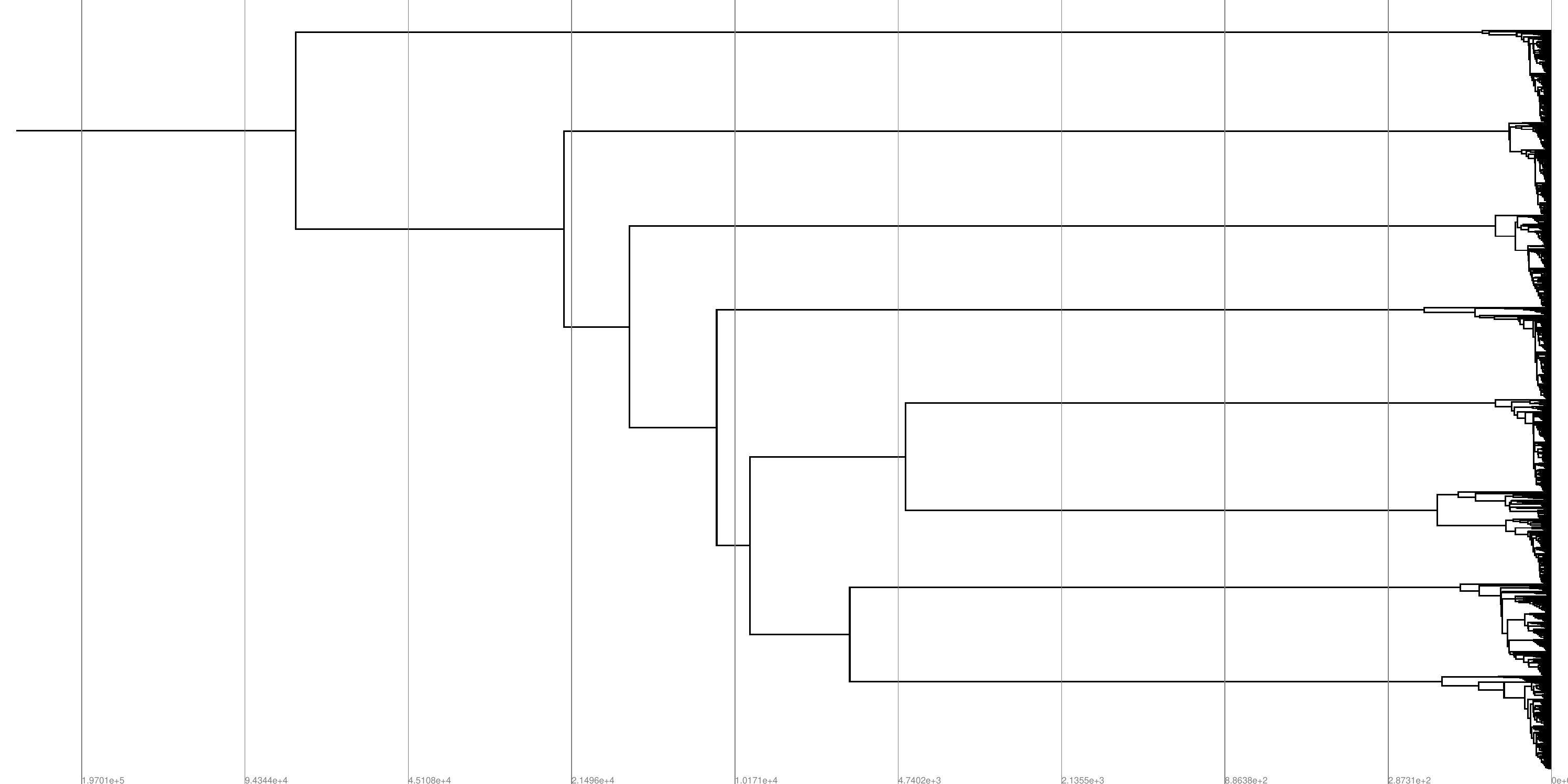}
    \caption{%
      8 niche ecology}
  \end{subfigure}
  \hfill
  \begin{subfigure}[b]{0.5\columnwidth}
    \includegraphics[height=0.12\textheight,width=\textwidth]{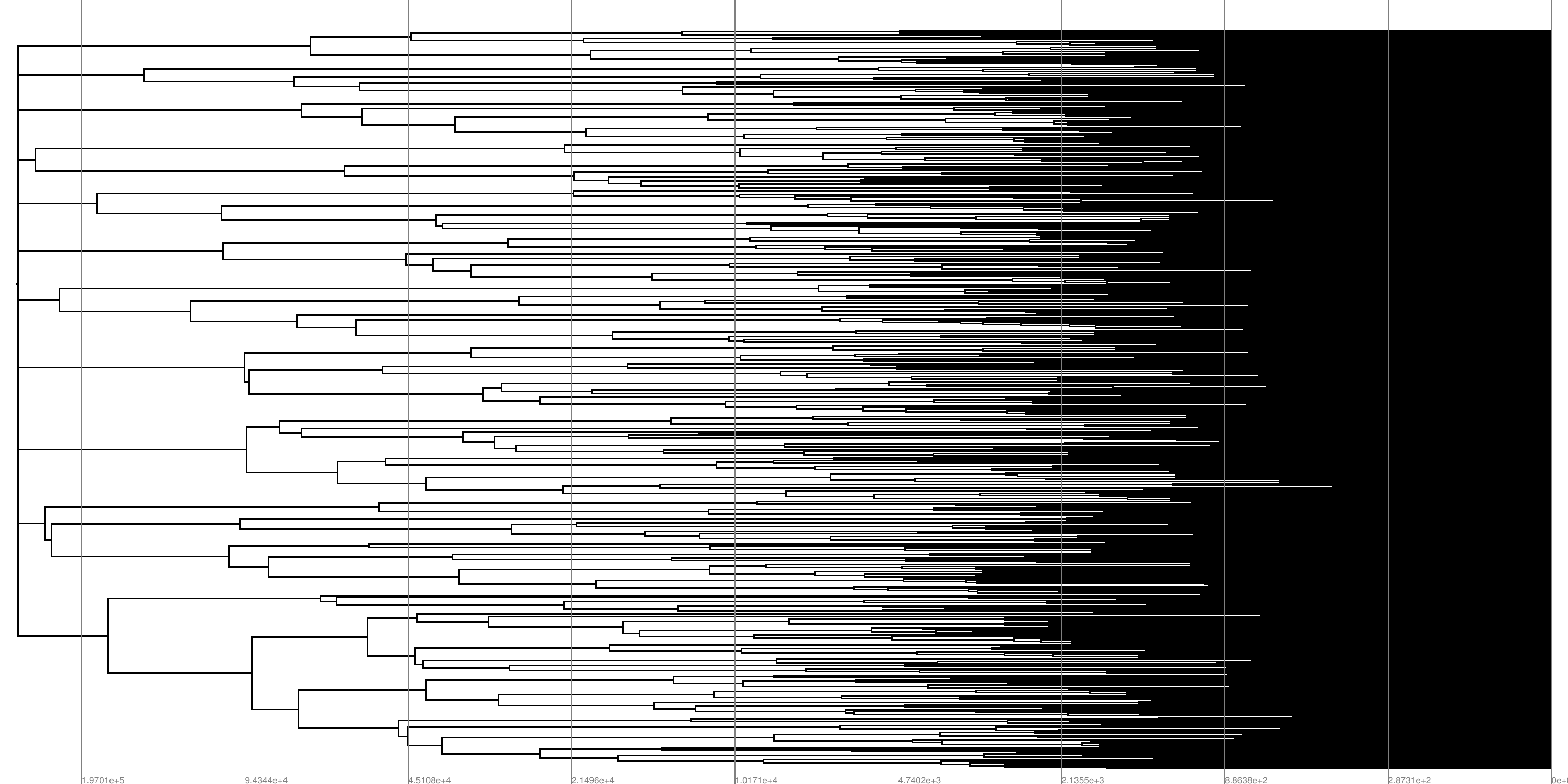}
    \caption{%
      8 niche ecology with spatial structure}
  \end{subfigure}
  \hfill
  \caption{%
    \textbf{Sample reference phylogenies from the simple model across evolutionary regimes.}
    Each phylogeny has 32,768 leaves.
    Note linear-scale $x$ axis.
  }
  \label{fig:perfect-tree-phylogenies-nonlog}
\end{figure*}

\begin{figure*}
  \centering
  \includegraphics[width=\textwidth]{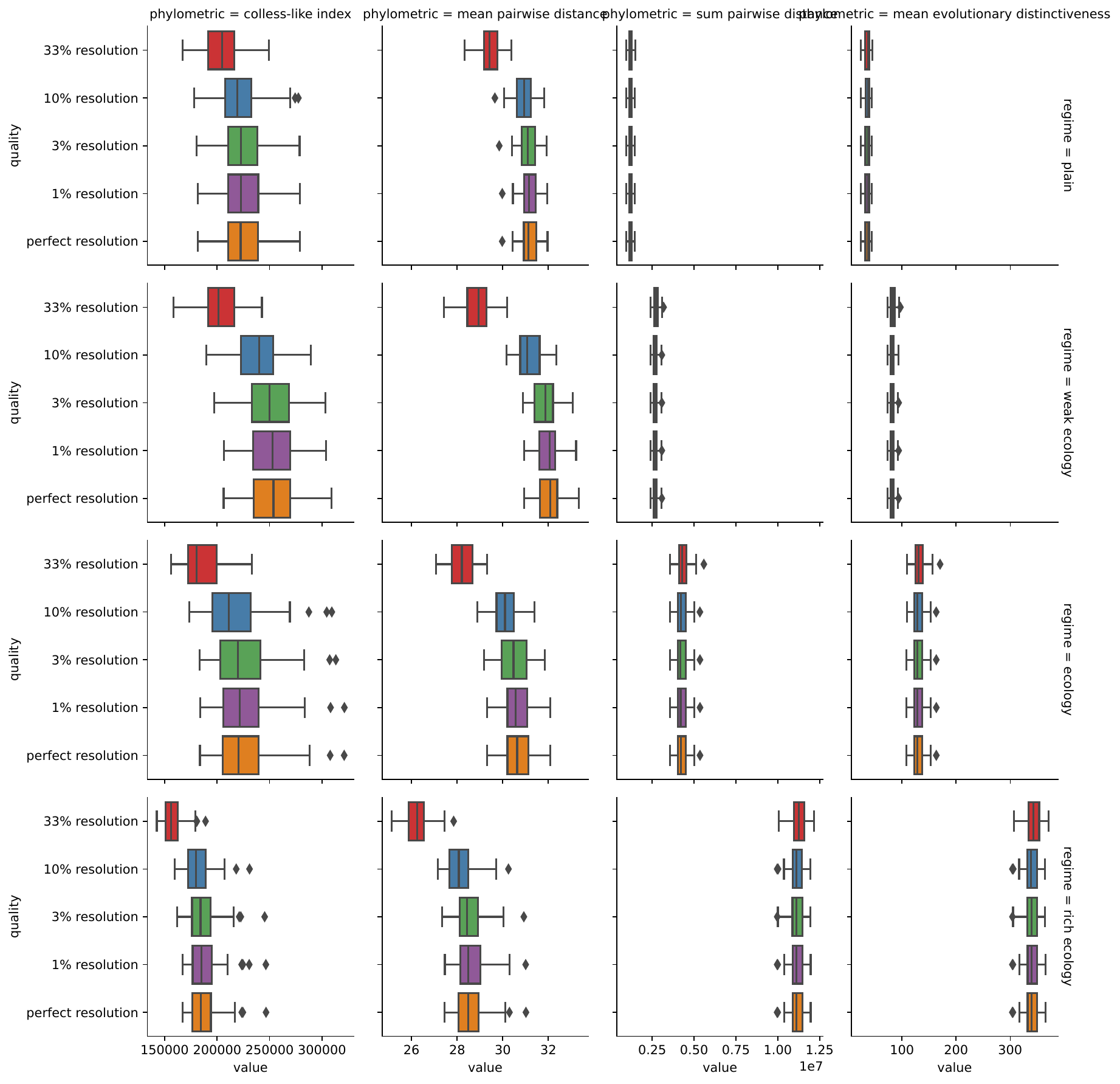}
  \caption{%
    Sensitivity analysis result for distributions of phylometrics across surveyed reconstruction fidelities for evolutionary regimes with underlying spatial structure (i.e., 1,024 niches).
    Sensitivity analysis condition is gaussian mutation distribution at epoch 0 (generation 32,768).
    Sample sizes of $n=50$ replicates define each depicted distribution.
  }
  \label{fig:reconstructed-tree-phylometrics-with-spatial-nuisance-epoch0}
\end{figure*}

\begin{figure*}
  \centering
  \includegraphics[width=\textwidth]{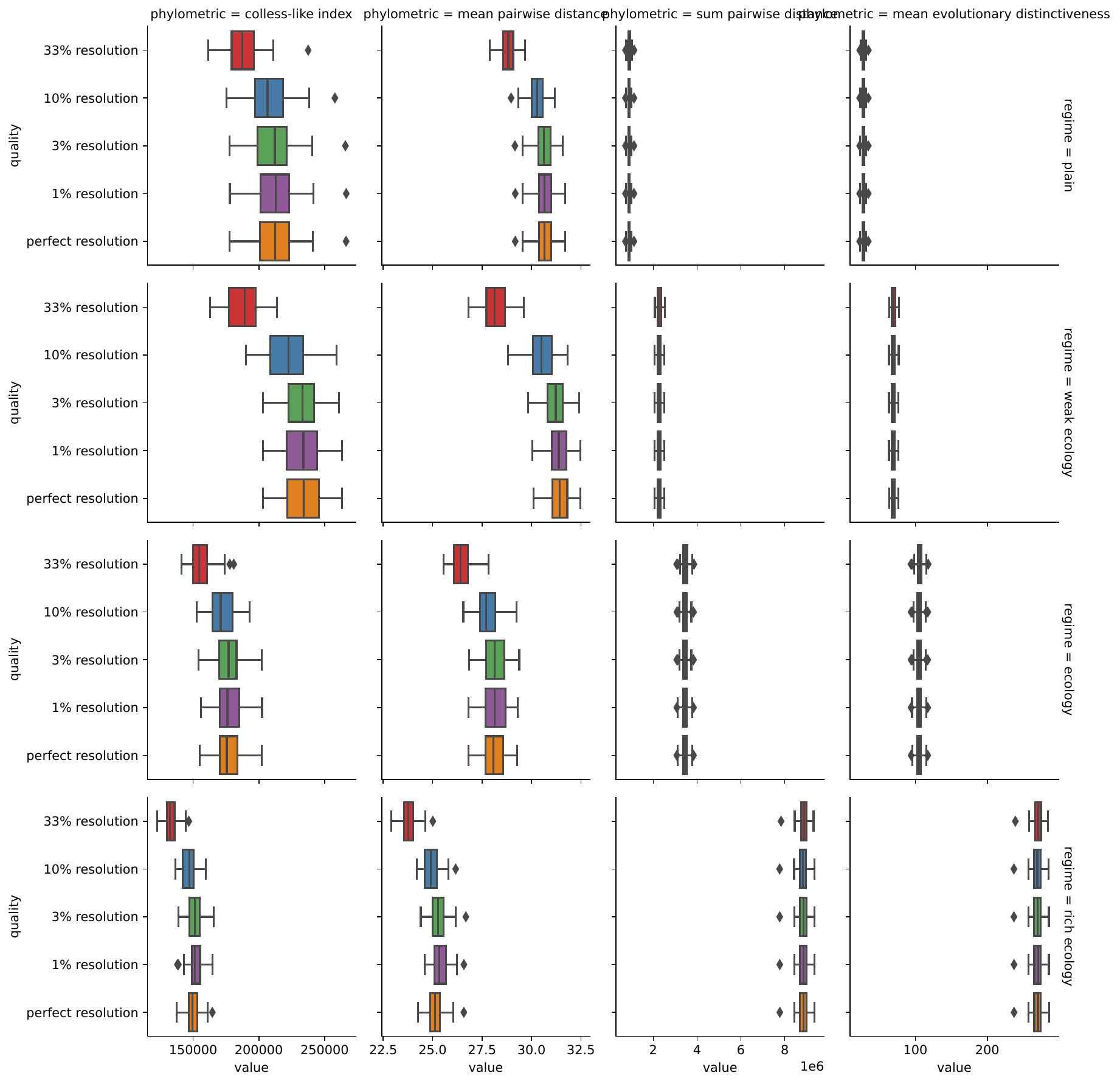}
  \caption{%
    Sensitivity analysis result for distributions of phylometrics across surveyed reconstruction fidelities for evolutionary regimes with underlying spatial structure (i.e., 1,024 niches).
    Sensitivity analysis condition is gaussian mutation distribution at epoch 2 (generation 98,304).
    Sample sizes of $n=50$ replicates define each depicted distribution.
  }
  \label{fig:reconstructed-tree-phylometrics-with-spatial-nuisance-epoch2}
\end{figure*}

\begin{figure*}
  \centering
  \includegraphics[width=\textwidth]{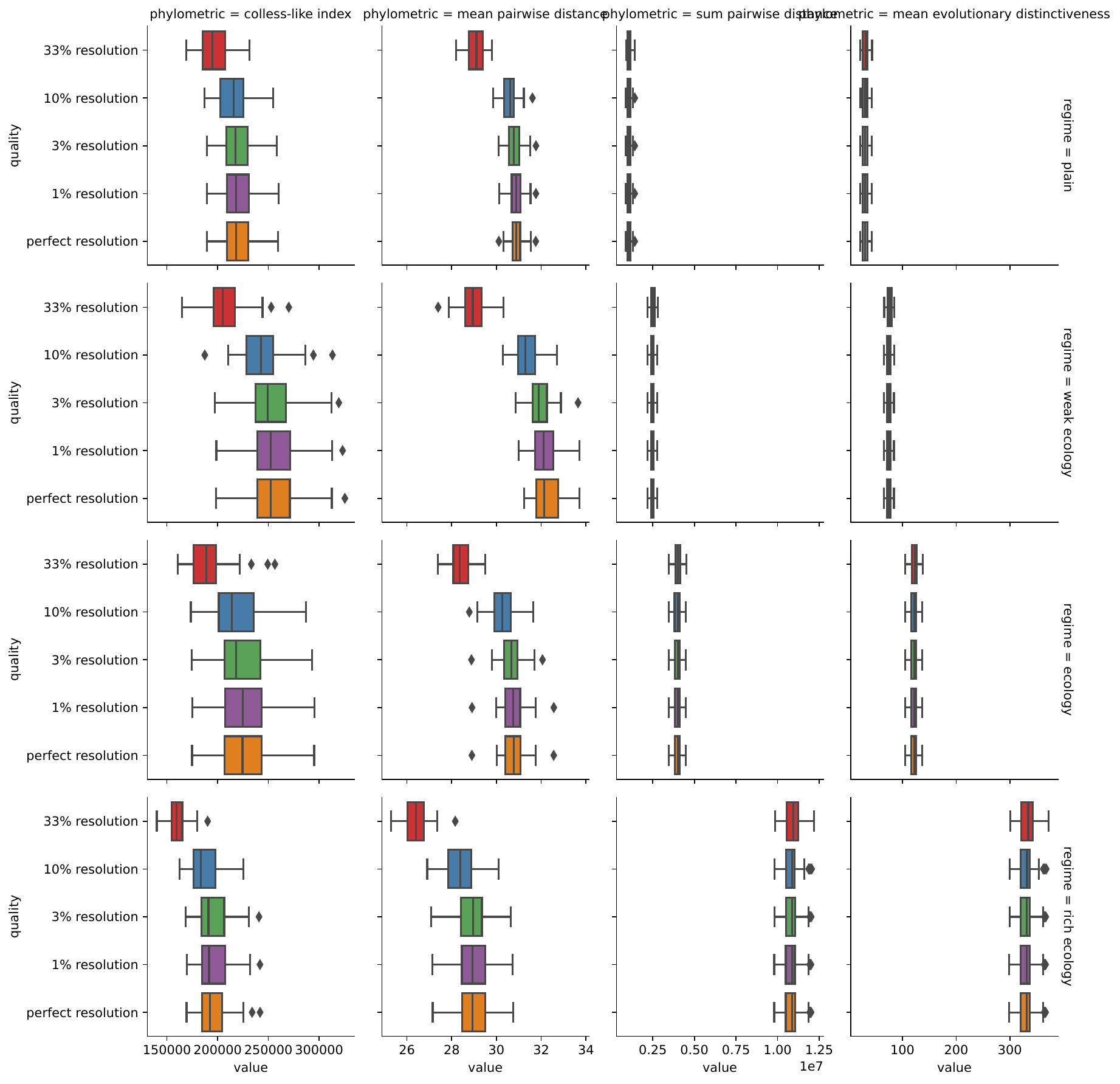}
  \caption{%
    Sensitivity analysis result for distributions of phylometrics across surveyed reconstruction fidelities for evolutionary regimes with underlying spatial structure (i.e., 1,024 niches).
    Sensitivity analysis condition is exponential mutation distribution at epoch 7 (generation 262,144).
    Sample sizes of $n=50$ replicates define each depicted distribution.
  }
  \label{fig:reconstructed-tree-phylometrics-with-spatial-nuisance-exponential}
\end{figure*}

\begin{sidewaysfigure*}
  \centering
  \includegraphics[height=\textwidth,angle=-90]{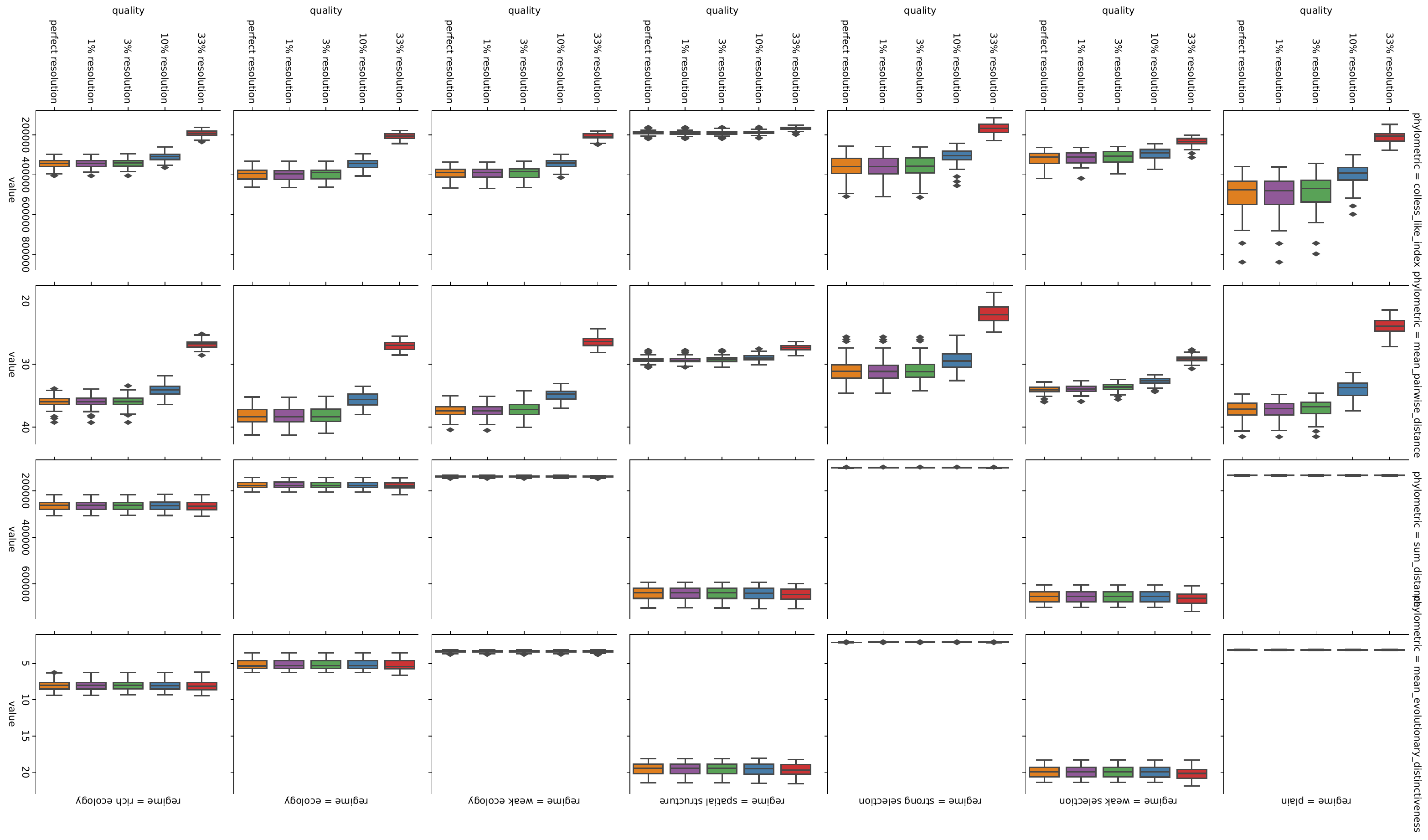}
  \caption{%
    Sensitivity analysis result for distributions of phylometrics across surveyed reconstruction fidelities and evolutionary regimes.
    Sensitivity analysis condition is gaussian mutation distribution at epoch 0 (generation 32,768).
    Sample sizes of $n=50$ replicates define each depicted distribution.
  }
  \label{fig:reconstructed-tree-phylometrics-epoch0}
\end{sidewaysfigure*}

\begin{sidewaysfigure*}
  \centering
  \includegraphics[height=\textwidth,angle=-90]{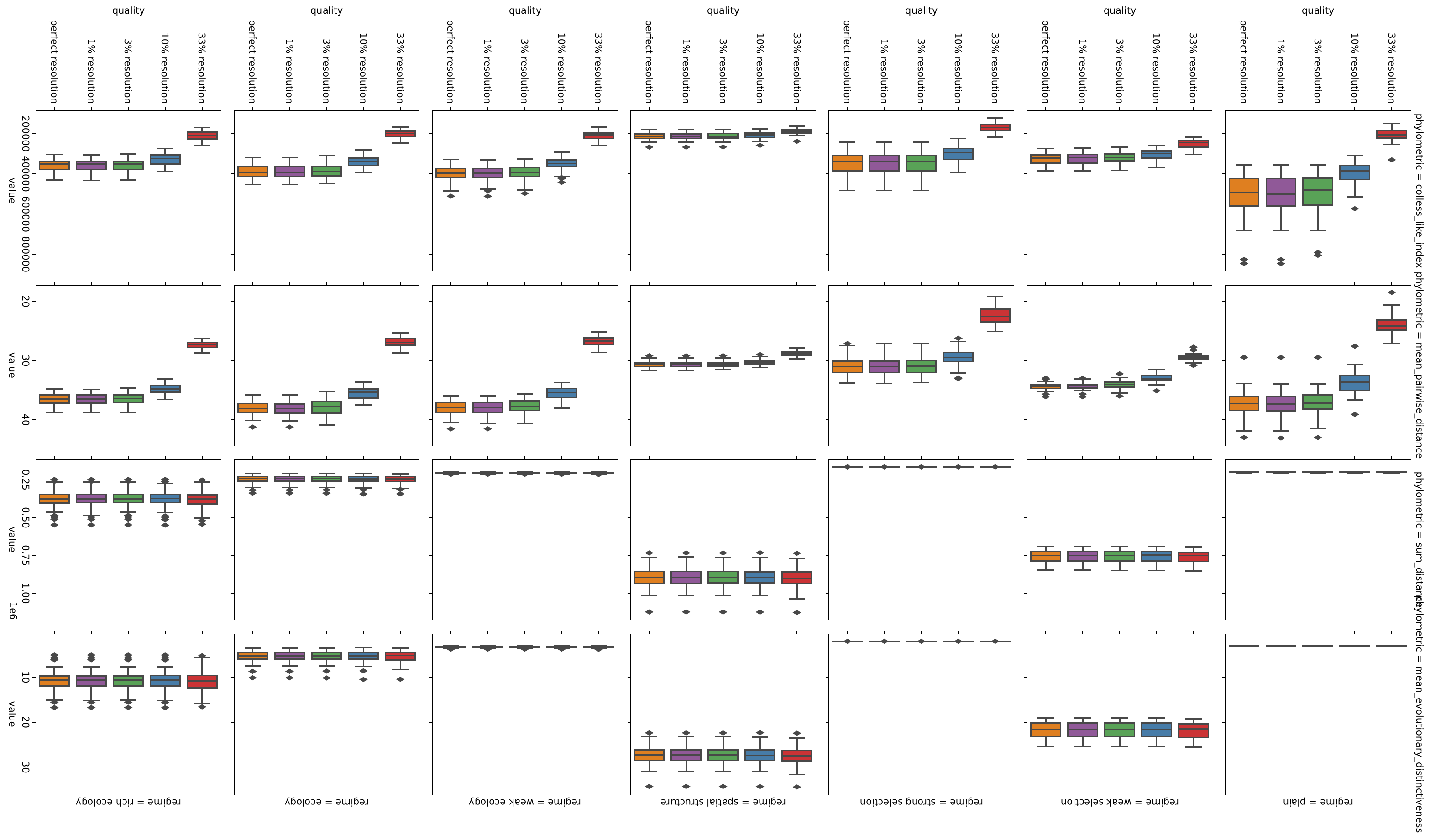}
  \caption{%
    Sensitivity analysis result for distributions of phylometrics across surveyed reconstruction fidelities and evolutionary regimes.
    Sensitivity analysis condition is gaussian mutation distribution at epoch 2 (generation 98,304).
    Sample sizes of $n=50$ replicates define each depicted distribution.
  }
  \label{fig:reconstructed-tree-phylometrics-epoch2}
\end{sidewaysfigure*}

\begin{sidewaysfigure*}
  \centering
  \includegraphics[height=\textwidth,angle=-90]{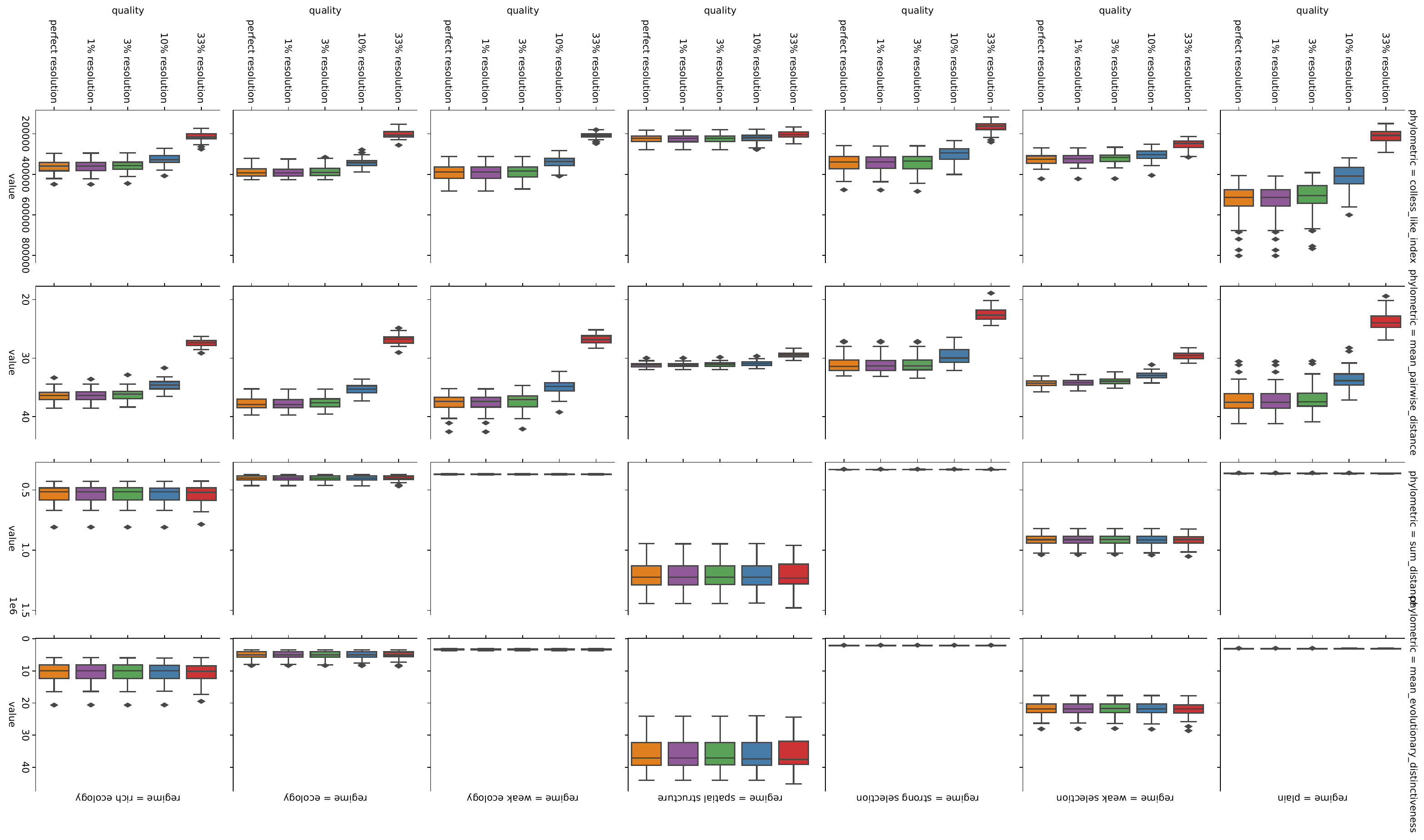}
  \caption{%
    Sensitivity analysis result for distributions of phylometrics across surveyed reconstruction fidelities and surveyed evolutionary regimes.
    Sensitivity analysis condition is exponential mutation distribution at epoch 7 (generation 262,144).
    Sample sizes of $n=50$ replicates define each depicted distribution.
  }
  \label{fig:reconstructed-tree-phylometrics-exponential}
\end{sidewaysfigure*}

\begin{figure*}
  \centering
  \begin{subfigure}[b]{\textwidth}
    \centering
    \includegraphics[width=\textwidth]{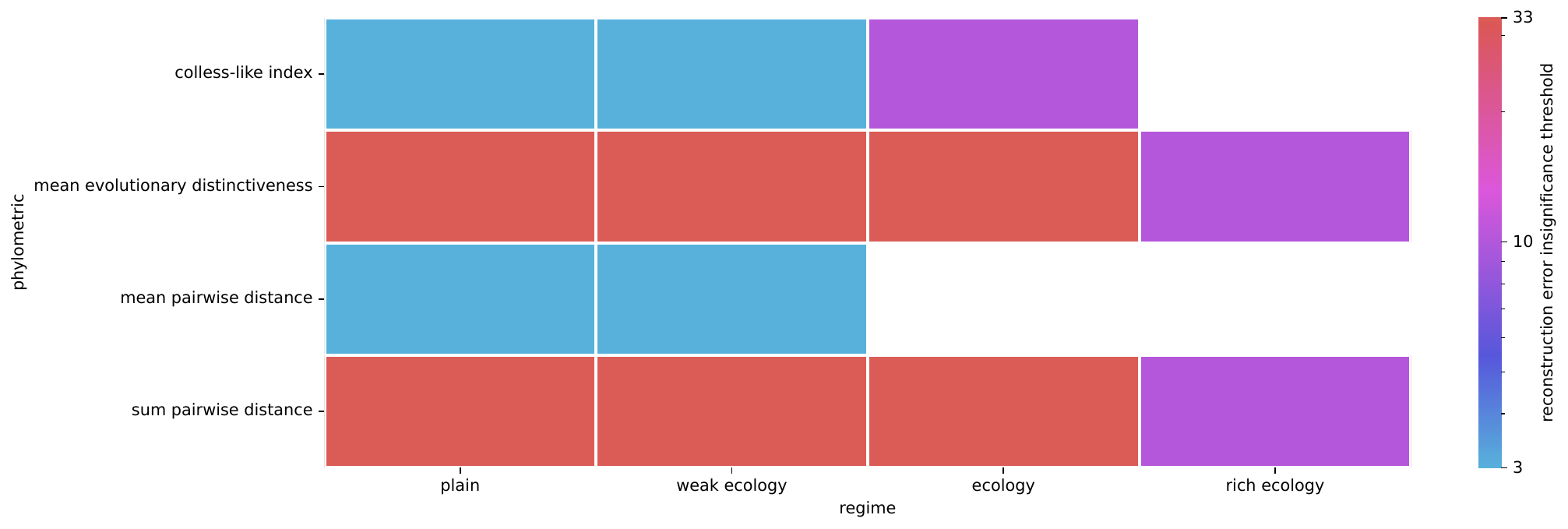}
    \caption{%
      gaussian mutation distribution at epoch 0 (generation 32,768)}
    \label{fig:reconstructed-tree-phylometrics-error-spatial-nuisance-sensitivity-analysis:epoch0}
  \end{subfigure}
  \begin{subfigure}[b]{\textwidth}
    \centering
    \includegraphics[width=\textwidth]{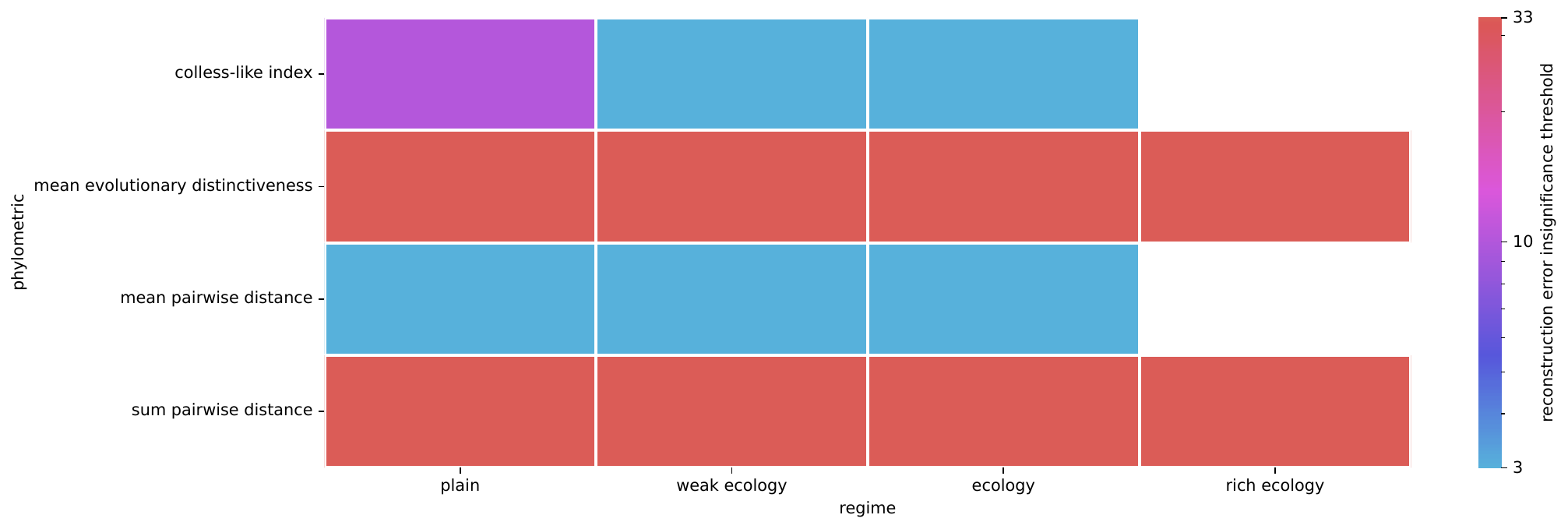}
    \caption{%
      gaussian mutation distribution at epoch 2 (generation 98,304)}
    \label{fig:reconstructed-tree-phylometrics-error-spatial-nuisance-sensitivity-analysis:epoch2}
  \end{subfigure}
  \begin{subfigure}[b]{\textwidth}
    \centering
    \includegraphics[width=\textwidth]{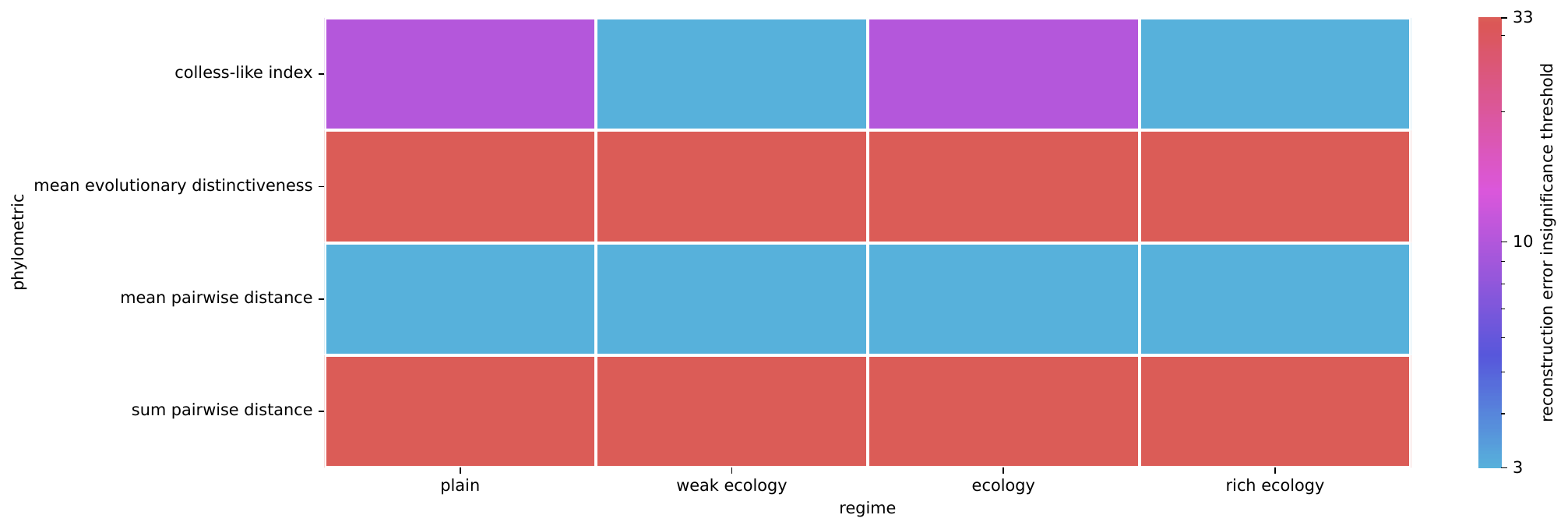}
    \caption{%
    exponential mutation distribution at epoch 7 (generation 262,144)}
    \label{fig:reconstructed-tree-phylometrics-error-spatial-nuisance-sensitivity-analysis:exponential}
  \end{subfigure}
  \caption{%
    Sensitivity analysis results for reconstruction resolutions required to achieve statistical indistinguishability between reconstructions corresponding reference trees for each phylometric across surveyed evolutionary conditions with spatial structure (i.e., island count 1,024).
    Significance level $p<0.05$ under the Wilcoxon signed-rank test between samples of 50 replicates each is used as the threshold for statistical distinguishability.
    Phylometrics with looser reconstruction resolution thresholds (i.e., higher resolution percentages) are less sensitive to reconstruction error.
    White heat map tiles indicate that no surveyed reconstruction resolution threshold was sufficient to achieve indistinguishability from the reference tree with respect a particular phylometric.
  }
  \label{fig:reconstructed-tree-phylometrics-error-spatial-nuisance-sensitivity-analysis}
\end{figure*}

\begin{sidewaysfigure*}
  \centering
  \includegraphics[height=\textwidth,angle=-90]{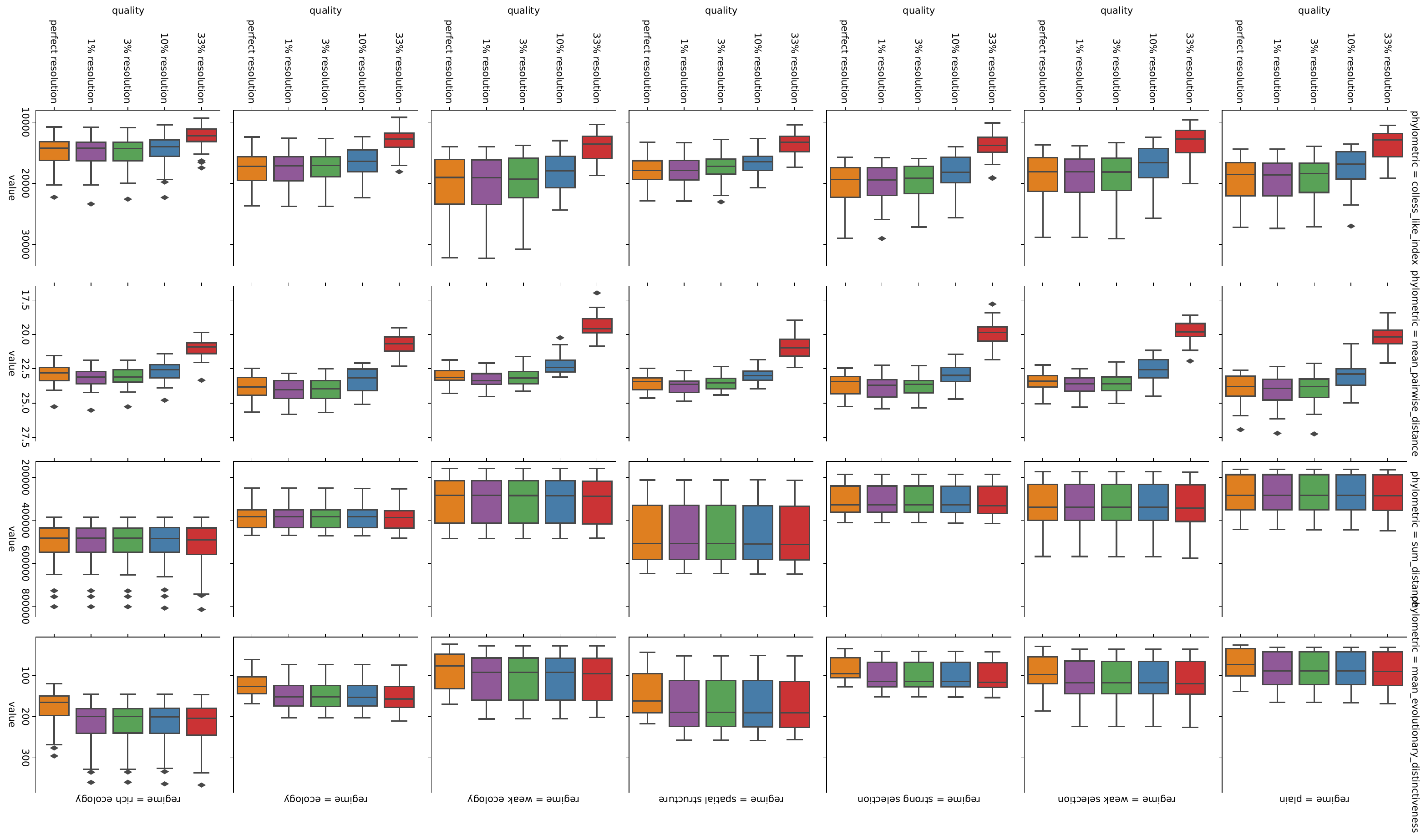}
  \caption{%
    Distributions of phylometrics across surveyed reconstruction fidelities and evolutionary regimes.
    Results are for standard experimental conditions: gaussian mutation distribution at epoch 7 (generation 262,144).
    See Figures \labelcref{fig:reconstructed-tree-phylometrics-epoch0,fig:reconstructed-tree-phylometrics-epoch2,fig:reconstructed-tree-phylometrics-exponential} for results under sensitivity analysis conditions.
    Sample sizes $n=50$.
  }
  \label{fig:reconstructed-tree-phylometrics}
\end{sidewaysfigure*}

\begin{figure*}
  \centering
  \includegraphics[width=\textwidth]{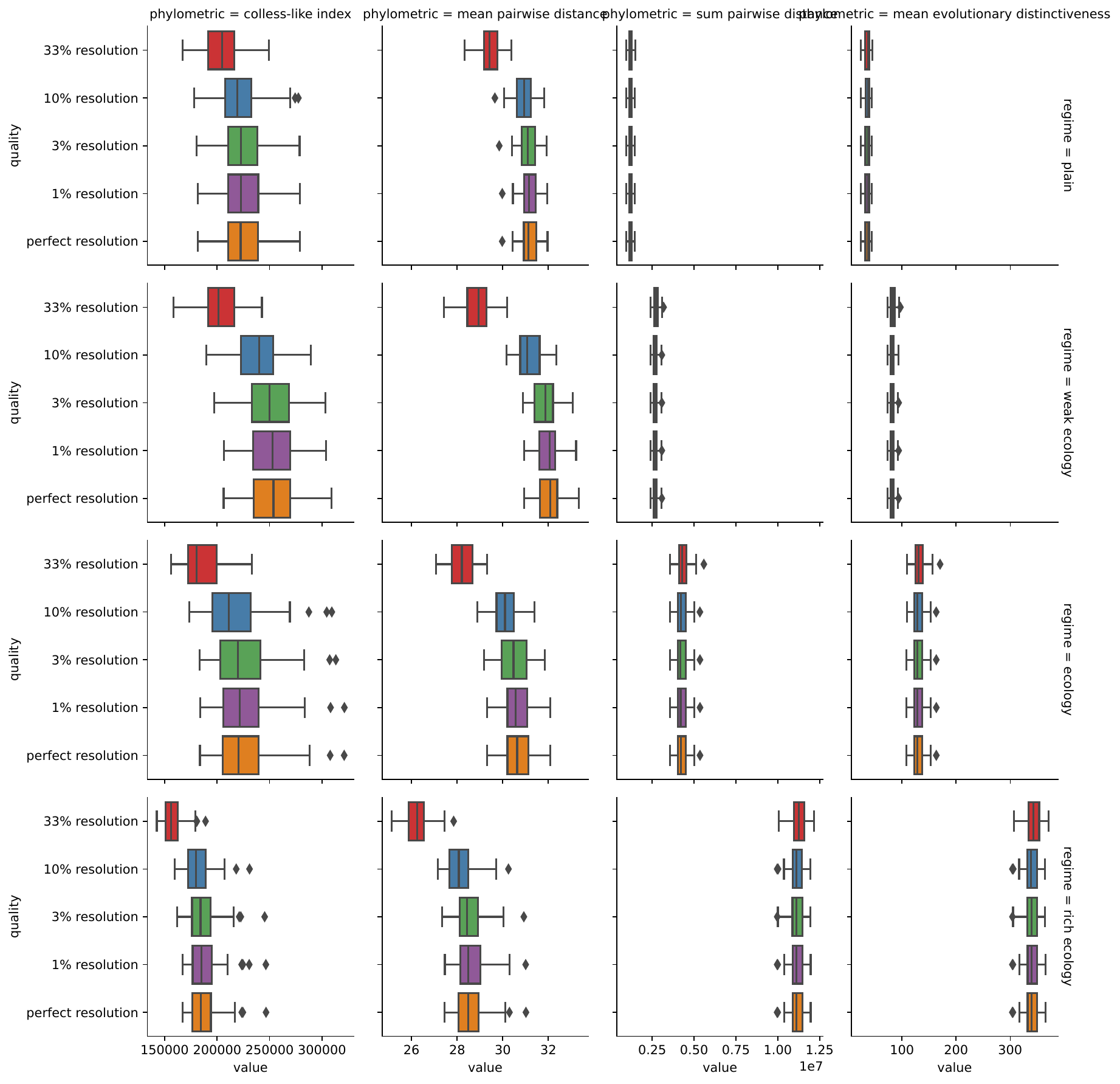}
  \caption{%
    Distributions of phylometrics across surveyed reconstruction fidelities for evolutionary regimes with underlying spatial structure (i.e., 1,024 niches).
    Results are for standard experimental conditions: gaussian mutation distribution at epoch 7 (generation 262,144).
    See Figures \labelcref{fig:reconstructed-tree-phylometrics-with-spatial-nuisance-epoch0,fig:reconstructed-tree-phylometrics-with-spatial-nuisance-epoch2,fig:reconstructed-tree-phylometrics-with-spatial-nuisance-exponential} for results under sensitivity analysis conditions.
    Sample sizes of $n=50$ replicates define each depicted distribution.
  }
  \label{fig:reconstructed-tree-phylometrics-with-spatial-nuisance}
\end{figure*}

\begin{figure*}
  \centering
  \includegraphics[width=\textwidth]{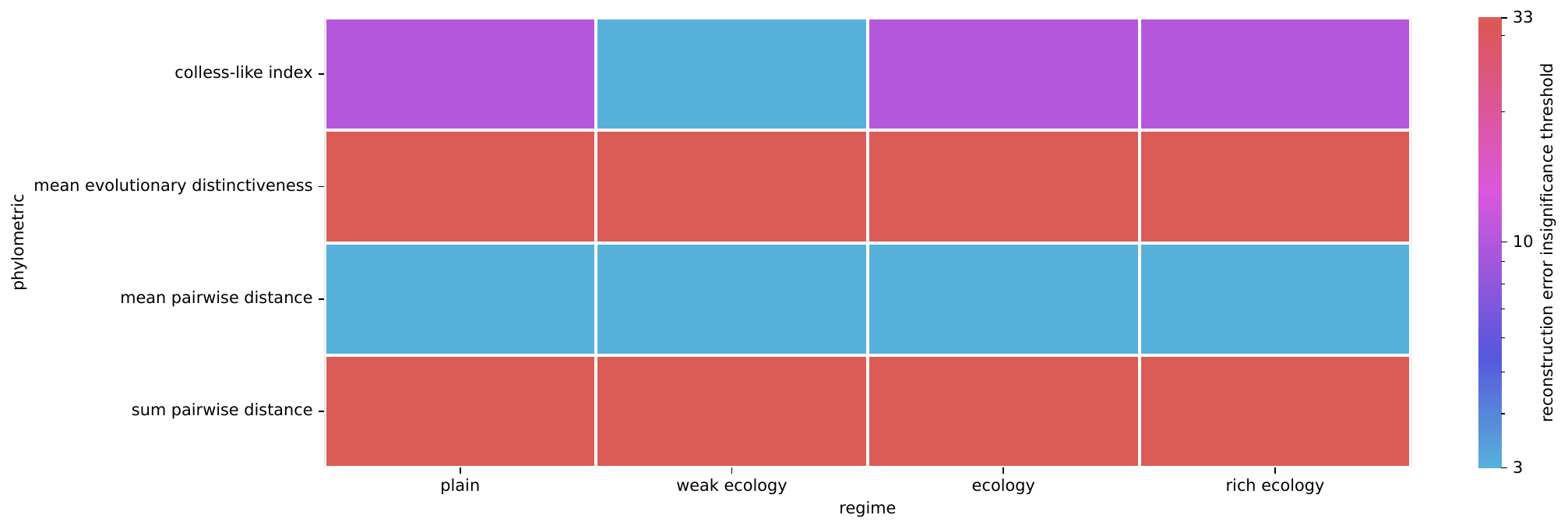}
  \caption{%
    Reconstruction resolutions required to achieve statistical indistinguishability between reconstructions corresponding reference trees for each phylometric across surveyed evolutionary conditions with spatial structure (i.e., island count 1,024).
    Significance level $p<0.05$ under the Wilcoxon signed-rank test between samples of 50 replicates each is used as the threshold for statistical distinguishability.
    Phylometrics with looser reconstruction resolution thresholds (i.e., higher resolution percentages) are less sensitive to reconstruction error.
    White heat map tiles indicate that no surveyed reconstruction resolution threshold was sufficient to achieve indistinguishability from the reference tree with respect a particular phylometric.
    See Supplementary Figure \ref{fig:reconstructed-tree-phylometrics-error-spatial-nuisance-sensitivity-analysis} for sensitivity analysis results.
  }
  \label{fig:reconstructed-tree-phylometrics-error-spatial-nuisance}
\end{figure*}

\providecommand{\thecaption}{
  Summary statistics for reconstruction error across reconstruction fidelity levels and evolutionary conditions, measured as quartet distance between reconstructions and corresponding reference trees.
}

\begin{landscape}
  \csvstyle{myTableStyle}{
    longtable=c c c c c c c c,
    before reading=\scriptsize,
    table head=\toprule evolutionary regime & reconstruction quality & sensitivity analysis conditions & quartet dist mean & quartet dist median & quartet dist std & quartet dist max\\ \midrule\endhead\bottomrule\\[-1em]
    \caption{%
      \thecaption
    }\endfoot\bottomrule\\[-1em]
    \caption{%
      \thecaption
    }
    \label{tab:tree-reconstruction-quality-quartet-summary-statistics}\endlastfoot,
    head to column names,
    respect all,
  }


\csvreader[myTableStyle]{binder/binder/outdata/a=tree-reconstruction-quality-quartet-summary-statistics.csv}{}{\csvlinetotablerow}
\end{landscape}

\providecommand{\thecaption}{
  Summary statistics for reconstruction error across reconstruction fidelity levels and evolutionary conditions, measured as triplet distance between reconstructions and corresponding reference trees.
}

\begin{landscape}
  \csvstyle{myTableStyle}{
    longtable=c c c c c c c c,
    before reading=\scriptsize,
    table head=\toprule evolutionary regime & reconstruction quality & sensitivity analysis conditions & triplet dist mean & triplet dist median & triplet dist std & triplet dist max\\ \midrule\endhead\bottomrule\\[-1em]
    \caption{%
      \thecaption
    }\endfoot\bottomrule\\[-1em]
    \caption{%
      \thecaption
    }
    \label{tab:tree-reconstruction-triplet-summary-statistics}\endlastfoot,
    head to column names,
    respect all,
  }


\csvreader[myTableStyle]{binder/binder/outdata/a=tree-reconstruction-quality-triplet-summary-statistics.csv}{}{\csvlinetotablerow}

\end{landscape}

\providecommand{\thecaption}{
  All-pairs two-tailed Wilcoxon signed-rank comparisons of phylometrics between surveyed evolutionary regimes with spatial structure (i.
  e., island count 1,024).
  Phylometrics were calculated on perfect-fidelity reference trees.
  All-pairs comparisons within each sensitivity analysis condition are provided.
  Sample size $n=50$ for each population.
  Reported $p$ values are not corrected for multiple comparisons.
  Instead, a Bonferoni-corrected significance threshold of $5.26 \times 10^{-4}$ was used.
}

\begin{landscape}
  \csvstyle{myTableStyle}{
    longtable=c c c c c c c c,
    before reading=\footnotesize,
    table head=\toprule evolutionary regime 1 & evolutionary regime 2 & $n$ & Phylometric & Statistic & $p$ & Epoch & Mutation Distribution\\ \midrule\endhead\bottomrule\\[-1em]
    \caption{%
      \thecaption
    }\endfoot\bottomrule\\[-1em]
    \caption{%
      \thecaption
    }
    \label{tab:phylostatistics-comparison-between-regimes-allpairs-spatial-nuisance-wilcox}\endlastfoot,
    head to column names,
    respect all,
  }

  \begin{filecontents*}{binder/binder/outdata/a=phylostatistics-comparison-between-regimes-allpairs-spatial-nuisance-wilcox+ext=.csv}
regime1,regime2,n,phylometric,statistic,p,epoch,mut_distn
ecology,plain,50,colless-like index,0.0,7.556929455863566e-10,0,np.random.exponential
ecology,plain,50,mean pairwise distance,0.0,7.556929455863566e-10,0,np.random.exponential
ecology,plain,50,sum pairwise distance,0.0,7.556929455863566e-10,0,np.random.exponential
ecology,plain,50,mean evolutionary distinctiveness,0.0,7.556929455863566e-10,0,np.random.exponential
ecology,rich ecology,50,colless-like index,1.0,8.031090907046913e-10,0,np.random.exponential
ecology,rich ecology,50,mean pairwise distance,0.0,7.556929455863566e-10,0,np.random.exponential
ecology,rich ecology,50,sum pairwise distance,0.0,7.556929455863566e-10,0,np.random.exponential
ecology,rich ecology,50,mean evolutionary distinctiveness,0.0,7.556929455863566e-10,0,np.random.exponential
ecology,weak ecology,50,colless-like index,0.0,7.556929455863566e-10,0,np.random.exponential
ecology,weak ecology,50,mean pairwise distance,0.0,7.556929455863566e-10,0,np.random.exponential
ecology,weak ecology,50,sum pairwise distance,0.0,7.556929455863566e-10,0,np.random.exponential
ecology,weak ecology,50,mean evolutionary distinctiveness,0.0,7.556929455863566e-10,0,np.random.exponential
plain,rich ecology,50,colless-like index,0.0,7.556929455863566e-10,0,np.random.exponential
plain,rich ecology,50,mean pairwise distance,0.0,7.556929455863566e-10,0,np.random.exponential
plain,rich ecology,50,sum pairwise distance,0.0,7.556929455863566e-10,0,np.random.exponential
plain,rich ecology,50,mean evolutionary distinctiveness,0.0,7.556929455863566e-10,0,np.random.exponential
weak ecology,plain,50,colless-like index,189.0,1.4945133405887308e-05,0,np.random.exponential
weak ecology,plain,50,mean pairwise distance,606.0,0.7610682552125038,0,np.random.exponential
weak ecology,plain,50,sum pairwise distance,0.0,7.556929455863566e-10,0,np.random.exponential
weak ecology,plain,50,mean evolutionary distinctiveness,0.0,7.556929455863566e-10,0,np.random.exponential
weak ecology,rich ecology,50,colless-like index,0.0,7.556929455863566e-10,0,np.random.exponential
weak ecology,rich ecology,50,mean pairwise distance,0.0,7.556929455863566e-10,0,np.random.exponential
weak ecology,rich ecology,50,sum pairwise distance,0.0,7.556929455863566e-10,0,np.random.exponential
weak ecology,rich ecology,50,mean evolutionary distinctiveness,0.0,7.556929455863566e-10,0,np.random.exponential
plain,ecology,50,colless-like index,0.0,7.556929455863566e-10,0,np.random.standard_normal
plain,ecology,50,mean pairwise distance,0.0,7.556929455863566e-10,0,np.random.standard_normal
plain,ecology,50,sum pairwise distance,0.0,7.556929455863566e-10,0,np.random.standard_normal
plain,ecology,50,mean evolutionary distinctiveness,0.0,7.556929455863566e-10,0,np.random.standard_normal
plain,rich ecology,50,colless-like index,0.0,7.556929455863566e-10,0,np.random.standard_normal
plain,rich ecology,50,mean pairwise distance,0.0,7.556929455863566e-10,0,np.random.standard_normal
plain,rich ecology,50,sum pairwise distance,0.0,7.556929455863566e-10,0,np.random.standard_normal
plain,rich ecology,50,mean evolutionary distinctiveness,0.0,7.556929455863566e-10,0,np.random.standard_normal
plain,weak ecology,50,colless-like index,451.0,0.07180752515234809,0,np.random.standard_normal
plain,weak ecology,50,mean pairwise distance,381.0,0.013283902663624031,0,np.random.standard_normal
plain,weak ecology,50,sum pairwise distance,0.0,7.556929455863566e-10,0,np.random.standard_normal
plain,weak ecology,50,mean evolutionary distinctiveness,0.0,7.556929455863566e-10,0,np.random.standard_normal
rich ecology,ecology,50,colless-like index,0.0,7.556929455863566e-10,0,np.random.standard_normal
rich ecology,ecology,50,mean pairwise distance,0.0,7.556929455863566e-10,0,np.random.standard_normal
rich ecology,ecology,50,sum pairwise distance,0.0,7.556929455863566e-10,0,np.random.standard_normal
rich ecology,ecology,50,mean evolutionary distinctiveness,0.0,7.556929455863566e-10,0,np.random.standard_normal
rich ecology,weak ecology,50,colless-like index,0.0,7.556929455863566e-10,0,np.random.standard_normal
rich ecology,weak ecology,50,mean pairwise distance,0.0,7.556929455863566e-10,0,np.random.standard_normal
rich ecology,weak ecology,50,sum pairwise distance,0.0,7.556929455863566e-10,0,np.random.standard_normal
rich ecology,weak ecology,50,mean evolutionary distinctiveness,0.0,7.556929455863566e-10,0,np.random.standard_normal
weak ecology,ecology,50,colless-like index,0.0,7.556929455863566e-10,0,np.random.standard_normal
weak ecology,ecology,50,mean pairwise distance,0.0,7.556929455863566e-10,0,np.random.standard_normal
weak ecology,ecology,50,sum pairwise distance,0.0,7.556929455863566e-10,0,np.random.standard_normal
weak ecology,ecology,50,mean evolutionary distinctiveness,0.0,7.556929455863566e-10,0,np.random.standard_normal
ecology,weak ecology,50,colless-like index,2.0,8.53422673646545e-10,2,np.random.exponential
ecology,weak ecology,50,mean pairwise distance,0.0,7.556929455863566e-10,2,np.random.exponential
ecology,weak ecology,50,sum pairwise distance,0.0,7.556929455863566e-10,2,np.random.exponential
ecology,weak ecology,50,mean evolutionary distinctiveness,0.0,7.556929455863566e-10,2,np.random.exponential
plain,ecology,50,colless-like index,29.0,4.253518335686794e-09,2,np.random.exponential
plain,ecology,50,mean pairwise distance,1.0,8.031090907046913e-10,2,np.random.exponential
plain,ecology,50,sum pairwise distance,0.0,7.556929455863566e-10,2,np.random.exponential
plain,ecology,50,mean evolutionary distinctiveness,0.0,7.556929455863566e-10,2,np.random.exponential
plain,rich ecology,50,colless-like index,0.0,7.556929455863566e-10,2,np.random.exponential
plain,rich ecology,50,mean pairwise distance,0.0,7.556929455863566e-10,2,np.random.exponential
plain,rich ecology,50,sum pairwise distance,0.0,7.556929455863566e-10,2,np.random.exponential
plain,rich ecology,50,mean evolutionary distinctiveness,0.0,7.556929455863566e-10,2,np.random.exponential
plain,weak ecology,50,colless-like index,122.0,6.482283621094172e-07,2,np.random.exponential
plain,weak ecology,50,mean pairwise distance,85.0,9.63712911922847e-08,2,np.random.exponential
plain,weak ecology,50,sum pairwise distance,0.0,7.556929455863566e-10,2,np.random.exponential
plain,weak ecology,50,mean evolutionary distinctiveness,0.0,7.556929455863566e-10,2,np.random.exponential
rich ecology,ecology,50,colless-like index,3.0,9.06805772149548e-10,2,np.random.exponential
rich ecology,ecology,50,mean pairwise distance,0.0,7.556929455863566e-10,2,np.random.exponential
rich ecology,ecology,50,sum pairwise distance,0.0,7.556929455863566e-10,2,np.random.exponential
rich ecology,ecology,50,mean evolutionary distinctiveness,0.0,7.556929455863566e-10,2,np.random.exponential
rich ecology,weak ecology,50,colless-like index,0.0,7.556929455863566e-10,2,np.random.exponential
rich ecology,weak ecology,50,mean pairwise distance,0.0,7.556929455863566e-10,2,np.random.exponential
rich ecology,weak ecology,50,sum pairwise distance,0.0,7.556929455863566e-10,2,np.random.exponential
rich ecology,weak ecology,50,mean evolutionary distinctiveness,0.0,7.556929455863566e-10,2,np.random.exponential
plain,ecology,50,colless-like index,2.0,8.53422673646545e-10,2,np.random.standard_normal
plain,ecology,50,mean pairwise distance,0.0,7.556929455863566e-10,2,np.random.standard_normal
plain,ecology,50,sum pairwise distance,0.0,7.556929455863566e-10,2,np.random.standard_normal
plain,ecology,50,mean evolutionary distinctiveness,0.0,7.556929455863566e-10,2,np.random.standard_normal
rich ecology,ecology,50,colless-like index,2.0,8.53422673646545e-10,2,np.random.standard_normal
rich ecology,ecology,50,mean pairwise distance,0.0,7.556929455863566e-10,2,np.random.standard_normal
rich ecology,ecology,50,sum pairwise distance,0.0,7.556929455863566e-10,2,np.random.standard_normal
rich ecology,ecology,50,mean evolutionary distinctiveness,0.0,7.556929455863566e-10,2,np.random.standard_normal
rich ecology,plain,50,colless-like index,0.0,7.556929455863566e-10,2,np.random.standard_normal
rich ecology,plain,50,mean pairwise distance,0.0,7.556929455863566e-10,2,np.random.standard_normal
rich ecology,plain,50,sum pairwise distance,0.0,7.556929455863566e-10,2,np.random.standard_normal
rich ecology,plain,50,mean evolutionary distinctiveness,0.0,7.556929455863566e-10,2,np.random.standard_normal
rich ecology,weak ecology,50,colless-like index,0.0,7.556929455863566e-10,2,np.random.standard_normal
rich ecology,weak ecology,50,mean pairwise distance,0.0,7.556929455863566e-10,2,np.random.standard_normal
rich ecology,weak ecology,50,sum pairwise distance,0.0,7.556929455863566e-10,2,np.random.standard_normal
rich ecology,weak ecology,50,mean evolutionary distinctiveness,0.0,7.556929455863566e-10,2,np.random.standard_normal
weak ecology,ecology,50,colless-like index,0.0,7.556929455863566e-10,2,np.random.standard_normal
weak ecology,ecology,50,mean pairwise distance,0.0,7.556929455863566e-10,2,np.random.standard_normal
weak ecology,ecology,50,sum pairwise distance,0.0,7.556929455863566e-10,2,np.random.standard_normal
weak ecology,ecology,50,mean evolutionary distinctiveness,0.0,7.556929455863566e-10,2,np.random.standard_normal
weak ecology,plain,50,colless-like index,121.0,6.166815818239549e-07,2,np.random.standard_normal
weak ecology,plain,50,mean pairwise distance,115.0,4.5628539597638217e-07,2,np.random.standard_normal
weak ecology,plain,50,sum pairwise distance,0.0,7.556929455863566e-10,2,np.random.standard_normal
weak ecology,plain,50,mean evolutionary distinctiveness,0.0,7.556929455863566e-10,2,np.random.standard_normal
ecology,plain,50,colless-like index,458.0,0.08313770889364551,7,np.random.exponential
ecology,plain,50,mean pairwise distance,456.0,0.0797617584041863,7,np.random.exponential
ecology,plain,50,sum pairwise distance,0.0,7.556929455863566e-10,7,np.random.exponential
ecology,plain,50,mean evolutionary distinctiveness,0.0,7.556929455863566e-10,7,np.random.exponential
ecology,weak ecology,50,colless-like index,178.0,9.178144213292979e-06,7,np.random.exponential
ecology,weak ecology,50,mean pairwise distance,25.0,3.366814035344517e-09,7,np.random.exponential
ecology,weak ecology,50,sum pairwise distance,0.0,7.556929455863566e-10,7,np.random.exponential
ecology,weak ecology,50,mean evolutionary distinctiveness,0.0,7.556929455863566e-10,7,np.random.exponential
rich ecology,ecology,50,colless-like index,103.0,2.473756574659273e-07,7,np.random.exponential
rich ecology,ecology,50,mean pairwise distance,2.0,8.53422673646545e-10,7,np.random.exponential
rich ecology,ecology,50,sum pairwise distance,0.0,7.556929455863566e-10,7,np.random.exponential
rich ecology,ecology,50,mean evolutionary distinctiveness,0.0,7.556929455863566e-10,7,np.random.exponential
rich ecology,plain,50,colless-like index,100.0,2.118385561420393e-07,7,np.random.exponential
rich ecology,plain,50,mean pairwise distance,0.0,7.556929455863566e-10,7,np.random.exponential
rich ecology,plain,50,sum pairwise distance,0.0,7.556929455863566e-10,7,np.random.exponential
rich ecology,plain,50,mean evolutionary distinctiveness,0.0,7.556929455863566e-10,7,np.random.exponential
rich ecology,weak ecology,50,colless-like index,0.0,7.556929455863566e-10,7,np.random.exponential
rich ecology,weak ecology,50,mean pairwise distance,0.0,7.556929455863566e-10,7,np.random.exponential
rich ecology,weak ecology,50,sum pairwise distance,0.0,7.556929455863566e-10,7,np.random.exponential
rich ecology,weak ecology,50,mean evolutionary distinctiveness,0.0,7.556929455863566e-10,7,np.random.exponential
weak ecology,plain,50,colless-like index,43.0,9.530799618618074e-09,7,np.random.exponential
weak ecology,plain,50,mean pairwise distance,0.0,7.556929455863566e-10,7,np.random.exponential
weak ecology,plain,50,sum pairwise distance,0.0,7.556929455863566e-10,7,np.random.exponential
weak ecology,plain,50,mean evolutionary distinctiveness,0.0,7.556929455863566e-10,7,np.random.exponential
ecology,plain,50,colless-like index,600.0,0.7173535621498253,7,np.random.standard_normal
ecology,plain,50,mean pairwise distance,216.0,4.7244338450388704e-05,7,np.random.standard_normal
ecology,plain,50,sum pairwise distance,0.0,7.556929455863566e-10,7,np.random.standard_normal
ecology,plain,50,mean evolutionary distinctiveness,0.0,7.556929455863566e-10,7,np.random.standard_normal
ecology,rich ecology,50,colless-like index,62.0,2.7687864179239132e-08,7,np.random.standard_normal
ecology,rich ecology,50,mean pairwise distance,5.0,1.0235189212610724e-09,7,np.random.standard_normal
ecology,rich ecology,50,sum pairwise distance,0.0,7.556929455863566e-10,7,np.random.standard_normal
ecology,rich ecology,50,mean evolutionary distinctiveness,0.0,7.556929455863566e-10,7,np.random.standard_normal
plain,rich ecology,50,colless-like index,33.0,5.365943352795552e-09,7,np.random.standard_normal
plain,rich ecology,50,mean pairwise distance,0.0,7.556929455863566e-10,7,np.random.standard_normal
plain,rich ecology,50,sum pairwise distance,0.0,7.556929455863566e-10,7,np.random.standard_normal
plain,rich ecology,50,mean evolutionary distinctiveness,0.0,7.556929455863566e-10,7,np.random.standard_normal
weak ecology,ecology,50,colless-like index,193.0,1.7796456927994258e-05,7,np.random.standard_normal
weak ecology,ecology,50,mean pairwise distance,11.0,1.4685612927356851e-09,7,np.random.standard_normal
weak ecology,ecology,50,sum pairwise distance,0.0,7.556929455863566e-10,7,np.random.standard_normal
weak ecology,ecology,50,mean evolutionary distinctiveness,0.0,7.556929455863566e-10,7,np.random.standard_normal
weak ecology,plain,50,colless-like index,139.0,1.4930817614885294e-06,7,np.random.standard_normal
weak ecology,plain,50,mean pairwise distance,40.0,8.02968346961793e-09,7,np.random.standard_normal
weak ecology,plain,50,sum pairwise distance,0.0,7.556929455863566e-10,7,np.random.standard_normal
weak ecology,plain,50,mean evolutionary distinctiveness,0.0,7.556929455863566e-10,7,np.random.standard_normal
weak ecology,rich ecology,50,colless-like index,0.0,7.556929455863566e-10,7,np.random.standard_normal
weak ecology,rich ecology,50,mean pairwise distance,0.0,7.556929455863566e-10,7,np.random.standard_normal
weak ecology,rich ecology,50,sum pairwise distance,0.0,7.556929455863566e-10,7,np.random.standard_normal
weak ecology,rich ecology,50,mean evolutionary distinctiveness,0.0,7.556929455863566e-10,7,np.random.standard_normal
\end{filecontents*}
\csvreader[myTableStyle]{binder/binder/outdata/a=phylostatistics-comparison-between-regimes-allpairs-spatial-nuisance-wilcox+ext=.csv}{}{\csvlinetotablerow}
\end{landscape}

\providecommand{\thecaption}{
  All-pairs two-tailed Wilcoxon signed-rank comparisons of phylometrics between surveyed evolutionary regimes.
  Phylometrics were calculated on perfect-fidelity reference trees.
  All-pairs comparisons within each sensitivity analysis condition are provided.
  Sample size $n=50$ for each population.
  Reported $p$ values are not corrected for multiple comparisons.
  Instead, a Bonferoni-corrected significance threshold of $1.49 \times 10^{-4}$ was used.
}

\begin{landscape}
  \csvstyle{myTableStyle}{
    longtable=c c c c c c c c,
    before reading=\footnotesize,
    table head=\toprule evolutionary regime 1 & evolutionary regime 2 & $n$ & Phylometric & Statistic & $p$ & Epoch & Mutation Distribution\\ \midrule\endhead\bottomrule\\[-1em]
    \caption{%
      \thecaption
    }\endfoot\bottomrule\\[-1em]
    \caption{%
      \thecaption
    }
    \label{tab:phylostatistics-comparison-between-regimes-allpairs-wilcox}\endlastfoot,
    head to column names,
    respect all,
  }


\csvreader[myTableStyle]{binder/binder/outdata/a=phylostatistics-comparison-between-regimes-allpairs-wilcox+ext=.csv}{}{\csvlinetotablerow}
\end{landscape}

\providecommand{\thecaption}{
  Two-tailed Wilcoxon signed-rank comparisons between phylometric distributions for reconstructed trees and corresponding perfect-fidelity reference trees.
  Comparisons were performed separately for each surveyed reconstruction fidelity, evolutionary regime, and phylometric.
  Evolutionary regimes described here all included spatial structure (i.e., island count 1,024).
  Sample size $n=50$ for each population.
  Reported $p$ values are not corrected for multiple comparisons, in order to conservatively detect possible phylometric bias from reconstruction error.
}

\begin{landscape}
  \csvstyle{myTableStyle}{
    longtable=c c c c c c c c,
    before reading=\footnotesize,
    table head=\toprule Reconstruction Fidelity & $n$ & Phylometric & Statistic & $p$ & Evolutionary Regime & Epoch & Mutation Distribution\\ \midrule\endhead\bottomrule\\[-1em]
    \caption{%
      \thecaption
    }\endfoot\bottomrule\\[-1em]
    \caption{%
      \thecaption
    }
    \label{tab:phylostatistics-comparison-between-resolutions-allpairs-wilcox-spatial-nuisance}\endlastfoot,
    head to column names,
    respect all,
  }


\csvreader[myTableStyle]{binder/binder/outdata/a=phylostatistics-comparison-between-resolutions-allpairs-wilcox-spatial-nuisance+ext=.csv}{}{\csvlinetotablerow}
\end{landscape}

\providecommand{\thecaption}{
  Two-tailed Wilcoxon signed-rank comparisons between phylometric distributions for reconstructed trees and corresponding perfect-fidelity reference trees.
  Comparisons were performed separately for each surveyed reconstruction fidelity, evolutionary regime, and phylometric.
  Sample size $n=50$ for each population.
  Reported $p$ values are not corrected for multiple comparisons, in order to conservatively detect possible phylometric bias from reconstruction error.
}

\begin{landscape}
  \csvstyle{myTableStyle}{
    longtable=c c c c c c c c,
    before reading=\footnotesize,
    table head=\toprule Reconstruction Fidelity & $n$ & Phylometric & Statistic & $p$ & Evolutionary Regime & Epoch & Mutation Distribution\\ \midrule\endhead\bottomrule\\[-1em]
    \caption{
      \thecaption
    }\endfoot\bottomrule\\[-1em]\caption{%
      \thecaption
    }
    \label{tab:phylostatistics-comparison-between-resolutions-allpairs-wilcox}\endlastfoot,
    head to column names,
    respect all,
  }


\csvreader[myTableStyle]{binder/binder/outdata/a=phylostatistics-comparison-between-resolutions-allpairs-wilcox+ext=.csv}{}{\csvlinetotablerow}
\end{landscape}

\providecommand{\thecaption}{
  Kruskal-Wallis one-way analysis of variance tests for nonequivalence of distributions for each phylometric among surveyed evolutionary regimes for Avida experiments.
  Phylometrics were calculated on perfect-fidelity reference trees.
  Kruskal-Wallis test results within each sensitivity analysis condition are provided.
  Sample size $n=50$ for each population, with $N=7$ populations (i.e., evolutionary regimes) compared within each sensitivity analysis condition.
}

\csvstyle{myTableStyle}{
  longtable=c c c c c c c c,
  before reading=\footnotesize,
  table head=\toprule $n$ & $N$ & Phylometric & Statistic & $p$ & Epoch & Mutation Distribution\\ \midrule\endhead\bottomrule\\[-1em]
  \caption{%
    \thecaption
  }\endfoot\bottomrule\\[-1em]
  \caption{%
    \thecaption
  }\endlastfoot,
  after reading=
  \label{tab:phylostatistics-comparison-between-regimes-kwallis-avida},
  head to column names,
  respect all,
}

\begin{filecontents*}{binder/binder/outdata/a=phylostatistics-comparison-between-regimes-kwallis+ext=.csv}
n,N,phylometric,statistic,p,epoch,mut_distn
50,7,colless-like index,265.97306373626384,1.5768832105180409e-54,0,np.random.exponential
50,7,mean pairwise distance,254.73968449328436,3.980180105925398e-52,0,np.random.exponential
50,7,sum pairwise distance,339.976537240537,2.187511968608002e-70,0,np.random.exponential
50,7,mean evolutionary distinctiveness,340.1784185592187,1.9798170393294457e-70,0,np.random.exponential
50,7,colless-like index,266.1814271062269,1.4230792052007214e-54,0,np.random.standard_normal
50,7,mean pairwise distance,290.99626666666654,6.944132320420735e-60,0,np.random.standard_normal
50,7,sum pairwise distance,334.96747055357065,2.599197187933233e-69,0,np.random.standard_normal
50,7,mean evolutionary distinctiveness,335.20224664224656,2.314532004690998e-69,0,np.random.standard_normal
50,7,colless-like index,256.6560351648352,1.5496417216869407e-52,2,np.random.exponential
50,7,mean pairwise distance,234.792634920635,7.262807031763435e-48,2,np.random.exponential
50,7,sum pairwise distance,335.0634432234433,2.478827864341954e-69,2,np.random.exponential
50,7,mean evolutionary distinctiveness,336.19878290598285,1.4145918841159894e-69,2,np.random.exponential
50,7,colless-like index,248.2970217338218,9.480601649823749e-51,2,np.random.standard_normal
50,7,mean pairwise distance,279.85718388278383,1.6856785307713092e-57,2,np.random.standard_normal
50,7,sum pairwise distance,339.0042183813818,3.536826551998292e-70,2,np.random.standard_normal
50,7,mean evolutionary distinctiveness,339.39189841269854,2.9202295921897736e-70,2,np.random.standard_normal
50,7,colless-like index,246.32398144078138,2.5026640715271667e-50,7,np.random.exponential
50,7,mean pairwise distance,239.36889670329674,7.656088623612794e-49,7,np.random.exponential
50,7,sum pairwise distance,337.6976374903542,6.745278149728936e-70,7,np.random.exponential
50,7,mean evolutionary distinctiveness,338.9171340659341,3.692343047605072e-70,7,np.random.exponential
50,7,colless-like index,263.5987184371186,5.0775260653086644e-54,7,np.random.standard_normal
50,7,mean pairwise distance,271.7057484737484,9.361542359908896e-56,7,np.random.standard_normal
50,7,sum pairwise distance,340.5841035409037,1.6201709195815552e-70,7,np.random.standard_normal
50,7,mean evolutionary distinctiveness,340.35205470085475,1.817025747699854e-70,7,np.random.standard_normal
\end{filecontents*}
\csvreader[myTableStyle]{binder/binder/outdata/a=phylostatistics-comparison-between-regimes-kwallis+ext=.csv}{}{\csvlinetotablerow}

\providecommand{\thecaption}{
  Kruskal-Wallis one-way analysis of variance tests for nonequivalence of distributions for each phylometric among surveyed evolutionary regimes.
  Phylometrics were calculated on perfect-fidelity reference trees.
  Kruskal-Wallis test results within each sensitivity analysis condition are provided.
  Sample size $n=30$ for each population, with $N=7$ populations (i.e., evolutionary regimes) compared within each sensitivity analysis condition.
}

\csvstyle{myTableStyle}{
  longtable=c c c c c c c c,
  before reading=\footnotesize,
  table head=\toprule $n$ & $N$ & Phylometric & Statistic & $p$ & Epoch & Mutation Distribution\\ \midrule\endhead\bottomrule\\[-1em]
  \caption{%
    \thecaption
  }\endfoot\bottomrule\\[-1em]
  \caption{%
    \thecaption
  }\endlastfoot,
  after reading=
  \label{tab:phylostatistics-comparison-between-regimes-kwallis},
  head to column names,
  respect all,
}

\begin{filecontents*}{binder/binder/avida-individual/outdata/a=phylostatistics-comparison-between-regimes-kwallis+ext=.csv}
n,N,phylometric,statistic,p,epoch,mut_distn
30,4,colless-like index,15.223195592286459,0.0016355134406377874,0,default
30,4,mean pairwise distance,20.933168044077092,0.00010869471996695262,0,default
30,4,sum pairwise distance,56.42110192837464,3.415676098277213e-12,0,default
30,4,mean evolutionary distinctiveness,59.55774104683195,7.306790461882688e-13,0,default
\end{filecontents*}
\csvreader[myTableStyle]{binder/binder/avida-individual/outdata/a=phylostatistics-comparison-between-regimes-kwallis+ext=.csv}{}{\csvlinetotablerow}

\providecommand{\thecaption}{
  Kruskal-Wallis one-way analysis of variance tests for nonequivalence of distributions for each phylometric among surveyed evolutionary regimes with spatial structure (i.
  e., island count 1,024).
  Phylometrics were calculated on perfect-fidelity reference trees.
  Kruskal-Wallis test results within each sensitivity analysis condition are provided.
  Sample size $n=50$ for each population, with $N=4$ populations (i.e., evolutionary regimes) compared within each sensitivity analysis condition.
}

\csvstyle{myTableStyle}{
  longtable=c c c c c c c c,
  before reading=\footnotesize,
  table head=\toprule $n$ & $N$ & Phylometric & Statistic & $p$ & Epoch & Mutation Distribution\\ \midrule\endhead\bottomrule\\[-1em]
  \caption{%
    \thecaption
  }\endfoot\bottomrule\\[-1em]
  \caption{%
    \thecaption
  }
  \label{tab:phylostatistics-comparison-between-regimes-spatial-nuisance-kwallis}\endlastfoot,
  head to column names,
  respect all,
}

\begin{filecontents*}{binder/binder/outdata/a=phylostatistics-comparison-between-regimes-spatial-nuisance-kwallis+ext=.csv}
n,N,phylometric,statistic,p,epoch,mut_distn
50,4,colless-like index,172.71356417910442,3.3023446905012562e-37,0,np.random.exponential
50,4,mean pairwise distance,167.92119402985077,3.576287470189254e-36,0,np.random.exponential
50,4,sum pairwise distance,186.56716417910445,3.3660619042710498e-40,0,np.random.exponential
50,4,mean evolutionary distinctiveness,186.56716417910445,3.3660619042710498e-40,0,np.random.exponential
50,4,colless-like index,168.73007761194026,2.3923069182144284e-36,0,np.random.standard_normal
50,4,mean pairwise distance,169.3564656716419,1.7522406039343323e-36,0,np.random.standard_normal
50,4,sum pairwise distance,186.56716417910445,3.3660619042710498e-40,0,np.random.standard_normal
50,4,mean evolutionary distinctiveness,186.56716417910445,3.3660619042710498e-40,0,np.random.standard_normal
50,4,colless-like index,167.4396298507462,4.5434571162024366e-36,2,np.random.exponential
50,4,mean pairwise distance,175.699247761194,7.484609832071976e-38,2,np.random.exponential
50,4,sum pairwise distance,186.56716417910445,3.3660619042710498e-40,2,np.random.exponential
50,4,mean evolutionary distinctiveness,186.56716417910445,3.3660619042710498e-40,2,np.random.exponential
50,4,colless-like index,173.64602985074634,2.0772845224899563e-37,2,np.random.standard_normal
50,4,mean pairwise distance,175.87354029850746,6.86333054269175e-38,2,np.random.standard_normal
50,4,sum pairwise distance,186.56716417910445,3.3660619042710498e-40,2,np.random.standard_normal
50,4,mean evolutionary distinctiveness,186.56716417910445,3.3660619042710498e-40,2,np.random.standard_normal
50,4,colless-like index,103.02394029850757,3.4769664208602374e-22,7,np.random.exponential
50,4,mean pairwise distance,156.73204776119417,9.2967706561219e-34,7,np.random.exponential
50,4,sum pairwise distance,186.56716417910445,3.3660619042710498e-40,7,np.random.exponential
50,4,mean evolutionary distinctiveness,186.56716417910445,3.3660619042710498e-40,7,np.random.exponential
50,4,colless-like index,108.65875820895519,2.1328778255398126e-23,7,np.random.standard_normal
50,4,mean pairwise distance,154.96925373134331,2.231992718618576e-33,7,np.random.standard_normal
50,4,sum pairwise distance,186.56716417910445,3.3660619042710498e-40,7,np.random.standard_normal
50,4,mean evolutionary distinctiveness,186.56716417910445,3.3660619042710498e-40,7,np.random.standard_normal
\end{filecontents*}
\csvreader[myTableStyle]{binder/binder/outdata/a=phylostatistics-comparison-between-regimes-spatial-nuisance-kwallis+ext=.csv}{}{\csvlinetotablerow}

\providecommand{\thecaption}{
  Kruskal-Wallis one-way analysis of variance tests for nonequivalence of distributions for each phylometric among surveyed evolutionary regimes with spatial structure (i.
  e., island count 1,024) for Avida experiments.
  Phylometrics were calculated on perfect-fidelity reference trees.
  Kruskal-Wallis test results within each sensitivity analysis condition are provided.
  Sample size $n=30$ for each population, with $N=4$ populations (i.e., evolutionary regimes) compared within each sensitivity analysis condition.
}

\csvstyle{myTableStyle}{
  longtable=c c c c c c c c,
  before reading=\footnotesize,
  table head=\toprule $n$ & $N$ & Phylometric & Statistic & $p$ & Epoch & Mutation Distribution\\ \midrule\endhead\bottomrule\\[-1em]
  \caption{%
    \thecaption
  }\endfoot\bottomrule\\[-1em]
  \caption{%
    \thecaption
  }
  \label{tab:phylostatistics-comparison-between-regimes-spatial-nuisance-kwallis-avida}\endlastfoot,
  head to column names,
  respect all,
}

\begin{filecontents*}{binder/binder/avida-individual/outdata/a=phylostatistics-comparison-between-regimes-spatial-nuisance-kwallis+ext=.csv}
n,N,phylometric,statistic,p,epoch,mut_distn
30,4,colless_like_index,15.223195592286459,0.0016355134406377874,0,default
30,4,mean_pairwise_distance,20.933168044077092,0.00010869471996695262,0,default
30,4,sum_distance,56.42110192837464,3.415676098277213e-12,0,default
30,4,mean_evolutionary_distinctiveness,59.55774104683195,7.306790461882688e-13,0,default
\end{filecontents*}
\csvreader[myTableStyle]{binder/binder/avida-individual/outdata/a=phylostatistics-comparison-between-regimes-spatial-nuisance-kwallis+ext=.csv}{}{\csvlinetotablerow}

\begin{figure*}
  \centering
  \begin{subfigure}[b]{\textwidth}
    \includegraphics[width=\textwidth]{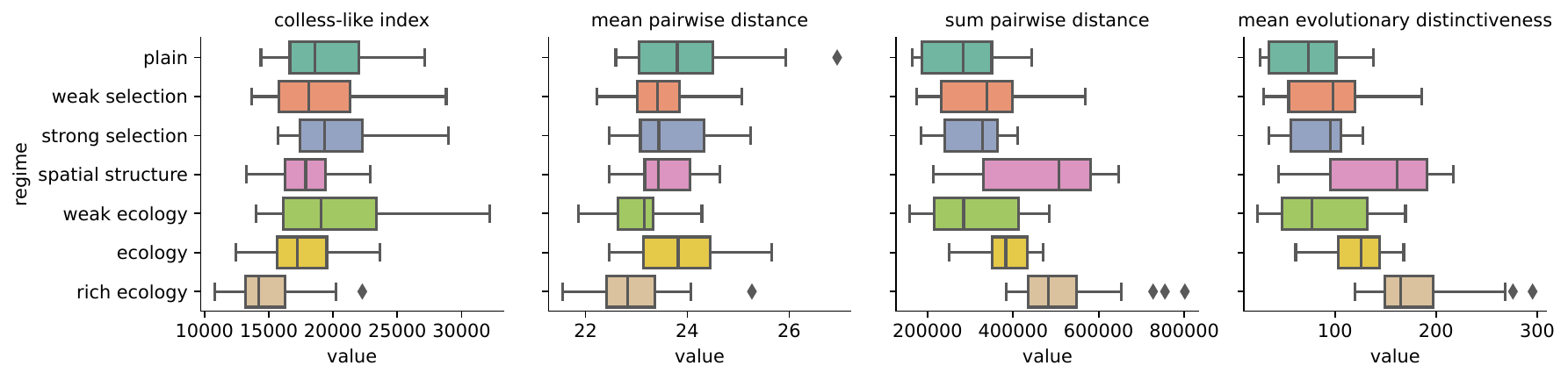}
    \caption{No spatial structure.}
  \end{subfigure}
  \begin{subfigure}[b]{\textwidth}
  \includegraphics[width=\textwidth]{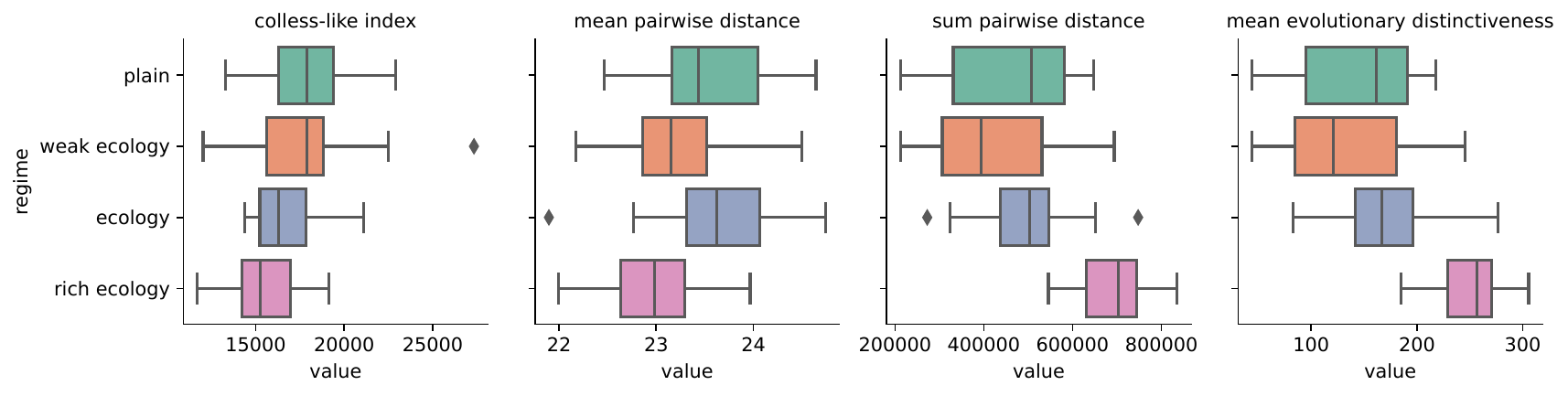}
    \caption{With spatial structure.}
  \end{subfigure}
  \caption{%
    Distribution of tree phylometrics measured with perfect phylogenetic tracking across surveyed evolutionary regimes in Avida.
    Sample size of $n=30$ per distribution.
  }
  \label{fig:perfect-tree-phylometrics-avida}
\end{figure*}

\begin{figure*}
  \centering
    \includegraphics[width=\textwidth]{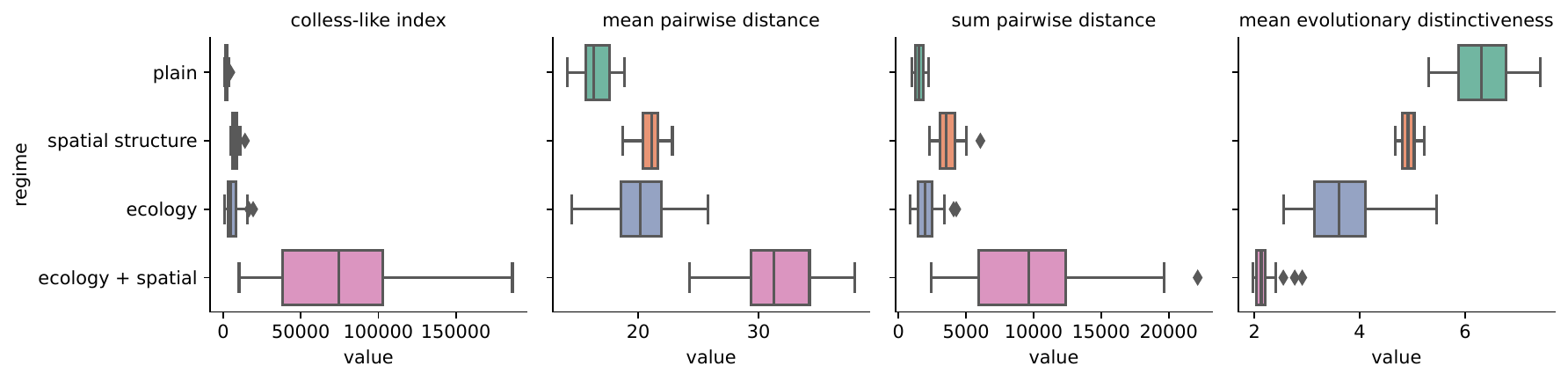}
  \caption{%
    Distribution of tree phylometrics measured with perfect phylogenetic tracking across surveyed evolutionary regimes under Gen3sis model.
    Sample size of $n=30$ per distribution.
  }
  \label{fig:perfect-tree-phylometrics-boxplot-gen3sis}
\end{figure*}

\FloatBarrier

\subsection{Trie-based Phylogenetic Tree Reconstruction Implementation}

This section provides source code and docstrings for trie-based phylogeny reconstruction.

The main reconstruction logic takes place in \texttt{build\_trie\_from\_artifacts} (Listing \ref{lst:build_tree_from_artifacts}).
First, hereditary stratigraphic annotations associated with each extant population member are sorted by the generational depth of their lineages in ascending order.
This ensures that higher-per-generation-density hereditary stratigraphic records of annotations from shorter lineages are encountered first.
(In this work, because generations are synchronous, this step had no effect.)

Next, annotations are inserted into a trie data structure one at a time.
All annotations begin at the root node of the trie, representing a universal common ancestor.
Each subsequent node in the trie denotes a particular ``fingerprint'' value associated with its generation of creation.
(This is referred to in the source code as an ``allele'').
Paths traced from the root node towards the tips therefore represent a particular sequence of allele generation events; organisms will share common allele generation histories for the over the evolutionary interval they share common ancestry.

Each annotation descends the trie along the path corresponding to the sequence of alleles contained within its own record.
The \texttt{GetDeepestCongruousAllele} member function of the \texttt{TrieInnerNode} class implements this process.
When no further consistent alleles are present in the trie, the remaining alleles within the annotation are unrolled to create a new unifurcating branch.
The label corresponding to the annotation is then appended as the leaf node on that branch.
The \texttt{InsertTaxon} member function of the \texttt{TrieInnerNode} class implements this process.
The \texttt{TrieInnerNode} and \texttt{TrieLeafNode} classes are provided in Listings \labelcref{lst:TrieInnerNode,lst:TrieLeafNode}.

The \texttt{build\_tree\_trie} method (Listing \ref{lst:build_tree_trie}) serves as the API entrypoint for the overall tree reconstruction routine.
Default options were used for all reconstructions in this work.
The \texttt{build\_tree\_trie} entrypoint delegates to \texttt{build\_tree\_trie\_ensemble} (Listing \ref{lst:build_tree_trie_ensemble}), which ultimately delegates to \texttt{build\_build\_trie\_from\_artifacts}.
These additional source files are included due to docstring parameter descriptions referenced in other files and information about minor trie postprocessing performed to perform ancestor origin time estimates and convert the trie structure into a standard phylogenetic output format.
Listing \ref{lst:AssignOriginTimeNaiveTriePostprocessor} provides the particular postprocessing implementation used to estimate ancestor origin times in reconstructions performed as part of this work.

The full source context for the files included in listings can be found at \url{https://github.com/mmore500/hstrat}.

\begin{lstlisting}[
  language=Python,
  label={lst:build_tree_from_artifacts},
  caption={\texttt{\_build\_trie\_from\_artifacts.py} source code},
  style=mypython
]
import typing

import opytional as opyt

from ...._auxiliary_lib import (
    HereditaryStratigraphicArtifact,
    argsort,
    give_len,
)
from ._TrieInnerNode import TrieInnerNode


def build_trie_from_artifacts(
    population: typing.Sequence[HereditaryStratigraphicArtifact],
    taxon_labels: typing.Optional[typing.Iterable],
    force_common_ancestry: bool,
    progress_wrap: typing.Callable,
) -> TrieInnerNode:
    """Implementation detail for build_tree_trie_ensemble.

    See `build_tree_trie` for parameter descriptions.
    """
    taxon_labels = opyt.or_value(
        taxon_labels,
        [*map(str, range(len(population)))],
    )

    root = TrieInnerNode(rank=None, differentia=None)

    is_perfectly_synchronous = all(
        artifact.GetNumStrataDeposited()
        == population[0].GetNumStrataDeposited()
        for artifact in population
    )

    sort_order = argsort([x.GetNumStrataDeposited() for x in population])
    sorted_labels = [taxon_labels[i] for i in sort_order]
    sorted_population = [population[i] for i in sort_order]
    for label, artifact in progress_wrap(
        give_len(zip(sorted_labels, sorted_population), len(population))
    ):

        if is_perfectly_synchronous:
            root.InsertTaxon(label, artifact.IterRankDifferentiaZip())
        else:
            res = root.GetDeepestCongruousAlleleOrigination(
                artifact.IterRankDifferentiaZip(copyable=True)
            )
            node, subsequent_allele_genesis_iter = res
            node.InsertTaxon(label, subsequent_allele_genesis_iter)

    return root
\end{lstlisting}

\begin{lstlisting}[
  language=Python,
  label={lst:TrieInnerNode},
  caption={\texttt{\_TrieInnerNode.py} source code},
  style=mypython
]
import copy
import random
import typing

import anytree
import opytional as opyt

from ...._auxiliary_lib import generate_n, render_to_base64url
from ._TrieLeafNode import TrieLeafNode


class TrieInnerNode(anytree.NodeMixin):
    """Inner node of a trie (a.k.a. prefix tree) of differentia
    sequences of hereditary stratigraphic artifacts within a population.
    Each inner node represents the hypothesized origination of a particular
    allele (i.e., differentia at a particular rank).

    Note that more than one` TrieInnerNode` representing a particular
    rank/differentia combination may be present. This is only the case when the
    colliding alleles are known to have independent origins due to previous
    disparities in the phylogenetic record.

    Only `TrieLeafNode` instances will occupy leaf node positions in the trie
    structure. Each `TrieLeafNode` represents a single member of the extant
    artifact population. An instance of `TrieInnerNode` may have child nodes of
    both types.
    """

    _rank: int  # rank of represented allele
    _differentia: int  # differentia of represented allele
    _tiebreaker: int  # random 128-bit integer used to break ties.

    def __init__(
        self: "TrieInnerNode",
        rank: typing.Optional[int] = None,
        differentia: typing.Optional[int] = None,
        parent: typing.Optional["TrieInnerNode"] = None,
    ) -> None:
        """Initialize a `TrieInnerNode` instance.

        Parameters
        ----------
        rank : int, optional
            The rank of the node, or None for shim "ancestor of all geneses" root node.

            Pass None for root node, int otherwise
        differentia : int, optional
            The fingerprint or differentia of the node, or None for shim
            "ancestor of all geneses" root node.
        parent : int, optional
            The parent node, or None for shim "ancestor of all geneses" root
            node.
        """
        self.parent = parent
        self._rank = rank
        self._differentia = differentia
        assert (self._rank is None) == (self._differentia is None)
        self._tiebreaker = random.getrandbits(128)  # uuid standard 128 bits

    def IsAnOriginationOfAllele(
        self: "TrieInnerNode",
        rank: int,
        differentia: int,
    ) -> bool:
        """Checks if the node represents an origination of a given allele."""
        return self._rank == rank and self._differentia == differentia

    def FindOriginationsOfAllele(
        self: "TrieInnerNode",
        rank: int,
        differentia: int,
    ) -> typing.Iterator["TrieInnerNode"]:
        """Searches subtree to find all possible `TrieInnerNodes` representing
        an origination of the specified `rank`-`differentia` allele."""
        assert rank is not None
        assert differentia is not None
        # note: assert handles root case
        assert (self._differentia is None) == (self._rank is None)

        search_stack = [self]
        while search_stack:
            current_node = search_stack.pop()

            # handle root node case
            if current_node._rank is None:
                search_stack.extend(current_node.inner_children)
            # If the current node matches the rank and differentia, yield it
            elif current_node.IsAnOriginationOfAllele(rank, differentia):
                yield current_node
            # If the target allele's rank is deeper than the current node,
            # keep searching
            elif rank > current_node._rank:
                search_stack.extend(current_node.inner_children)

    def GetDeepestCongruousAlleleOrigination(
        self: "TrieInnerNode",
        taxon_allele_iter: typing.Iterator[typing.Tuple[int, int]],
    ) -> (
        typing.Tuple["TrieInnerNode", typing.Iterator[typing.Tuple[int, int]]]
    ):
        """Descends the subtrie to retrieve the deepest prefix consistent with
        the hereditary stratigraphic record of a query taxon.

        In the case where more than one possible originations are found for an
        allele, all will be searched recursively for further matches with the
        next alleles in the focal taxon's hereditary stratigraphic record.
        (This scenario occurs when the focal taxon has discarded a strata that
        differentiates subsequently colliding alleles among earlier taxa).
        Nodes' `_tiebreaker` fields determinstically resolve any ambiguities
        for the deepest consistent prefix that may aries in such cases.

        Parameters
        ----------
        taxon_allele_iter : typing.Iterator[typing.Tuple[int, int]]
            An iterator over the taxon alleles (rank/differentia pairs) in rank-
            ascending order.

            Must be copyable.

        Returns
        -------
        typing.Tuple["TrieInnerNode", typing.Iterator[typing.Tuple[int, int]]]
            A tuple containing the retrieved deepest prefix tree match and an
            iterator over the taxon alleles remaining past the retrieved allele.
        """
        # iterator must be copyable
        assert [*copy.copy(taxon_allele_iter)] == [
            *copy.copy(taxon_allele_iter)
        ]

        node_stack = [self]
        allele_origination_stack = [taxon_allele_iter]
        deepest_origination = self
        deepest_taxon_allele_iter = copy.copy(taxon_allele_iter)
        while True:

            candidate_origination = node_stack.pop()
            taxon_allele_iter = allele_origination_stack.pop()

            # Update the deepest origination if the candidate is deeper
            if (
                opyt.or_value(candidate_origination._rank, -1),
                candidate_origination._tiebreaker,
            ) > (
                opyt.or_value(deepest_origination._rank, -1),
                deepest_origination._tiebreaker,
            ):
                deepest_origination = candidate_origination
                deepest_taxon_allele_iter = copy.copy(taxon_allele_iter)

            # If taxon has susbsequent allele, search its origination
            next_allele = next(taxon_allele_iter, None)
            if next_allele is not None:
                node_stack.extend(
                    candidate_origination.FindOriginationsOfAllele(
                        *next_allele
                    )
                )
                allele_origination_stack.extend(
                    generate_n(
                        lambda: copy.copy(taxon_allele_iter),
                        len(node_stack) - len(allele_origination_stack),
                    )
                )

            assert len(node_stack) == len(allele_origination_stack)

            if not node_stack:  # no more nodes to explore
                break

        return deepest_origination, deepest_taxon_allele_iter

    def InsertTaxon(
        self: "TrieInnerNode",
        taxon_label: str,
        taxon_allele_iter: typing.Iterator[typing.Tuple[int, int]],
    ) -> TrieLeafNode:
        """Inserts a taxon into the trie, ultimately resulting in the creation
        of an additional `TrieLeafNode` leaf and --- if necessary --- a
        unifurcating chain of `TrieInnerNode`'s subtending it.

        Parameters
        ----------
        taxon_label : str
            The label of the taxon to be inserted.
        taxon_allele_iter : typing.Iterator[typing.Tuple[int, int]]
            An iterator over the taxon's remaining allele sequence that has not
            yet been accounted for thus deep into the trie.

        Notes
        -----
        Should only be called on the node corresponding to the deepest congrous
        allele origination found via `GetDeepestCongruousAlleleOrigination`.
        However, may be called on the trie root node when the entire artifact
        population has identical deposition counts. (Because, given a
        deterministic stratum retention strategy, insertion is greatly
        simplified because no colliding allele originations can arise due to
        the impossibility of preceding divergences not retained by the taxon
        being inserted.)

        Returns
        -------
        TrieLeafNode
            The leaf node representing the inserted taxon.
        """
        cur_node = self
        for next_rank, next_differentia in taxon_allele_iter:
            assert next_rank is not None
            assert next_differentia is not None
            for child in cur_node.inner_children:
                # check immediate children for next allele
                #
                # common allele origination trace is for special condition
                # optimization where GetDeepestCongruousAlleleOrigination
                # isn't needed
                if child.IsAnOriginationOfAllele(next_rank, next_differentia):
                    cur_node = child
                    break
            else:
                # if no congruent node exists, create a new TrieInnerNode
                cur_node = TrieInnerNode(
                    next_rank, next_differentia, parent=cur_node
                )

        # create a TrieLeafNode representing the inserted taxon
        return TrieLeafNode(parent=cur_node, taxon_label=taxon_label)

    @property
    def taxon_label(self: "TrieInnerNode") -> str:
        """Programatically-generated unique identifier for internal node,
        intended to translate into a unique label for internal taxa after
        conversion to a phylogenetic reconstruction."""
        if self.parent is None:
            return "Root"
        else:
            # numpy ints cause indexing errors; convert to native int
            # uid field necessary to distinguish colliding allele originations
            return f"""Inner+r={self._rank}+d={
                render_to_base64url(int(self._differentia))
            }+uid={
                render_to_base64url(int(self._tiebreaker))
            }"""

    @property
    def taxon(self: "TrieInnerNode") -> str:
        """Alias for taxon_label."""
        return self.taxon_label

    @property
    def rank(self: "TrieInnerNode") -> int:
        return opyt.or_value(self._rank, 0)

    @property
    def inner_children(
        self: "TrieInnerNode",
    ) -> typing.Iterator["TrieInnerNode"]:
        """Returns iterator over non-leaf child nodes."""
        return filter(
            lambda child_: isinstance(child_, TrieInnerNode),
            self.children,
        )

    @property
    def outer_children(
        self: "TrieInnerNode",
    ) -> typing.Iterator["TrieLeafNode"]:
        """Returns iterator over leaf child nodes."""
        return filter(
            lambda child_: isinstance(child_, TrieLeafNode),
            self.children,
        )

    def __repr__(self: "TrieInnerNode") -> str:
        return f"""(rank {
            self._rank
        }, diff {
            self._differentia
        }) @ {
            render_to_base64url(id(self) % 8192)
        }"""
\end{lstlisting}

\begin{lstlisting}[
  language=Python,
  label={lst:TrieLeafNode},
  caption={\texttt{\_TrieLeafNode.py} source code},
  style=mypython
]
import anytree

from ...._auxiliary_lib import render_to_base64url


class TrieLeafNode(anytree.NodeMixin):
    """A leaf node in a  phylogenetic reconstruction trie.

    Represents an single, specific extant annotation among the population
    subject to phylogenetic inference.

    Parameters
    ----------
    parent : TrieInnerNode
        The parent node of this leaf node.
    taxon_label : str
        The taxon label for this leaf node.

    Attributes
    ----------
    taxon_label : str
        The taxon label for this leaf node.
    taxon : str
        Synonym for taxon_label.
    """

    taxon_label: str

    def __init__(
        self: "TrieLeafNode",
        parent: "TrieInnerNode",
        taxon_label: str,
    ) -> None:
        """Initialize a new TrieLeafNode.

        Parameters
        ----------
        parent : TrieInnerNode
            The parent node of this leaf node.
        taxon_label : str
            The taxon label for this leaf node.
        """
        self.parent = parent
        self.taxon_label = taxon_label

    @property
    def taxon(self: "TrieLeafNode") -> str:
        """Return the taxon label for this leaf node."""
        return self.taxon_label

    @property
    def rank(self: "TrieLeafNode") -> int:
        """Return the origin time for this leaf node."""
        return self.parent.rank

    def __repr__(self: "TrieLeafNode") -> str:
        """Return a string representation of this leaf node."""
        return f"""{self.taxon_label} @ {
            render_to_base64url(id(self) % 8192)
        }"""
\end{lstlisting}

\begin{lstlisting}[
  language=Python,
  label={lst:build_tree_trie},
  caption={\texttt{\_build\_tree\_trie.py} source code},
  style=mypython
]
import typing

from iterpop import iterpop as ip
import pandas as pd

from . import trie_postprocess
from ..._auxiliary_lib import (
    HereditaryStratigraphicArtifact,
    alifestd_make_empty,
)
from ._build_tree_trie_ensemble import build_tree_trie_ensemble


def build_tree_trie(
    population: typing.Sequence[HereditaryStratigraphicArtifact],
    taxon_labels: typing.Optional[typing.Iterable] = None,
    force_common_ancestry: bool = False,
    progress_wrap: typing.Callable = lambda x: x,
    seed: typing.Optional[int] = 1,
    bias_adjustment: typing.Union[str, object, None] = None,
) -> pd.DataFrame:
    """Estimate the phylogenetic history among hereditary stratigraphic
    columns by building a trie (a.k.a. prefix tree) of the differentia
    sequences of hereditary stratigraphic artifacts within a population.

    Exhibits time complexity at most `O(nlog(n))` for population size `n`.

    Parameters
    ----------
    population: Sequence[HereditaryStratigraphicArtifact]
        Hereditary stratigraphic columns corresponding to extant population members.

        Each member of population will correspond to a unique leaf node in the
        reconstructed tree.
    taxon_labels: Optional[Iterable], optional
        How should leaf nodes representing extant hereditary stratigraphic
        columns be named?

        Label order should correspond to the order of corresponding hereditary
        stratigraphic columns within `population`. If None, taxons will be
        named according to their numerical index.
    force_common_ancestry: bool, default False
        How should columns that definively share no common ancestry be handled?

        If set to True, treat columns with no common ancestry as if they
        shared a common ancestor immediately before the genesis of the
        lineages. If set to False, columns within `population` that
        definitively do not share common ancestry will raise a ValueError.
    progress_wrap : Callable, default identity function
        Pass tqdm or equivalent to display progress bars.
    seed : int, default 1
        Controls tiebreaking decisions in the algorithm.

        Pass an int for reproducible output across multiple function calls. The
        default value, 1, ensures reproducible output. Pass None to use global
        RNG context.
    bias_adjustment : "sample_ancestral_rollbacks" or prior, optional
        How should bias toward overestimation of relatedness due to differentia
        collisions be corrected for?

        If "sample_ancestral_rollbacks", the trie topology will be adjusted as
        if the expected number of collisions had occured. Targets for
        "unzipping" to reverse the effect of a speculated collision are
        chosen randomly from within the tree. See
        `SampleAncestralRollbacksTriePostprocessor` for details.

        If a prior functor is passed, the origin time for each trie node will
        be calculated as the expected origin time over the distribution of
        possible differentia collisions. Correction recursively takes into
        account the possibility of multiple collisions. See
        `hstrat.phylogenetic_inference.priors` for available prior
        distributions. A custom prior distribution may also be supplied. See
        `AssignOriginTimeExpectedValueTriePostprocessor` for details.

        If a prior functor is passed, correction for guaranteed-spurious
        collision between most-recent strata will also be performed. See
        `PeelBackConjoinedLeavesTriePostprocessor` for details.

        If None, no correction will be performed. The origin time for each trie
        node will be assigned using a naive strategy, calculated as the average
        of the node's rank and the minimum rank among its children. See
        `AssignOriginTimeNaiveTriePostprocessor` for details.

    Returns
    -------
    pd.DataFrame
        The reconstructed phylogenetic tree in alife standard format.

    Notes
    -----
    Unifurcations in the reconstructed tree are collapsed.

    However, polytomies are not resolved. In addition to any true polytomies,
    ancestry sequences that cannot be resolved due to missing information
    appear as polytomies in the generated reconstruction. Therefore, polytomies
    are generally overrepresented in reconstructions, especially when low
    hereditary stratigraphic resolution is available. If overestimation of
    polytomies is problematic, external tools can be used to decompose
    polytomies into arbitrarily-arranged bifurcations.
    """

    # for simplicity, return early for this special case
    if len(population) == 0:
        return alifestd_make_empty()

    if bias_adjustment is None:
        trie_postprocessor = (
            trie_postprocess.AssignOriginTimeNaiveTriePostprocessor()
        )
    elif (
        isinstance(bias_adjustment, str)
        and bias_adjustment == "sample_ancestral_rollbacks"
    ):

        trie_postprocessor = trie_postprocess.CompoundTriePostprocessor(
            postprocessors=[
                trie_postprocess.SampleAncestralRollbacksTriePostprocessor(
                    seed=1
                ),
                trie_postprocess.AssignOriginTimeNaiveTriePostprocessor(),
            ],
        )

    else:
        trie_postprocessor = trie_postprocess.CompoundTriePostprocessor(
            postprocessors=[
                trie_postprocess.PeelBackConjoinedLeavesTriePostprocessor(),
                trie_postprocess.AssignOriginTimeExpectedValueTriePostprocessor(
                    prior=bias_adjustment,
                ),
            ],
        )

    return ip.popsingleton(
        build_tree_trie_ensemble(
            population=population,
            taxon_labels=taxon_labels,
            force_common_ancestry=force_common_ancestry,
            progress_wrap=progress_wrap,
            seed=seed,
            trie_postprocessors=[trie_postprocessor],
        )
    )
\end{lstlisting}

\begin{lstlisting}[
  language=Python,
  label={lst:build_tree_trie_ensemble},
  caption={\texttt{\_build\_tree\_trie\_ensemble.py} source code},
  style=mypython
]
import contextlib
import typing

from iterpop import iterpop as ip
import opytional as opyt
import pandas as pd

from ..._auxiliary_lib import (
    HereditaryStratigraphicArtifact,
    RngStateContext,
    alifestd_collapse_unifurcations,
    alifestd_make_empty,
    anytree_tree_to_alife_dataframe,
    flag_last,
)
from ...juxtaposition import calc_probability_differentia_collision_between
from ._impl import TrieInnerNode, build_trie_from_artifacts


def _finalize_trie(trie: TrieInnerNode) -> pd.DataFrame:
    "Convert to alifestd dataframe and collapse unifurcations."
    return alifestd_collapse_unifurcations(
        anytree_tree_to_alife_dataframe(trie), mutate=True
    )


def _build_tree_trie_ensemble(
    population: typing.Sequence[HereditaryStratigraphicArtifact],
    trie_postprocessors: typing.Iterable[typing.Callable],
    taxon_labels: typing.Optional[typing.Iterable],
    force_common_ancestry: bool,
    progress_wrap: typing.Callable,
) -> typing.List[pd.DataFrame]:
    """Implementation detail for build_tree_trie_ensemble.

    See `build_tree_trie_ensemble` for parameter descriptions.
    """
    # for simplicity, return early for this special case
    if len(population) == 0:
        return alifestd_make_empty()

    root = build_trie_from_artifacts(
        population=population,
        taxon_labels=taxon_labels,
        force_common_ancestry=force_common_ancestry,
        progress_wrap=progress_wrap,
    )
    if not force_common_ancestry:
        try:
            root = ip.popsingleton(root.children)
            root.parent = None
        except:
            raise ValueError

    p_differentia_collision = calc_probability_differentia_collision_between(
        population[0], population[0]
    )

    res = []
    for is_last, postprocessor in flag_last(trie_postprocessors):
        res.append(
            _finalize_trie(
                postprocessor(
                    root,
                    p_differentia_collision=p_differentia_collision,
                    mutate=is_last,
                    progress_wrap=progress_wrap,
                )
            )
        )

    return res


def build_tree_trie_ensemble(
    population: typing.Sequence[HereditaryStratigraphicArtifact],
    trie_postprocessors: typing.Sequence[typing.Callable],
    taxon_labels: typing.Optional[typing.Iterable] = None,
    force_common_ancestry: bool = False,
    progress_wrap: typing.Callable = lambda x: x,
    seed: typing.Optional[int] = 1,
) -> typing.List[pd.DataFrame]:
    """Estimate the phylogenetic history among hereditary stratigraphic
    columns by building a trie (a.k.a. prefix tree) of the differentia
    sequences of hereditary stratigraphic artifacts within a population.

    Returns phylogeny reconstruction outcomes from alternate postprocessing
    schemes applied between trie contruction and conversion to an alife
    standard data frame, including ancestor taxon origin time estimation.
    Because the underlying pre-postprocess trie is only constructed once, this
    method allows for efficient comparison of potprocessing schemes.

    Unless comparing alternate postprocessing schemes or applying a custom
    postprocessing option, end users should likely prefer `build_tree_trie`.
    This interface applies a single postprocess, set to a generally-appropriate
    default with a few other curated postprocesses specifiable by optional
    argument.

    Parameters
    ----------
    population: Sequence[HereditaryStratigraphicArtifact]
        Hereditary stratigraphic columns corresponding to extant population
        members.

        Each member of population will correspond to a unique leaf node in the
        reconstructed tree.
    trie_postprocessors: Iterable[Callable]
        Tree postprocess functors.

        Must take `trie` of type `TrieInnerNode`, `p_differentia_collision` of
        type `float`, `mutate` of type `bool`, and `progress_wrap` of type
        `Callable` params. Must returned postprocessed trie (type
        `TrieInnerNode`). Several

        Each postprocess will be called indpendently to produce a returned
        postprocessed variant. Use `CompoundTriePostprocessor` to chain
        postprocess functors that should be applied in succession.
    taxon_labels: Optional[Iterable], optional
        How should leaf nodes representing extant hereditary stratigraphic
        columns be named?

        Label order should correspond to the order of corresponding hereditary
        stratigraphic columns within `population`. If None, taxons will be
        named according to their numerical index.
    force_common_ancestry: bool, default False
        How should columns that definively share no common ancestry be handled?

        If set to True, treat columns with no common ancestry as if they
        shared a common ancestor immediately before the genesis of the
        lineages. If set to False, columns within `population` that
        definitively do not share common ancestry will raise a ValueError.
    progress_wrap : Callable, default identity function
        Wrapper applied around generation iterator and row generator for final
        phylogeny compilation process.

        Pass tqdm or equivalent to display progress bars.
    seed: int, default 1
        Controls tiebreaking decisions in the algorithm.

        Pass an int for reproducible output across multiple function calls. The
        default value, 1, ensures reproducible output. Pass None to use
        existing RNG context directly.

    Returns
    -------
    typing.List[pd.DataFrame]
        Reconstructed phylogenetic trees with each postprocessor applied, in
        alife standard format.
    """
    with opyt.apply_if_or_value(
        seed,
        RngStateContext,
        contextlib.nullcontext(),
    ):
        return _build_tree_trie_ensemble(
            population=population,
            trie_postprocessors=trie_postprocessors,
            taxon_labels=taxon_labels,
            force_common_ancestry=force_common_ancestry,
            progress_wrap=progress_wrap,
        )
\end{lstlisting}

\begin{lstlisting}[
  language=Python,
  label={lst:AssignOriginTimeNaiveTriePostprocessor},
  caption={\texttt{\_AssignOriginTimeNaiveTriePostprocessor.py} source code},
  style=mypython
]
import typing

from ...._auxiliary_lib import (
    AnyTreeFastPreOrderIter,
    anytree_iterative_deepcopy,
)
from ...priors import ArbitraryPrior
from .._impl import TrieInnerNode, TrieLeafNode


class AssignOriginTimeNaiveTriePostprocessor:
    """Functor to assign origin time property to trie nodescalculated as the
    average of the node's rank and the minimum rank among its children.

    Optionally calculates origin time expected value over this interval for
    a provided prior distribution.
    """

    _assigned_property: str  # property name for assigned origin time
    _prior: object  # prior expectation for ancestor origin times

    def __init__(
        self: "AssignOriginTimeNaiveTriePostprocessor",
        prior: object = ArbitraryPrior(),  # ok as kwarg; immutable
        assigned_property: str = "origin_time",
    ) -> None:
        """Initialize functor instance.

        Parameters
        ----------
        prior : object, default ArbitraryPrior()
            Prior distribution of ancestor origin times.

            Used to calculate interval means.
        assigned_property : str, default "origin_time"
            The property name for the assigned origin time.
        """
        self._assigned_property = assigned_property
        self._prior = prior

    def __call__(
        self: "AssignOriginTimeNaiveTriePostprocessor",
        trie: TrieInnerNode,
        p_differentia_collision: float,
        mutate: bool = False,
        progress_wrap: typing.Callable = lambda x: x,
    ) -> TrieInnerNode:
        """Assign origin times to trie nodes.

        Parameters
        ----------
        trie : TrieInnerNode
            The input trie to be postprocessed.
        p_differentia_collision : float
            Probability of a randomly-generated differentia matching an
            existing differentia.

            Not used in the current implementation.
        mutate : bool, default False
            Are side effects on the input argument `trie` allowed?
        progress_wrap : typing.Callable, optional
            Pass tqdm or equivalent to report progress.

        Returns
        -------
        TrieInnerNode
            The postprocessed trie with assigned origin times.
        """
        if not mutate:
            trie = anytree_iterative_deepcopy(
                trie, progress_wrap=progress_wrap
            )

        for node in progress_wrap(AnyTreeFastPreOrderIter(trie)):
            if node.is_leaf:
                setattr(node, self._assigned_property, node.rank)
                assert isinstance(node, TrieLeafNode)
            elif node.parent is None:
                setattr(node, self._assigned_property, 0)
            else:
                interval_mean = self._prior.CalcIntervalConditionedMean(
                    node.rank,
                    min(
                        (
                            child.rank
                            for child in node.children
                            if not child.is_leaf
                        ),
                        default=node.rank + 1,
                    ),  # endpoint is exclusive
                )
                setattr(
                    node,
                    self._assigned_property,
                    min(
                        interval_mean,
                        min(
                            (child.rank for child in node.children),
                            default=interval_mean,
                        ),
                    ),
                )

            assert hasattr(node, self._assigned_property)

        return trie
\end{lstlisting}

\end{bibunit}

\end{document}